\documentclass[ aip, jmp, reprint,showkeys,showpacs]{revtex4}
\usepackage{amsfonts}
\usepackage{url}
\usepackage{amssymb}
\usepackage{amsmath}
\usepackage{graphicx}
\usepackage{dcolumn}
\usepackage{bm}

\begin{document}

\title[]{\textbf{The Lagrangian dynamics of thermal tracer particles in
Navier-Stokes fluids}}
\author{Massimo Tessarotto}
\affiliation{Department of Mathematics and Informatics, University of Trieste, Italy}
\affiliation{Consortium for Magnetofluid Dynamics, University of Trieste, Italy}
\author{Claudio Asci}
\affiliation{Department of Mathematics and Informatics, University of Trieste, Italy}
\affiliation{Consortium for Magnetofluid Dynamics, University of Trieste, Italy}
\author{Claudio Cremaschini}
\affiliation{International School for Advanced Studies (SISSA) and INFN, Trieste, Italy}
\affiliation{Consortium for Magnetofluid Dynamics, University of Trieste, Italy}
\author{Alessandro Soranzo}
\affiliation{Department of Mathematics and Informatics, University of Trieste, Italy}
\affiliation{Consortium for Magnetofluid Dynamics, University of Trieste, Italy}
\author{Marco Tessarotto}
\affiliation{Civil Protection Agency, Regione Friuli Venezia-Giulia, Palmanova (Udine),
Italy}
\affiliation{Consortium for Magnetofluid Dynamics, University of Trieste, Italy}
\author{Gino Tironi}
\affiliation{Department of Mathematics and Informatics, University of Trieste, Italy}
\affiliation{Consortium for Magnetofluid Dynamics, University of Trieste, Italy}
\date{\today }

\begin{abstract}
A basic issue for Navier-Stokes (NS) fluids is their
characterization in terms of the so-called NS phase-space
classical dynamical system, which provides a mathematical model
for the description of the dynamics of infinitesimal (or
i\textit{deal}) tracer particles in these fluids. The goal of this
paper is to analyze the properties of a particular subset of
solutions of the NS dynamical system, denoted as \textit{thermal
tracer particles} (TTPs), whose states are determined uniquely by
the NS fluid fields. Applications concerning both deterministic
and stochastic NS fluids are pointed out. In particular, in both
cases it is shown that in terms of the ensemble of TTPs a
statistical description of NS fluids can be formulated. In the
case of stochastic fluids this feature permits to uniquely
establish the corresponding Langevin and Fokker-Planck dynamics.
Finally, the relationship with the customary statistical treatment
of hydrodynamic turbulence (HT) is analyzed and a solution to the
closure problem for the statistical description of HT is proposed.
\end{abstract}

\maketitle

\bigskip



\section{INTRODUCTION}

A fundamental aspect of theoretical fluid dynamics is represented by the
discovery of the \emph{thermal tracer particles }(TTPs) for Navier-Stokes
(NS) fluids recently reported (see Ref.\cite{Tessarotto2011}). The latter
represent a suitable subset of the so-called \emph{ideal tracer particles }%
(ITPs \cite{Tessarotto2009}) and are defined in such a way that their states
are uniquely dependent, in a sense to be specified below, on the local state
of the fluid. A basic implication of the result is that an appropriate
statistical ensemble of TTPs should reproduce exactly the dynamics of the
fluid. In other words, it should be possible to determine the fluid fields
characterizing the fluid state by means of suitable statistical averages on
the ensemble of TTPs, and in particular performed so that they satisfy \emph{%
identically} a required set of fluid equations. The conclusion is expected
to apply, in principle, to arbitrary NS fluids described as mesoscopic,
i.e., continuous fluids, which can be either viscous or inviscid,
compressible or incompressible, thermal or isothermal, isentropic or
non-isentropic.

We shall assume, for this purpose, that the state of these fluids is
represented by an ensemble of observables $\left\{ Z(\mathbf{r},t)\right\} $
$\equiv \left\{ Z_{i}(\mathbf{r},t),i=1,..,n\right\} $ (with $n$ an integer $%
\geq 1$), i.e., \emph{fluid fields,} which can be \emph{unambiguously
prescribed }as continuous and suitably smooth functions, respectively, in $%
\overline{\Omega }\times I$ and in the open set $\Omega \times I$, with $%
\Omega \subset
\mathbb{R}
^{3}$ and $I\equiv
\mathbb{R}
$ being the configuration space and time axis respectively. We
intend to show that, as a basic consequence, the Newtonian state
of each TTP,
namely $\mathbf{x}(t)\equiv \mathbf{x}=(\mathbf{r},\mathbf{v})$, with $%
\mathbf{r}$ and $\mathbf{v}$ denoting respectively the particle position and
velocity, is advanced in time in terms of a suitable acceleration field $%
\mathbf{F=F}(\mathbf{x},t)$ which can be defined in such a way to depend
\emph{only on the state of the same particle} (\emph{mean-field} \emph{%
acceleration}). Remarkably, it is found that $\mathbf{F}$ can be \emph{%
uniquely} prescribed in such a way to determine \emph{self-consistently} the
time evolution of the complete set of fluid equations characterizing the
fluid. This implies that TTPs must reproduce \emph{exactly} the dynamics of
the fluid. In other words, by means of appropriate statistical averages on
the ensemble of the TTPs, it is possible to uniquely determine the
time-evolution of the fluid state, in such a way that it \emph{satisfies
identically} the required set of fluid equations.

\subsection{Lagrangian dynamics of ideal tracer particles}

A key aspect of fluid dynamics is the proper definition of the \emph{%
phase-space Lagrangian dynamics} for continuous fluid systems, whereby
possibly \emph{all the fluid fields} characterizing the actual fluid state $%
\left\{ Z(\mathbf{r},t)\right\} $ can be identified with suitable
statistical averages on appropriate ensembles of (fictitious) particles.
Thus, for example, in the case of an incompressible NS fluid, this would
require to represent both the fluid velocity $\mathbf{V}(\mathbf{r},t)$ and
the fluid pressure $p\mathbf{(r},t)$ in terms of suitable statistical
averages of an appropriate probability density. This goal can be realized by
means of the inverse kinetic theory (IKT) developed by Ellero and Tessarotto
(see Refs.\cite{Ellero2000,Tessarotto2004,Ellero2005}). This refers, in
particular, to the phase-space\emph{\ }dynamics of \emph{ideal tracer
particles}, namely rigid\textbf{\ }extended classical particles immersed in
the fluid, all having the same support and infinitesimal size such that
during their motion they \emph{do} \emph{not mutually interact} and \emph{do
not perturb} \emph{the state of the fluid}. Depending on their inertial mass
$m_{P}$, ITPs can belong to different species of particles; thus, in
general, their mass can differ from that of the corresponding displaced
fluid element $m_{F}.$ On the other hand, ITPs carrying the mass $%
m_{P}\equiv m_{F}$ will be denoted as the \emph{NS ideal tracer particles} (%
\emph{NS-ITPs})\emph{. }In the following, in order to characterize the
Lagrangian dynamics of NS fluids, ITPs will be identified only with NS-ITPs.
In this framework, it follows that ITPs can undergo, by assumption, only
\textquotedblleft unary\textquotedblright\ interactions with external
force-fields and with the continuum fluid. Namely, in both cases they are
subject only to the action of a continuum mean-field acceleration which
depends only on the local state of each particle. As a consequence, ITPs can
be treated as Newtonian point-like particles characterized by a Newtonian
state $\mathbf{x}=(\mathbf{r},\mathbf{v})$ spanning the phase-space $\Gamma
\equiv \Omega \times U,$ with the position $\mathbf{r}$ and the \emph{%
kinetic velocity} $\mathbf{v}$ belonging respectively to the configuration
space of the fluid $\Omega $ (in the following to be identified with a
bounded subset of $%
\mathbb{R}
^{3}$) and the velocity space $U\equiv
\mathbb{R}
^{3}$.

\subsection{The Navier-Stokes dynamical system}

By assumption, the state $\mathbf{x}$ of a generic ITP advances in time by
means of a Newtonian classical dynamical system (DS) defined in terms of the
vector field $\mathbf{X}(\mathbf{x},t)\equiv \left\{ \mathbf{v},\frac{1}{%
m_{P}}\mathbf{\mathbf{K}}\equiv \mathbf{\mathbf{F}}\right\} ,$ with $\frac{1%
}{m_{P}}\mathbf{K}\equiv \mathbf{F}$ a suitable mean-field acceleration.
This is identified with the flow ($T_{t_{o},t}$) generated by the initial
value problem associated to the deterministic equations of motion \ (\emph{%
Newton's equations})%
\begin{equation}
\left\{
\begin{array}{c}
\frac{d}{dt}\mathbf{x}=\mathbf{X}(\mathbf{x},t), \\
\mathbf{x}(t_{o})=\mathbf{x}_{o}.%
\end{array}%
\right.  \label{eqq-1}
\end{equation}%
Such flow is referred to as \emph{Navier-Stokes dynamical system} (NS--DS)
and is a homeomorphism in $\Gamma $ with existence domain $\Gamma \times I,$
of the type
\begin{equation}
T_{t_{o},t}:\mathbf{x}_{o}\rightarrow \mathbf{x}(t)=T_{t_{o},t}\mathbf{x}%
_{o},  \label{DYN-S}
\end{equation}%
with $t\in I\subseteq
\mathbb{R}
$ and $T_{t_{o},t}$ being a measure-preserving evolution operator associated
to $\mathbf{X}(\mathbf{x},t)$. \emph{Thus, by definition, the NS-DS is
uniquely prescribed by the couple }$\left\{ \mathbf{x},\mathbf{X}(\mathbf{x}%
,t)\right\} $, with $\mathbf{x}(t)\equiv \mathbf{x}=(\mathbf{r},\mathbf{v})$
to be identified with the instantaneous Newtonian state of a generic ITP.

\subsection{The relative-dynamics NS-DS}

The state of a TTP can be equivalently represented in terms of the \emph{%
relative-dynamics }Newtonian state $\mathbf{y}=(\mathbf{r},\mathbf{u}),$
with $\mathbf{u}(t)=\mathbf{v}(t)-\mathbf{V}(\mathbf{r}(t),t)$ denoting the
\emph{relative kinetic velocity }defined with respect to the local fluid
velocity $\mathbf{V}(\mathbf{\mathbf{r}},t).$ As a consequence, by
introducing the local phase-space diffeomorphism $\mathbf{x}=(\mathbf{r},%
\mathbf{v})\rightarrow \mathbf{y}=(\mathbf{r},\mathbf{u})$, the dynamical
system (\ref{DYN-S}) can be cast in terms of $\mathbf{y}=(\mathbf{r},\mathbf{%
u})$. Then, (\ref{DYN-S}) can be equivalently represented by the
homeomorphism in $\Gamma $:%
\begin{equation}
T_{t_{o},t}^{(RD)}:\mathbf{y}_{o}\rightarrow \mathbf{y}(t)=T_{t_{o},t}^{(RD)}%
\mathbf{y}_{o},  \label{RD-NS-DS}
\end{equation}%
to be identified with the \emph{relative-dynamics NS dynamical system} (%
\emph{RD-NS-DS}), $T_{t_{o},t}^{(RD)}$ being the flow generated by the
initial-value problem%
\begin{equation}
\left\{
\begin{array}{c}
\frac{d\mathbf{r}}{dt}=\mathbf{u}+\mathbf{V}(\mathbf{r},t), \\
\frac{d\mathbf{u}}{dt}=\mathbf{F}_{u}(\mathbf{x},t), \\
\mathbf{y}(t_{o})=\mathbf{y}_{o}.%
\end{array}%
\right.  \label{eqq-2}
\end{equation}%
Here, for a prescribed form of the mean-field $\mathbf{F}(\mathbf{x},t)$
(see below), $\mathbf{F}_{u}(\mathbf{x},t)$ denotes the \emph{kinetic
relative acceleration} which is defined as%
\begin{equation}
\mathbf{F}_{u}(\mathbf{x},t)\equiv \mathbf{F}(\mathbf{x},t)-\mathbf{F}_{H}(%
\mathbf{r},t)-\mathbf{u}\cdot \nabla \mathbf{V}(\mathbf{\mathbf{r}},t),
\label{kinetic acceleration}
\end{equation}%
with $\mathbf{F}_{H}(\mathbf{r},t)$ being the \emph{NS} \emph{fluid
acceleration} defined by Eq.(\ref{fluid-acceleration}) (see Appendix A). To
establish on rigorous grounds the connection with the corresponding fluid
description, the vector field $\mathbf{F}(\mathbf{x},t)$ must be suitably
determined. For definiteness, we shall consider here the case of a
Navier-Stokes thermofluid described either by the compressible or
incompressible Navier-Stokes-Fourier equations, requiring that the fluid
fields $\left\{ Z(\mathbf{r},t)\right\} $ are strong solutions either of the
compressible or incompressible Navier-Stokes-Fourier problems (CNSFE or
INSFE problems respectively; see Appendix A). In the case of CNSFE the state
of the fluid is defined by the ensemble of smooth fluid fields%
\begin{equation}
\left\{ Z(\mathbf{r},t)\right\} =\left\{ \rho (\mathbf{r},t),\mathbf{V}(%
\mathbf{r},t),p(\mathbf{r},t),T(\mathbf{r},t),S_{T}(t)\right\} ,
\label{fluid fields}
\end{equation}%
with $\rho (\mathbf{r},t)\geq 0,\mathbf{V}(\mathbf{r},t),$ $p(\mathbf{r}%
,t)\geq 0,T(\mathbf{r},t)\geq 0$ and $S_{T}(t)$ denoting the \emph{fluid}
\emph{mass} \emph{density}, the \emph{fluid velocity}, the \emph{fluid}
\emph{scalar pressure,} the\emph{\ fluid temperature} and the \emph{global
thermodynamic entropy }respectively. In particular, by introducing an
arbitrary reference mass $m>0,$ for example to be identified with the
particle mass $m_{F}$ defined above, the notion of \emph{fluid number
density }$n(\mathbf{r},t)\equiv \frac{1}{m}\rho (\mathbf{r},t)$ can be
introduced. As an alternative, in the following the set $\left\{ Z(\mathbf{r}%
,t)\right\} $ will be replaced by the \emph{reduced set of fluid fields}%
\begin{equation}
\left\{ Z_{1}(\mathbf{r},t)\right\} =\left\{ \rho (\mathbf{r},t),\mathbf{V}(%
\mathbf{r},t),p_{1}(\mathbf{r},t),S_{T}(t)\right\} .
\label{REDUCED-FLUID FIELDS}
\end{equation}%
\emph{\ }Here $p_{1}(\mathbf{r},t)$ denotes the \emph{kinetic pressure}$,$
i.e., a \emph{strictly positive scalar observable }defined as
\begin{equation}
p_{1}(\mathbf{r},t)=p_{0}(t)+p(\mathbf{r},t)-\phi (\mathbf{r},t)+n(\mathbf{r}%
,t)T(\mathbf{r},t),  \label{kinetic pressure}
\end{equation}%
with $p_{0}(t)$ being the \emph{pseudo-pressure} and $\phi (\mathbf{r},t)$
the potential associated to the conservative part of the volume force (see
Appendix A)\footnote{%
We notice that, in principle, alternative possible definitions of $p_{1}$
might be obtained replacing the contribution $nT$ on the r.h.s. of Eq.(\ref%
{kinetic pressure}) (thermal energy density) either with $n(\mathbf{r}%
,t)\varepsilon (\mathbf{r},t)$ (internal energy density) or $n(\mathbf{r}%
,t)T(\mathbf{r},t)s(\mathbf{r},t)$ (thermodynamic energy density), where $%
\varepsilon (\mathbf{r},t)$ and $s(\mathbf{r},t)$ are respectively the
internal energy and the local entropy (see Appendix A). However, contrary to
$T(\mathbf{r},t),$ both $\varepsilon (\mathbf{r},t)$ and $s(\mathbf{r},t)$
are not observables.}. For later use we introduce here also the notion of
\emph{specific kinetic pressure} $\widehat{p}_{1}(\mathbf{r},t)$\emph{:}%
\begin{equation}
\widehat{p}_{1}(\mathbf{r},t)\equiv p_{1}(\mathbf{r},t)/\rho (\mathbf{r},t).
\label{specific kinetic pressure}
\end{equation}%
In particular, we shall assume that in $\overline{\Omega }\times I$ both $%
\phi (\mathbf{r},t)$, $p_{0}(t)$ and $S_{T}(t)$ are uniquely defined. In
particular they are such that:

\begin{itemize}
\item $\phi (\mathbf{r},t)$ is bounded and such that, for $\mathbf{r}\equiv
\mathbf{R}_{o}$
\begin{equation}
\left. \phi (\mathbf{R}_{o},t)\right\vert =0,  \label{INITIAL
CONSTANT}
\end{equation}%
with $\mathbf{R}_{o}$ being a suitable vector belonging to $\Omega $.

\item $p_{0}(t)$ is a suitably-prescribed smooth real function defined so
that $p_{1}(\mathbf{r},t)$ remains strictly positive in $\overline{\Omega }%
\times I$ (see Axiom \#2, Section 4).

\item The initial value $p_{0}(t_{o})$ is in principle an arbitrary constant
to be prescribed in such a way that $p_{0}\left( t_{0}\right) >p_{0\inf },$
with $p_{0\inf }$ being such that $p_{1}(\mathbf{r},t_{o})$ vanishes locally
in $\overline{\Omega }$ (see Axiom \#2, Section 4).
\end{itemize}

\subsection{The IKT statistical description}

It must be stressed that the vectors $\mathbf{v}$ and $\mathbf{u}$
defined above can be interpreted as stochastic variables [see
Appendix B], and the corresponding equations of motion,
Eqs.(\ref{eqq-1}) and (\ref{eqq-2}) viewed as \emph{Langevin}
(i.e., stochastic) \emph{equations}, provided a \emph{statistical
description} in terms of a suitable probability density is
introduced for them. Statistical descriptions of this type, based
on classical statistical mechanics (CSM), can be adopted in
principle both for classical and quantum fluids (see for example
Refs.\cite{Ellero2000,Tessarotto2004,Ellero2005,Tessarotto2006,Tessarotto2009A}
and \cite{Tessarotto2007a}) characterized either by deterministic
or stochastic flows \cite{Tessarotto20083,Tessarotto20082}. This
is realized by introducing an appropriate axiomatic approach
denoted as \textit{statistical
model, }represented by a set $\left\{ f,\Gamma \right\} ,$ with $f(\mathbf{x}%
,t)$ denoting a suitable \emph{kinetic distribution function} (KDF) - or a
\emph{probability density function} (PDF), to be identified with the
so-called \emph{1-point PDF} - which is defined in the phase-space $\Gamma $%
. Its is worth noting that such a type of approach can be determined in
accordance with the GENERIC dynamical model $\left\{ f,\Gamma \right\} $
developed by Grmela and Ottinger \cite{Grmela1997,Oettnger1997}. This is
defined in such a way to prescribe:

\begin{itemize}
\item A \emph{bundle structure} on $\Gamma $ \cite{Grmela1997}, i.e., a
mapping between $\Gamma $ itself and the ensemble of fluid fields $\left\{
Z\right\} $ (or $\left\{ Z_{1}\right\} $), generated via appropriate
statistical averages, i.e., phase-space moments, of $f(\mathbf{x},t)$.

\item A \emph{functional class} $\left\{ f\right\} $ to which $f(\mathbf{x}%
,t)$ belongs \cite{Tessarotto2004,Ellero2005}. This should be based \emph{%
exclusively} on the knowledge of the same set of fluid fields. The
prescription of $\left\{ f\right\} $ includes also the initial condition on $%
f(\mathbf{x},t)$ at $t=t_{o}.$

\item A suitable \emph{phase-space dynamics: }this is introduced in terms of
the \emph{NS-DS} [see Eq.(\ref{DYN-S})]. This should inherit the basic
properties of the fluid system, i.e., in particular, the conservation of
mass and momentum, the energy balance equation and the entropy law. Thus, in
the case it is described by deterministic, dissipative and irreversible
fluid equations, it should be a deterministic, non-conservative and
irreversible dynamical system.
\end{itemize}

The problem of its construction (Frisch, 1995 \cite{Frisch1995}), i.e., the
actual definition of the vector field $\mathbf{F}$, has remained for long
time unsolved (see for example, Vishik and Fursikov, 1988 \cite{Vishik1988}
and Ruelle, 1989 \cite{Ruelle1989}, where approaches based on kinetic theory
were attempted). Nevertheless, in the past various models for the dynamics
of tracer particles in incompressible fluids have actually been developed
\cite{Tchen1947,Corrsin1956,Buevich1966,Riley 1971,Maxey}, which should
manifestly apply, at least in principle, also to ITPs. As a consequence,
their dynamics should be controlled only by the mean-field force $\mathbf{F}$
produced by the unperturbed fluid fields. Since the original
Basset-Boussinesq-Oseen approach \cite{Basset,Boussinesq,Oseen} formulated
in the case of a uniform flow, several attempts to evaluate the form of $%
\mathbf{F},$ as well as the vector field $\mathbf{F}_{u}$, have appeared
\cite{Tchen1947,Corrsin1956,Buevich1966,Riley 1971,Maxey}. All such
approaches propose \textquotedblleft ad hoc\textquotedblright\ modifications
or corrections of the same equation, exclusively based on phenomenological
arguments, in order to adapt it for the treatment of non-uniform flows in NS
fluids. A popular form (for $\mathbf{F}$) frequently adopted for fluid
simulations is the one developed by Maxey and Riley (1982 \cite{Maxey}).
These treatments appear questionable because of the critical common
assumption on which they are based. Precisely, the requirement that the
tracer-particle velocity $\mathbf{v}(t)$ (\emph{kinetic velocity}) remains
always suitably close to the local velocity of the fluid $\mathbf{V}(\mathbf{%
r},t)$ evaluated at the position of the moving particle $\mathbf{r=r}(t)$.
This implies that for all ITPs the asymptotic condition%
\begin{equation}
\left\vert \mathbf{u}(t)\right\vert \ll \left\vert \mathbf{V}(\mathbf{r}%
(t),t)\right\vert  \label{assumption1}
\end{equation}%
should hold at any time $t$. As a consequence, in the determination of $%
\mathbf{F}$ all contributions proportional to the relative velocity should
be considered negligible. The constraint (\ref{assumption1}) imposes,
however, potentially serious limitations on particle dynamics. In fact, it
may easily be violated either due to the arbitrariness of the particle
initial conditions [in Eq.(\ref{eqq-1})] - in fact ITPs can be injected in a
fluid with arbitrary initial velocity - or because ITPs initially at rest
with respect to the fluid might develop finite relative velocities, thus
causing Eq.(\ref{assumption1}) to fail. On the other hand, the determination
of the exact ITP dynamics is manifestly of fundamental importance in order
to obtain detailed quantitative theoretical and numerical predictions in
fluid dynamics.

A first-principle solution of this problem has recently been proposed in Ref.%
\cite{Tessarotto2009} adopting a suitable statistical description for
incompressible Navier-Stokes thermofluids (see Appendix A), i.e., the
representation of the dynamical system (\ref{DYN-S}) in terms of the \emph{%
IKT-statistical model }$\left\{ f(\mathbf{x},t),\Gamma \right\} $\emph{, }%
obtained in the framework of the so-called inverse kinetic theory (IKT; see
Refs.\cite{Tessarotto2004,Ellero2005}). In such a formulation,
\begin{equation}
f(t)\equiv f(\mathbf{x},t)  \label{KDF}
\end{equation}%
is identified with a KDF whose velocity and phase-space moments are
prescribed in terms of a suitable subset of the fluid fields $\left\{
Z\right\} $, to be identified with the ensemble $\left\{ Z_{1}\right\} $.
Hence, provided $\int\limits_{U}d\mathbf{v}f(\mathbf{x},t)>0$, the
corresponding velocity PDF is%
\begin{equation}
\widehat{f}(\mathbf{x},t)=f(\mathbf{x},t)/\int\limits_{U}d\mathbf{v}f(%
\mathbf{x},t).  \label{PDF}
\end{equation}%
The KDF $f$ is required to satisfy in $\Gamma \times I$ the statistical
equation%
\begin{equation}
Lf=0,  \label{IKE}
\end{equation}%
denoted as \emph{inverse kinetic equation }(IKE \cite{Ellero2005}) in \emph{%
Eulerian form, }with $L\equiv \frac{\partial }{\partial t}f+\mathbf{v}\cdot
\nabla f+\frac{\partial }{\partial \mathbf{v}}\cdot \left( \mathbf{F}%
f\right) $ being the Liouville streaming operator.\ The same equation can be
equivalently cast in terms of\emph{\ }the \emph{integral Lagrangian IKE }%
\begin{equation}
f(\mathbf{x},t)=f_{o}(T_{t,t_{o}}\mathbf{x})\left\vert \frac{\partial
T_{t,t_{o}}\mathbf{x}}{\partial \mathbf{x}}\right\vert .
\label{INTEGRAL LIOUVILLE EQ}
\end{equation}%
Here $f_{o}(\mathbf{x}_{o})$ is a suitable initial KDF, while
\begin{equation}
\left\vert \frac{\partial T_{t,t_{o}}\mathbf{x}}{\partial \mathbf{x}}%
\right\vert =\exp \left\{ -\int\limits_{t_{o}}^{t}dt^{\prime }\frac{\partial
}{\partial \mathbf{v}(t^{\prime })}\cdot \mathbf{F}(\mathbf{x}(t^{\prime
}),t^{\prime })\right\}
\end{equation}%
(\emph{Liouville theorem}). In addition, the following assumptions are
introduced (see points a,b,c,d below):

a) A particular solution of Eq.(\ref{IKE}) is provided by the local Gaussian
distribution function (\emph{kinetic equilibrium})
\begin{equation}
f_{M}(\mathbf{x},t)=\frac{\rho (\mathbf{r},t)}{\left( \pi \right) ^{\frac{3}{%
2}}v_{th}^{3}}\exp \left\{ -X^{2}\right\} ,  \label{Maxwellian}
\end{equation}%
where $X^{2}=\frac{u^{2}}{vth^{2}}$ and $v_{th}=\sqrt{\frac{2p_{1}}{\rho }}$
denotes the \emph{thermal velocity.}

b) In terms of\ a suitable set of the velocity and phase-space moments, the
complete set of fluid equations are constructed from IKE.

c) Let us assume that $f(\mathbf{x},t)$ is strictly positive in $\Gamma $
and admits for all $(\mathbf{x},t)\in \Gamma \times I$ the \emph{%
Boltzmann-Shannon }(BS) \emph{statistical entropy} \cite{Tessarotto2006}
(also known as \emph{differential entropy}). This is defined as the
functional%
\begin{equation}
S(f(t))\equiv -\alpha _{1}^{2}\int\limits_{\Gamma }d\mathbf{x}f(\mathbf{x,}%
t)\ln f(\mathbf{x,}t)+c_{1},  \label{B-S ENTROPY}
\end{equation}%
where $\alpha _{1}\neq 0$ and $c_{1}$ are arbitrary real constants
independent of $(\mathbf{x},t)$, to be suitably defined. In information
theory the Boltzmann-Shannon entropy $S(f(t))$ can be intended as a \emph{%
measure of the ignorance} on $f(t)$. Here we remark that, by suitable
definition of the constants $\alpha _{1}$ and $c_{1}$, $S(f(t))$ can be
represented in terms of the corresponding BS entropy for the phase-space PDF
$\overline{f}(t)\equiv f(t)/M$, where $M=\int\limits_{\Gamma }d\mathbf{x}f(%
\mathbf{x},t).$ In fact, letting%
\begin{equation}
S(\overline{f}(t))\equiv -\int\limits_{\Gamma }d\mathbf{x}\overline{f}(%
\mathbf{x,}t)\ln \overline{f}(\mathbf{x},t),
\end{equation}%
it follows that%
\begin{equation}
S(\overline{f}(t))=\frac{1}{M}S(f(t))+\ln M,
\end{equation}%
which is again of the form (\ref{B-S ENTROPY}). In the following we shall
set in particular $\alpha _{1}=1$ and $c_{1}=0$ in Eq.(\ref{B-S ENTROPY}).

d) The time-derivative of the pseudopressure $dp_{0}(t)/dt$ is determined by
suitably prescribing the entropy production rate $\frac{\partial }{\partial t%
}S(f(t))$ \cite{Tessarotto2006}.

As a basic consequence of the previous assumptions it is follows
that \cite{Ellero2005,Tessarotto2006}:

\begin{enumerate}
\item The mean-field $\mathbf{F}$ is generally functionally dependent on $f$%
, i.e., it is of the form $\mathbf{F}=\mathbf{F}(\mathbf{x},t;f)$.

\item $\mathbf{F}$ is defined up to an arbitrary real gauge vector-field $%
\Delta \mathbf{F}$ obeying the gauge condition%
\begin{equation}
\frac{\partial }{\partial \mathbf{v}}\cdot \left[ \Delta \mathbf{F}(\mathbf{x%
},t;f)f(\mathbf{x},t)\right] =0.  \label{gauge-condition}
\end{equation}

\item The choice of the gauge field $\Delta \mathbf{F}$ \emph{does not
affect the time-evolution of} $f(x,t)$. Consequently, the corresponding
velocity and phase-space moments of the IKE are in all cases necessarily
\emph{unique}.

\item In the case $f\equiv f_{M}$ and up to the gauge field $\Delta \mathbf{%
F,}$ the functional form of the mean field $\mathbf{F}(\mathbf{x},t;f_{M})$
is uniquely determined.

\item In the case $f\neq f_{M}$ the determination of $\mathbf{F}$, again up
to the gauge field $\Delta \mathbf{F}$, requires suitable \emph{kinetic
closure conditions}.\emph{\ }In fact, in principle, in such a case $\mathbf{F%
}$ might depend on arbitrary higher-order phase-space moments of\
$f$ which vanish in the case $f\equiv f_{M}.$ To this end, in the
case of incompressible NS fluids, in Refs.\cite{Ellero2005,Tessarotto2006,Tessarotto2009}
it was assumed that $\mathbf{F}(%
\mathbf{x},t;f)$ (and $\Delta \mathbf{F}$) can be represented as polynomials
of lowest possible degree with respect to the relative kinetic velocity $%
\mathbf{u}$ and depend on the lowest-order and minimal number of velocity
moments of the KDF.
\end{enumerate}

We remark that, although in the context of IKT the unique specification of $%
\Delta \mathbf{F}$ (and hence of $\mathbf{F}$) is superfluous, its
determination is, instead, manifestly required in order to uniquely
prescribe the dynamics of ITPs. In particular in Ref.\cite{Tessarotto2009},
based on the analogy with extended thermodynamics \cite{Tessarotto2006},\ $%
\Delta \mathbf{F}$ was identified with a first-degree polynomial with
respect to $\mathbf{u}$ of the form $\Delta \mathbf{F}=\frac{1}{2}\nabla
\mathbf{V}\cdot \mathbf{u}-\frac{1}{2}\mathbf{u}\cdot \nabla \mathbf{V}$. In
the following{\ we intend to propose a generalization of IKT for \emph{%
compressible Navier-Stokes thermofluids} (Section 2) and in which }$\Delta
\mathbf{F}$ is\emph{\ uniquely determined,} based on suitable physical
assumptions{. }For this purpose, leaving initially unspecified the form of
the gauge field $\Delta \mathbf{F}$, we intend to prove that $\Delta \mathbf{%
F}$ is uniquely prescribed imposing the requirements stemming from the
following Gedanken (\emph{conceptual}) experiment (GDE).

\bigskip

\section{Gedanken experiment}

For a prescribed continuous fluid system, such as a
compressible/incompressible NS thermofluid, the problem arises
whether there might exist a subset of the ensemble of ideal tracer
particles (ITPs) for the dynamical system (\ref{DYN-S}) such that
their Newtonian state and corresponding time evolution depend only
on the state of the fluid $\left\{ Z\right\} $. In the following
the subset of ITPs which exhibit these properties are referred to
as thermal tracer particles (TTPs). Such a result was reached for
incompressible and isothermal NS fluids in
Ref.\cite{Tessarotto2011}. Here we claim that it should be
possible to extend the same conclusion to \emph{arbitrary
compressible and non-isothermal NS fluids} by performing a
conceptual experiment (\emph{Gedanken experiment}) on such a type
of fluid, i.e., looking at the properties of the IKT-statistical
models $\left\{ f,\Gamma \right\} $. The conjecture is suggested
by the following arguments:

\begin{itemize}
\item The state of the fluid is solely dependent on the fluid fields, which
in the case of a compressible NS thermofluid can be identified with the set $%
\left\{ Z_{1}\right\} $.

\item The time-evolution of $\left\{ Z_{1}\right\} $ as determined by CNSFE
is necessarily independent of the KDF $f(\mathbf{x},t)$ and of the NS-DS (%
\ref{DYN-S}). In fact, obviously the CNSFE (or INSE) problem cannot depend
on the functional form of $f(\mathbf{x},t)$.

\item On the other hand, in the context of IKT, the time evolution
of the KDF is determined by the Liouville operator $L$, which
enters the corresponding Liouville equation. As pointed out in
Refs.\cite{Ellero2005,Tessarotto2009}, this generally contains a vector field $\mathbf{%
F}$ whose form can depend functionally also on the same KDF $f(\mathbf{x},t)$%
, namely is of the form $\mathbf{F}=\mathbf{F}\left( \mathbf{x},t;f\right) $%
. Despite this, the time evolution of the fluid fields $\left\{
Z_{1}\right\} $ generated in terms of the KDF by the Lagrangian IKE (\ref%
{INTEGRAL LIOUVILLE EQ}) through the NS-DS (\ref{DYN-S}), can easily be
shown to remain independent of the functional form of the same KDF.
\end{itemize}

\subsection{GDE requirements}

On the basis of these considerations, here we conjecture that TTPs should
exist as a subset of ITPs and fulfill the following properties:

\begin{enumerate}
\item \emph{GDE-requirement \#1:} their time evolution, as determined by the
vector field $\mathbf{F}$ in terms of the NS-DS (\ref{DYN-S}), should remain
at all times $t\in I$ independent of the particular form of the KDF $f(%
\mathbf{x},t)$. As a consequence, for them the form of the mean-field force $%
\mathbf{F}$ should be also \emph{independent} of the KDF $f(\mathbf{x},t)$
[introduced in the IKT-statistical model $\left\{ f,\Gamma \right\} $],
namely simply of the form $\mathbf{F}=\mathbf{F}\left( \mathbf{x},t\right) $.

\item \emph{GDE-requirement \#2:} for prescribed initial conditions, their
Newtonian states $\mathbf{x}(t)\equiv \mathbf{x}=(\mathbf{r},\mathbf{v})$,
and equivalently also $\mathbf{y}(t)\equiv \mathbf{y}=(\mathbf{r},\mathbf{u}%
),$ should depend solely on the fluid fields $\left\{ Z_{1}\right\} $.

In addition, one should expect that for all TTPs:

\item \emph{GDE-requirement \#3 - Local magnitude of }$\mathbf{u}(t)$: the
magnitude of their instantaneous relative velocity $\left\vert \mathbf{u}%
(t)\right\vert \equiv \left\vert \mathbf{u}(\mathbf{r},t)\right\vert $
remains\ at all times $t\in I$ proportional to the local thermal velocity $%
v_{th}(\mathbf{r},t),$ i.e., of the form
\begin{equation}
\left\vert \mathbf{u}(t)\right\vert =\beta v_{th}(\mathbf{r},t),
\label{CONSTRAINT}
\end{equation}%
with $p_{1}(\mathbf{r},t)>0,\mathbf{r\equiv r}(t)$ and $\beta $ denoting
respectively the \emph{kinetic pressure} (\ref{kinetic pressure}), the
instantaneous position of the same particle and an appropriate non-vanishing
constant, i.e., a function independent of $(\mathbf{r},t)$. This means that $%
\beta $ is necessarily determined by the TTP initial condition (see
discussion below, after THM.2);

\item \emph{GDE-requirement \#4 - Kinetic constraint on the local direction
of }$\mathbf{u}(t)$: let us introduce for $\mathbf{u}(t)$ the representation
\begin{equation}
\mathbf{u}(t)=\beta v_{th}(\mathbf{r},t)\mathbf{n}(\mathbf{r},t),
\label{REPRESENTATION of u}
\end{equation}%
with $\mathbf{n}(\mathbf{r},t)$ being the unit vector prescribing the local
direction of $\mathbf{u}(t)$. Then, if at a given point the constraint (\ref%
{CONSTRAINT}) is satisfied, in order to warrant that $\mathbf{u}(t)$
satisfies it also at time $t+dt$ (with $dt$ being infinitesimal), it is
necessary to require that the unit vector $\mathbf{n}(\mathbf{r},t)$ be
tangent to the \emph{local isobaric surface} $\widehat{p}_{1}(\mathbf{r}%
,t)=const.$ Therefore, for a non-uniform kinetic pressure satisfying locally
$\nabla \widehat{p}_{1}(\mathbf{r},t)\neq 0$, the unit vector $\mathbf{n}(%
\mathbf{r},t)$ must satisfy the kinetic constraint:%
\begin{equation}
\mathbf{n}(\mathbf{r},t)\cdot \nabla \widehat{p}_{1}(\mathbf{r},t)=0.
\label{TANGENCY CONDITION-2}
\end{equation}%
As a consequence, the direction of $\mathbf{u}(t)$ is necessarily uniquely
determined, once the initial conditions (\ref{eqq-1}) and consequently its
initial direction%
\begin{equation}
\mathbf{n}(\mathbf{r},t_{o})\equiv \mathbf{n}_{o}(\mathbf{r})
\label{INITIAL DIRECTION}
\end{equation}%
have been set. Hence, for a non-uniform specific kinetic pressure $\widehat{p%
}_{1}$, the unit vector $\mathbf{n}(\mathbf{r},t)$ must be orthogonal to the
unit vector%
\begin{equation}
\mathbf{b}(\mathbf{r},t)=\frac{\nabla \widehat{p}_{1}(\mathbf{r},t)}{%
\left\vert \nabla \widehat{p}_{1}(\mathbf{r},t)\right\vert },  \label{Omega2}
\end{equation}%
i.e., the \emph{kinetic constraint}%
\begin{equation}
\mathbf{n}(\mathbf{r},t)\cdot \mathbf{b}(\mathbf{r},t)=0
\label{TANGENCY CONDITION}
\end{equation}%
must hold identically for all $\left( \mathbf{r},t\right) \in \Omega \times
I $.

\item \emph{GDE-requirement \#5 - Time evolution of }$\mathbf{n}(\mathbf{r}%
,t)$: the unit vector $\mathbf{n}(\mathbf{r},t)$ satisfies an initial-value
problem of the form%
\begin{equation}
\left\{
\begin{array}{c}
\frac{d\mathbf{n}(\mathbf{r},t)}{dt}=\mathbf{\Omega }(\mathbf{r},t)\times
\mathbf{n}(\mathbf{r},t), \\
\mathbf{n}(\mathbf{r}(t_{o}),t_{o})=\mathbf{n}(\mathbf{r}_{o},t_{o}),%
\end{array}%
\right.  \label{EVOLUTION EQUATION for n}
\end{equation}%
with $\mathbf{\Omega }(\mathbf{r},t)$ denoting a suitable pseudo-vector.
Without loss of generality we shall require that $\mathbf{\Omega }(\mathbf{r}%
,t)$ is a smooth real vector function defined in $\overline{\Omega }\times I$
and that it is defined also in the limit $p_{1}(\mathbf{r},t)\rightarrow
0^{+}$.

\item \emph{GDE-requirement \#6 - Rotation dynamics of }$\mathbf{n}(\mathbf{r%
},t)$: we require that the unit vector $\mathbf{n}(\mathbf{r},t)$ exhibits a
rotation motion with respect to the direction $\mathbf{b}(\mathbf{r},t)$
which is determined by the parallel component of the vorticity. In other
words, the $\mathbf{\Omega }(\mathbf{r},t)$ is required to satisfy the
constraint%
\begin{equation}
\mathbf{\Omega }(\mathbf{r},t)\cdot \mathbf{b}(\mathbf{r},t)=-\mathbf{\xi }(%
\mathbf{r},t)\cdot \mathbf{b}(\mathbf{r},t),  \label{NO-PRECESSION}
\end{equation}%
where $\mathbf{\xi }(\mathbf{r},t)\equiv \nabla \times \mathbf{V}(\mathbf{r}%
,t)$ is the vorticity field. As a consequence of Eqs.(\ref{EVOLUTION
EQUATION for n}) and (\ref{NO-PRECESSION}), the particle relative velocity
exhibits a rotation caused by two distinct physical mechanisms. The first
one is due to the rotation of the unit vector $\mathbf{b}(\mathbf{r},t)$
characterizing the isobaric surfaces, while the second one is due to the
intrinsic rotation of the parallel component of $\mathbf{\Omega }(\mathbf{r}%
,t)$ around $\mathbf{b}(\mathbf{r},t)$ as determined by the local vorticity
field. In particular, the constraint placed by Eq.(\ref{NO-PRECESSION})
establishes a direct connection between TTP dynamics and fluid vorticity and
has important implications for the treatment of strong turbulence in the
framework of TTP statistics (see related discussion in subsection VII.B
below).
\end{enumerate}

\section{Goals of the investigation}

In this paper we point out that IKT can be determined in such a way to
satisfy the requirements dictated by the GDE. For this purpose, first the
IKT statistical description $\left\{ f,\Gamma \right\} $ earlier pointed out
in Refs. \cite{Ellero2005,Tessarotto2006,Tessarotto2009} is extended to the
treatment of compressible and non-isothermal NS fluids. Next, the NS
dynamical system is shown to admit particular solutions which are of the
form of TTPs, namely ITPs for which the particle state takes the form
indicated above [see Eqs.(\ref{CONSTRAINT}) and (\ref{TANGENCY CONDITION})-(%
\ref{NO-PRECESSION})].\ More precisely, extending the results pointed out in
Ref.\cite{Tessarotto2011}, and holding in the case of incompressible and
isothermal NS fluids, here we intend to prove that for compressible,
non-isothermal fluids satisfying the\ CNSFE Problem (see THM.1 in Section 4):

\begin{itemize}
\item Goal \#1 - \emph{TTPs are particular solutions of the NS dynamical
system }(see Section 5, THM.2).
\end{itemize}

In such a setting, we claim that the following additional properties are
fulfilled:

\begin{itemize}
\item Goal \#2 -\emph{\ For all TTPs a unique realization exists for }$%
\mathbf{F}+\Delta \mathbf{F}$\emph{\ satisfying the requirements of GDE.}

\item Goal \#3 - \emph{In terms of the ensemble of TTPs a
reduced-dimensional statistical model }$\left\{ f_{1},\Gamma _{1}\right\} $ (%
\emph{TTP-statistical model})\emph{\ is introduced }(Section 6, THM.3).
\end{itemize}

Another interesting application concerns the treatment of stochastic flows.
For this purpose the fluid fields (\ref{REDUCED KDF h}) are assumed to
admit, in terms of suitable stochastic variables $\mathbf{\alpha }=\left\{
\alpha _{i},i=1,k\right\} \in V_{\mathbf{\alpha }}\subseteq \mathbf{R}^{k},$
with $k\geq 1,\mathbf{\ }$a\textit{\ }\emph{stochastic representation} of
the form \cite{Tessarotto20082,Tessarotto2009}%
\begin{equation}
\left\{ Z_{1}\right\} =\left\{ Z_{1}(\mathbf{r},t,\mathbf{\alpha })\right\} ,
\label{stochastic functions}
\end{equation}%
to be defined in terms of a suitable \emph{stochastic model} $\left\{ g(%
\mathbf{r},t,\mathbf{\alpha }),V_{\mathbf{\alpha }}\right\} $ (see Appendix
B). We intend to show that for incompressible fluids:

\begin{itemize}
\item Goal \#4 -\emph{\ The Langevin equations associated to TTPs dynamics
provides a possible mathematical model for tracer-particle motion in the
presence of fluctuating fluid fields }(Section 7, subsection 7.1).

\item Goal \#5 - \emph{The stochastic-averaged KDF of the TTP-statistical
model} $\left\{ f_{1},\Gamma _{1}\right\} $ \emph{satisfies a Fokker-Planck
statistical equation and a H-theorem }(Section 7, subsections 7.2 and 7.3).

\item Goal \#6 - \emph{The IKT and TTP statistical models }$\left\{ f,\Gamma
\right\} $\emph{\ and }$\left\{ f_{1},\Gamma _{1}\right\} $\emph{\ are
suitably related to the customary statistical treatment of turbulence due to
Hopf, Rosen and Edwards }(HRE approach \cite{Hopf1952,Rosen1960,Edwards1964}%
) (see Section 7, subsection 7.4).

\item Goal \#7 - \emph{The TTP-statistical model }$\left\{ f_{1},\Gamma
_{1}\right\} $\emph{\ provides a solution to Closure Problem for the
statistical description of hydrodynamic turbulence (HT) }(Section 7,
subsection 7.4).
\end{itemize}

\section{IKT for compressible NS thermofluids}

\subsection{Axiomatic formulation}

The basic requirements of the \emph{IKT statistical model }$\left\{ f(%
\mathbf{x},t),\Gamma \right\} $ have been discussed elsewhere
\cite{Ellero2005,Tessarotto2009}. In the case of a compressible NS
thermofluids, these can be re-formulated as follows. First we
require that the KDF $f(t)$, solution of IKE [see Eq.(\ref{IKE})],
uniquely determines the complete set of fluid fields $\left\{
Z_{1}\right\} $, in terms of suitable phase-space moments of the
same KDF. This is obtained imposing the following axiom:

\emph{Axiom \#1 - Correspondence principle:} For compressible (or
incompressible) NS thermofluids - in the closure of the fluid domain $%
\overline{\Omega }$, where by definition $\rho (\mathbf{\mathbf{r}},t)>0$ in
$\overline{\Omega }\times I$ - the following functional constraints hold%
\begin{equation}
\left\{
\begin{array}{c}
\int\limits_{U}d\mathbf{v}f(\mathbf{x},t)=\rho (\mathbf{\mathbf{r}},t), \\
\frac{1}{\rho (\mathbf{\mathbf{r}},t)}\int\limits_{U}d\mathbf{vv}f(\mathbf{x}%
,t)=\mathbf{V}(\mathbf{\mathbf{r}},t), \\
\left. \int\limits_{U}d\mathbf{v}\frac{1}{3}u^{2}f(\mathbf{x},t)=p_{1}(%
\mathbf{r},t),\right. \\
S(f(t))=S_{T}(t),%
\end{array}%
\right.  \label{MOMENTS}
\end{equation}%
which are referred to as \emph{correspondence principle for }$\left\{
Z_{1}\right\} $. Here we remark that due to the arbitrariness of the
constants $\alpha _{1}$ and $c_{1}$ appearing in the definition of the BS
entropy [see Eq.(\ref{B-S ENTROPY})] the last equation can also be replaced
by $S(f(t))=\alpha _{2}^{2}S_{T}(t)+c_{2}$, with $\alpha _{2}\neq 0$ and $%
c_{2}$ being two arbitrary real constants independent of $\left( \mathbf{x,}%
t\right) $. Furthermore, we shall require that\emph{\ }$f(t)$ admits also
the higher-order moments
\begin{equation}
\left\{
\begin{array}{c}
\mathbf{Q}(\mathbf{r,}t)\ =\int d\mathbf{vu}\frac{u^{2}}{3}f(\mathbf{r,u},t),
\\
\underline{\underline{{\Pi }}}(\mathbf{r,}t)=\int d\mathbf{vuu}f(\mathbf{r,u}%
,t),%
\end{array}%
\right.  \label{EXTENDED FLUID FIELDS}
\end{equation}%
to be denoted as \emph{extended fluid fields}. As a consequence, by an
appropriate definition of the mean-field,\ the correspondence principle and
IKE [see Eq.(\ref{IKE})] must deliver the complete set of fluid equations,
i.e., respectively CNSFE or INSFE (\emph{fluid closure condition}){.}

Second, consistent with CSM and the second principle of thermodynamics, the
initial KDF $f(t_{o})\equiv f_{o}(\mathbf{x})$ and the pseudopressure $%
p_{0}(t)$ are uniquely prescribed. This is obtained introducing the second
axiom:

\emph{Axiom \#2 - Entropic principle:}

This consists in the following three requirements:

A) At the initial time $t_{o}\in I$ the initial KDF $f(t_{o})\equiv f_{o}(%
\mathbf{x})$ is determined in such a way to maximize the BS-entropy $%
S(f(t_{o}))$ in a suitable functional class $\left\{ f(t_{o})\right\} $.\
This coincides with the axiom of CSM known as \emph{principle of entropy
maximization} (PEM, Jaynes 1957 \cite{Jaynes1957}).

B) The time derivative of pseudo-pressure $p_{0}(t)$ is prescribed for all $%
t\in I$ in such a way that the entropy law (\ref{5B}) [see Appendix A] is
identically fulfilled.

C) The initial condition $p_{0}(t_{o})$ is determined by suitably
prescribing the initial value of the BS entropy $S(f(t_{o}))$.

The first requirement is met as follows. Denoting by $\delta $ the Frechet
functional derivative operator, let us introduce the first and second
variations of $S(f(t_{o}))$, $\delta S(f(t_{o}))$ and $\delta
^{2}S(f(t_{o})) $. Then, the initial KDF $f_{o}\in \left\{ f(t_{o})\right\} $
is determined imposing the variational equation%
\begin{equation}
\left. \delta S(f(t_{o}))\right\vert _{_{f_{o}}}=0
\end{equation}%
subject to the inequality%
\begin{equation}
\left. \delta ^{2}S(f(t_{o}))\right\vert _{_{f_{o}}}<0.
\end{equation}%
The determination of $dp_{0}(t)/dt$ is obtained, instead, in such a way to
warrant - for consistency with the correspondence principle and the entropy
law - that the weak H-theorem%
\begin{equation}
\frac{\partial }{\partial t}S(f(t))=\frac{\partial }{\partial t}S_{T}(t)\geq
0  \label{H-THEOREM-1}
\end{equation}%
holds for all $t\in I$. In the case of isentropic flows this reduces to the
constant H-theorem:
\begin{equation}
\frac{\partial }{\partial t}S(f(t))=\frac{\partial }{\partial t}S_{T}(t)=0.
\label{CONSTANT-H-THEOREM-0}
\end{equation}%
Furthermore, the initial kinetic pressure $p_{0}\left( t_{0}\right) $ is
prescribed in such a way that the initial BS entropy $S(f(t_{o}))$ vanishes,
i.e.,%
\begin{equation}
S(f(t_{o}))=0.  \label{INITIAL BS ENTROPY}
\end{equation}%
The initial condition (\ref{INITIAL BS ENTROPY}) on the BS entropy is
equivalent to demand that the measure of ignorance $S(f(t_{o}))$ is zero.\
Since $S(f(t_{o}))$ $\leq $ $S(f_{M}(t_{o}))$ this means that the Gaussian
KDF $f_{M}(t_{o})$ must admit the BS entropy and hence that the kinetic
pressure is necessarily $p_{1}(\mathbf{r},t_{o})\geq 0.$ Due to the
arbitrariness of $p_{0}\left( t_{0}\right) $ this requirement can always be
satisfied.

Finally, suitable kinetic closure conditions are introduced to determine $%
\mathbf{F}(\mathbf{x},t;f)$ in the non-Gaussian case [$f(\mathbf{x},t)\neq
f_{M}(\mathbf{x},t)$]:

\emph{Axiom \#3 - Kinetic closure conditions:\ }For this purpose, in analogy
with INSE \cite{Ellero2005}, we shall assume that $\mathbf{F}(\mathbf{r},%
\mathbf{u},t;f)$ is a polynomial of lowest possible degree with respect to
the relative kinetic velocity $\mathbf{u}$ and depends on the lowest-order
and minimal number of velocity moments of the KDF. In particular:

- \emph{Axiom \#3a})\emph{\ }$\mathbf{F}(\mathbf{r},\mathbf{u},t;f)$ can
depend, besides $\left\{ Z_{1}\right\} $, only on the \emph{minimal set of
extended fluid fields} $\left\{ \mathbf{Q},\underline{\underline{\mathbf{\Pi
}}}\right\} $;

- \emph{Axiom \#3b})\emph{\ }$\mathbf{F}(\mathbf{r},\mathbf{u},t;f)$ depends
only \emph{linearly} with respect to $\mathbf{Q}$ and $\underline{\underline{%
\mathbf{\Pi }}}$.

Let us briefly comment on these requirements. In principle the vector field $%
\mathbf{F}(\mathbf{r},\mathbf{u},t;f)$ might depend on arbitrary
higher-order moments of the KDF. In fact, due to the arbitrariness in its
definition (see discussion above), it is always possible to include an
additive contribution which vanishes identically in the case $f=f_{M}$
(Gaussian KDF) and which does not contribute the velocity moments of the
Liouville equations corresponding to the weight functions $G=\left( 1,%
\mathbf{v},u^{2}/3\right) $. Hence the above closure conditions warrant that
such contributions are excluded, so that Axioms \#3a and \#3b actually
realize the minimal requirements when $f$ is non-Gaussian.

The motivation for the precise choice of Axiom \#3 is mathematical
simplicity. In fact, within the framework of IKT, the mean-field $\mathbf{F}$
is not a physical observable, both for Gaussian and non-Gaussian KDFs, and
therefore it remains intrinsically non-unique. Its indeterminacy in the case
of a non-Gaussian KDF arises because the vector field $\mathbf{F}(\mathbf{r},%
\mathbf{u},t;f)$ might include in principle higher-order velocity moments of
the KDF, besides the fluid fields $\left\{ Z_{1}\right\} $ which are by
construction the only observables. These additional moments have no physical
meaning (i.e., they are not observables), and therefore remain completely
undetermined in the framework of IKT. Hence, $\mathbf{F}(\mathbf{r},\mathbf{u%
},t;f)$ should depend only on a minimum finite number of fluid fields which
are required for the validity of the theory. This means that there must
exist a finite subset of moment equations which coincide with CNSFE. To
further clarify the issue, we notice that $\mathbf{F}(\mathbf{r},\mathbf{u}%
,t;f)$ can always be given a polynomial representation in terms of the
relative velocity $\mathbf{u}$. Such terms would necessarily depend on
higher-order velocity moments which vanish in the case of the Gaussian KDF
and can always be prescribed in such a way not to contribute to the same
moment equations. Unless Axioms \#3a and \#3b are introduced, such
additional contributions to $\mathbf{F}(\mathbf{r},\mathbf{u},t;f)$,
depending, besides the CNSFE fluid fields, also on $\mathbf{Q}$ and $%
\underline{\underline{\mathbf{\Pi }}}$, would remain undetermined. Due to
the intrinsic freedom of their choice, in the following they will be set
identically equal to zero. As clarified below, such an assumption is
equivalent to require Axioms \#3a and \#3b. This choice does not affect the
validity of the CNSFE problem and does not constraint in any way its
solutions. In conclusion, in view of these considerations, Axiom \#3 can be
viewed as a set of kinetic closure condition which are needed for the
prescription of the mean-field $\mathbf{F}(\mathbf{r},\mathbf{u},t;f)$ and
the related kinetic equation (\ref{IKE}).

\subsection{The IKT statistical model for compressible thermofluids}

Based on the axiomatic formulation given above [Axioms \#1-\#3], we can now
proceed to the explicit determination of the mean-field\ $\mathbf{F}=\mathbf{%
F}(\mathbf{x},t,f)$ appropriate for a compressible thermofluid satisfying
the CNSFE problem [defined by Eqs.(\ref{1A})-(\ref{5B}) and the
initial-boundary conditions (\ref{initial-boundary conditions}); see
Appendix A]. It is immediate to show that in such a case the form of the
mean-field $\mathbf{F}(\mathbf{x},t,f)$ can be determined analytically. The
result is summarized by the following theorem:

\textbf{Theorem 1 - IKT statistical model for CNSFE - }\emph{Let us require
that the IKT statistical model }$\left\{ f(\mathbf{x},t),\Gamma \right\} $
\emph{satisfies} \emph{Axioms \#1-\#3 and furthermore that:}

\emph{1) The CNSFE problem admits a smooth strong solution in }$\Gamma
\times I$.

\emph{2) The mean-field }$\mathbf{F}(\mathbf{x},t;f)$ \emph{is defined as}

\begin{equation}
\mathbf{F}(\mathbf{x},t;f)=\mathbf{F}_{H}(\mathbf{r},t)+\mathbf{F}_{u}(%
\mathbf{x},t;f)+\mathbf{F}_{a}(\mathbf{r},\mathbf{u},t;f)+\mathbf{u}\cdot
\nabla \mathbf{V+}\Delta \mathbf{F}(\mathbf{x},t;f).  \label{FF-1}
\end{equation}%
\emph{Here }$\mathbf{F}_{H}$\emph{\ is the Navier-Stokes acceleration, given
by Eq.(\ref{fluid-acceleration}) }[see Appendix A],\emph{\ }$\mathbf{F}_{u}$
\emph{is the relative kinetic acceleration defined as}%
\begin{equation}
\left. \mathbf{F}_{u}(\mathbf{x},t;f)=\frac{v_{th}^{2}}{2}\nabla \ln \rho +%
\frac{\mathbf{u}}{2p_{1}}A+\frac{v_{th}^{2}}{2}\nabla \ln \left( \frac{p_{1}%
}{\rho }\right) \left( X^{2}-\frac{1}{2}\right) ,\right.  \label{FF-2}
\end{equation}%
\emph{with}%
\begin{equation}
A=\rho \frac{D}{Dt}\left( \frac{p_{1}}{\rho }\right) \equiv \rho \frac{D}{Dt}%
\left( \frac{p_{o}+p-\phi }{\rho }\right) +nK(\mathbf{r},t)  \label{FF-3}
\end{equation}%
\emph{and with} $K(\mathbf{r},t)$ \emph{being prescribed by Eq.(\ref%
{TEMPERATURE SOURCE TERM}) in Appendix A. Finally, }$\mathbf{F}_{a}(\mathbf{r%
},\mathbf{u},t;f)$ \emph{is defined as}%
\begin{equation}
\mathbf{F}_{a}(\mathbf{r},\mathbf{u},t;f)\equiv \frac{1}{\rho }\left[
\mathbf{\nabla \cdot }\underline{\underline{{\Pi }}}-\mathbf{\nabla }p_{1}%
\right] +\frac{\mathbf{u}}{2p_{1}}\left[ \mathbf{\nabla \cdot Q}-\nabla \ln
\left( \frac{p_{1}}{\rho }\right) \cdot \mathbf{Q}\right] .  \label{FF-4}
\end{equation}

\emph{3) }$\mathbf{F}(\mathbf{x},t;f)$\emph{\ is defined up to an arbitrary
real gauge field }$\Delta \mathbf{F}(\mathbf{x},t;f)$\emph{\ satisfying the
gauge condition (\ref{gauge-condition}). We shall require that }$\Delta
\mathbf{F}(\mathbf{x},t;f)$ \emph{is a smooth vector field analytic with
respect to the Newtonian velocity vector }$\mathbf{v}\in U$.

\emph{4) The BS entropy and the velocity moments} $\int\limits_{U}d\mathbf{v}%
G(\mathbf{x},t)f(\mathbf{x},t)$ \emph{evaluated for} $\left\{ G(\mathbf{x}%
,t)\right\} =\left\{ 1,\mathbf{v},\frac{1}{3}u^{2},\mathbf{uu},\frac{1}{3}%
u^{2}\mathbf{u}\right\} $ \emph{exist for all }$\left( \mathbf{r},t\right)
\in \overline{\Omega }\times I.$

\emph{5) In }$\overline{\Omega }\times I$ \emph{the KDF }$f(\mathbf{x},t)$
\emph{admits the correspondence principle defined by Eqs.(\ref{MOMENTS}).}

\emph{6) Let us introduce the decomposition}%
\begin{eqnarray}
&&\left. \frac{\partial }{\partial t}S(f(t))=P_{1}(f(t))+\frac{\partial }{%
\partial t}S_{T}(t),\right.  \label{ASSUMPTION 6a} \\
&&\left. P_{1}(f(t))\equiv P(f(t))-\frac{\partial }{\partial t}%
S_{T}(t),\right.  \label{ASSUMPTION 6b}
\end{eqnarray}%
\emph{where}%
\begin{equation}
P(f(t))\equiv \int\limits_{\Omega }d\mathbf{r}\left[ \frac{3\rho (\mathbf{r}%
,t)}{2p_{1}}A+\rho (\mathbf{r,}t)\nabla \cdot \mathbf{V}+\frac{3\rho (%
\mathbf{r},t)}{2p_{1}}\left\{ \mathbf{\nabla \cdot Q}-\nabla \ln \left(
\frac{p_{1}}{\rho }\right) \cdot \mathbf{Q}\right\} \right]
\label{ASSUMPTION 6c}
\end{equation}%
\emph{and }$\frac{\partial }{\partial t}S_{T}(t)$\emph{\ denotes the global
thermodynamic entropy production rate} [defined by (\ref{THERMODYNAMIC
ENTROPY PRODICTION RATE})]. \emph{Then, we require that for} \emph{all }$%
t\in I$\emph{\ the pseudo-pressure }$p_{0}(t)$\emph{\ is determined so that }%
\begin{equation}
P_{1}(f(t))=0.  \label{ASSUMPTION 6d}
\end{equation}

\emph{It follows that:}

T1$_{1}$\emph{) The local Gaussian distribution function (\ref{Maxwellian})
is a particular solution of the IKE (\ref{IKE}) if and only if the fluid
fields }$\left\{ Z\right\} $\emph{\ satisfy the CNSFE problem }[see Appendix
A]\emph{. For a generic KDF }$f(\mathbf{x},t)$\emph{, introducing the
representation}%
\begin{equation}
f(\mathbf{x},t)=f_{M}(\mathbf{x},t)h(\mathbf{x},t),  \label{REDUCED KDF h}
\end{equation}%
\emph{it follows that the reduced KDF }$h(\mathbf{x},t)$ \emph{satisfies the
integral IKE}%
\begin{equation}
h(\mathbf{x},t)=h_{o}(T_{t,t_{o}}\mathbf{x})\exp \left\{
-\int\limits_{t_{o}}^{t}dt^{\prime }\frac{\partial }{\partial \mathbf{v}%
(t^{\prime })}\cdot \mathbf{F}_{a}(\mathbf{x}(t^{\prime }),t^{\prime
};f)\right\} ,  \label{LAGRANGIAN-IKE FOR h}
\end{equation}%
$h_{o}(\mathbf{x}_{o})$ \emph{being a suitable initial KDF.}

T1$_{2}$\emph{) In the case of a general non-Gaussian KDF }$f(\mathbf{x},t)$%
, \emph{the velocity-moment equations obtained by taking the weighted
velocity integrals of Eq.(\ref{IKE}) with the weights }$\left\{ G(\mathbf{x}%
,t)\right\} =\left\{ 1,\mathbf{v},u^{2}/3\right\} $\emph{\ deliver
identically the fluid equations (\ref{1A})- (\ref{2B}).}

T1$_{3}$\emph{) For all }$t\in I$\emph{\ the pseudo-pressure }$p_{0}(t)$%
\emph{\ must satisfy the ODE:}%
\begin{equation}
\left. \frac{d}{dt}p_{0}=\frac{1}{\frac{3}{2}\int\limits_{\Omega }d\mathbf{r}%
\frac{\rho }{p_{1}}}\left[ S_{p}(t)+Q(t)+\frac{\partial }{\partial t}S_{T}(t)%
\right] ,\right.  \label{constraint on p_0}
\end{equation}%
\emph{where}%
\begin{equation}
S_{p}(t)\equiv -\frac{3}{2}\int\limits_{\Omega }d\mathbf{r}\frac{\rho }{p_{1}%
}\left[ \frac{\partial }{\partial t}\left( p-\phi +nT\right) +\mathbf{V}%
\cdot \nabla \left( p-\phi +nT\right) \right] -\int\limits_{\Omega }d\mathbf{%
r}\rho \nabla \cdot \mathbf{V},  \label{S_p}
\end{equation}%
\begin{equation}
Q(t)\equiv -\int\limits_{\Omega }d\mathbf{r}\frac{3\rho }{2p_{1}}\left[
\nabla \cdot \mathbf{Q}-\nabla \ln \left( \frac{p_{1}}{\rho }\right) \cdot
\mathbf{Q}\right] .  \label{Q}
\end{equation}

T1$_{4}$\emph{) The BS entropy }$S\left( f(t)\right) $\emph{\ satisfies for
all} $t\in I$ \emph{the H-theorem (\ref{H-THEOREM-1}). Instead, for an
isothermal fluid the constant H-theorem (\ref{CONSTANT-H-THEOREM-0}) holds.}

T1$_{5}$\emph{) In validity of Eq.(\ref{constraint on p_0}), for all }$(%
\mathbf{r},t)\in \overline{\Omega }\times I$\emph{\ the kinetic pressure} $%
p_{1}(\mathbf{r},t)$ \emph{is strictly positive. Furthermore, the initial
value }$p_{0}(t_{o})$\emph{\ is uniquely determined by prescribing the
condition of vanishing of the\ initial BS entropy in the case }$%
f(t_{o})\equiv f_{M}(t_{o})$ [see Axiom \#2, entropic principle].

\emph{Proof - }First, it is immediate to prove that, in validity of Eqs.(\ref%
{FF-1})-(\ref{FF-4}), $f_{M}(\mathbf{x},t)$ is a particular solution of the
inverse kinetic equation (\ref{IKE}). The proof is analogous to that given
in Refs.\cite{Ellero2005,Tessarotto2009} and it follows by direct
substitution of the distribution $f_{M}(\mathbf{x},t)$ in the same equation
(Proposition T1$_{1}$). This implies that $f_{M}(\mathbf{x},t)$ satisfies
necessarily the integral Liouville equation (\ref{INTEGRAL LIOUVILLE EQ}),
so that
\begin{equation}
f_{M}(\mathbf{x},t)=f_{M}(T_{t,t_{o}}\mathbf{x},t_{o})\exp \left\{
-\int\limits_{t_{o}}^{t}dt^{\prime }\left[ \frac{\partial }{\partial \mathbf{%
v}(t^{\prime })}\cdot \mathbf{F}_{u}(\mathbf{x}(t^{\prime }),t^{\prime };f)+%
\frac{\partial }{\partial \mathbf{r}(t^{\prime })}\cdot \mathbf{V}(\mathbf{r}%
(t^{\prime }),t^{\prime })\right] \right\} .  \notag
\end{equation}%
Therefore, in case of a non-Gaussian KDF the same equation manifestly
implies also Eq.(\ref{LAGRANGIAN-IKE FOR h}).

Instead, if we assume that in $\Gamma \times I,$ $f(\mathbf{x},t)$ is a
particular solution of the inverse kinetic equation, it follows that the
fluid fields $\left\{ \rho ,\mathbf{V},p,T\right\} $ are necessarily
solutions of the CNSFE equations. This can be proved either in the case $%
f\equiv f_{M}(\mathbf{x},t)$ by direct substitution in Eq.(\ref{IKE}) or, in
the general case in which $f\neq f_{M}(\mathbf{x},t)$ is an arbitrary smooth
and strictly positive particular solution, by direct calculation of the
velocity moments\ of the same equation, evaluated with respect to the
weight-functions $G(\mathbf{x},t)=\left( 1,\mathbf{v},u^{2}/3\right) $
(Proposition T1$_{2}$). In fact, in validity of Axiom \#2 and Eqs.(\ref{FF-1}%
)-(\ref{FF-3}), the moments equations corresponding to Eqs.(\ref{MOMENTS})
yield respectively:%
\begin{equation}
\frac{\partial \rho (\mathbf{r},t)}{\partial t}+\nabla \cdot \left[ \rho (%
\mathbf{r},t)\mathbf{V}(\mathbf{r},t)\right] =0,
\end{equation}%
\begin{equation}
\frac{\partial }{\partial t}\rho (\mathbf{r},t)\mathbf{V}(\mathbf{r}%
,t)+\nabla \cdot \left[ \rho (\mathbf{r},t)\mathbf{V}(\mathbf{r},t)\mathbf{V}%
(\mathbf{r},t)\right] \equiv \rho \frac{D}{Dt}\mathbf{\mathbf{V}}(\mathbf{%
\mathbf{r}},t)=\rho \mathbf{F}_{H},
\end{equation}%
\begin{eqnarray}
&&\left. \frac{\partial }{\partial t}p_{1}(\mathbf{r,}t)+\nabla \cdot \left[
\mathbf{V}(\mathbf{r},t)p_{1}(\mathbf{\mathbf{r}},t)+\mathbf{Q}\right]
=\right.  \notag \\
&&\left. =A+\left[ \nabla \cdot \mathbf{Q}-\nabla \ln \left( \frac{p_{1}}{%
\rho }\right) \cdot \mathbf{Q}\right] +\mathbf{Q}\cdot \nabla \ln \left(
\frac{p_{1}}{\rho }\right) .\right.
\end{eqnarray}%
The first two equations coincide, respectively, with the continuity and
Navier-Stokes equations while the third one, thanks to Eq.(\ref{FF-3}),
recovers the Fourier equation [see respectively Eqs.(\ref{1A}),(\ref{1B})
and (\ref{2B}) in Appendix A]. Let us now evaluate the entropy production
rate $\frac{\partial }{\partial t}S\left( f(t)\right) $ (Proposition T1$_{3}$%
). First we notice that thanks to the Brillouin Lemma \cite{Brillouin1959}:%
\begin{eqnarray}
&&\left. S(f(t))=-\int\limits_{\Gamma }d\mathbf{x}f(\mathbf{x},t)\ln f(%
\mathbf{x},t)\leq \right.  \label{INEQUALITY-ALFA} \\
&&\left. \leq -\int\limits_{\Gamma }d\mathbf{x}f(\mathbf{x},t)\ln f_{M}(%
\mathbf{x},t)=-\int\limits_{\Gamma }d\mathbf{x}f_{M}(\mathbf{x},t)\ln f_{M}(%
\mathbf{x},t)\equiv S(f_{M}(t)),\right.  \notag
\end{eqnarray}%
which implies%
\begin{eqnarray}
&&\left. \frac{\partial }{\partial t}S(f(t))\equiv \int\limits_{\Gamma }d%
\mathbf{x}f(\mathbf{x},t)\frac{\partial }{\partial \mathbf{v}}\cdot \mathbf{F%
}(\mathbf{x},t;f)\leq \right.  \notag \\
&&\left. \leq \frac{\partial }{\partial t}S(f_{M}(t))\equiv
\int\limits_{\Gamma }d\mathbf{x}f_{M}(\mathbf{x},t)\frac{\partial }{\partial
\mathbf{v}}\cdot \mathbf{F}(\mathbf{x},t;f_{M}),\right.
\end{eqnarray}%
where respectively:%
\begin{eqnarray}
&&\left. \frac{\partial }{\partial \mathbf{v}}\cdot \mathbf{F}(\mathbf{x}%
,t;f)=\frac{3}{2p_{1}}A+\frac{3}{2p_{1}}\left[ \nabla \cdot \mathbf{Q}%
-\nabla \ln \left( \frac{p_{1}}{\rho }\right) \cdot \mathbf{Q}\right]
+\nabla \cdot \mathbf{V},\right. \\
&&\left. \frac{\partial }{\partial \mathbf{v}}\cdot \mathbf{F}(\mathbf{x}%
,t;f_{M})=\frac{3}{2p_{1}}A+\nabla \cdot \mathbf{V}.\right.
\end{eqnarray}

It follows that $P(f(t))\leq P(f_{M}(t))$, with $P(f(t))$ being defined by
Eq.(\ref{ASSUMPTION 6c}), implying in turn that the inequality%
\begin{equation}
Q(t)\geq 0  \label{INEQUALITY FOR Q(t)}
\end{equation}%
is necessarily fulfilled for all $t\in I$. Therefore, introducing the
decomposition (\ref{ASSUMPTION 6a})-(\ref{ASSUMPTION 6b}), the constraint (%
\ref{ASSUMPTION 6d}) manifestly implies that Proposition T1$_{3}$ must hold.
In addition, the entropy production rate fulfills identically the constraint%
\begin{equation}
\frac{\partial }{\partial t}S(f(t))=\frac{\partial }{\partial t}S_{T}(t).
\label{CONSTRAINT-NEW}
\end{equation}%
Hence, thanks to the entropy law (\ref{4b}), necessarily the BS entropy $%
S\left( f(t)\right) $\ satisfies the H-theorem (\ref{H-THEOREM-1})
(Proposition T1$_{4}$). Let us now prove Proposition T1$_{5}$, namely that
the constraint (\ref{constraint on p_0}) requires the kinetic pressure $%
p_{1}(\mathbf{r},t)$ to be strictly positive in $\Omega \times I$. For this
purpose, we first consider the case $f=f_{M}$ and impose $\forall t\in
I_{o}\equiv \left\{ t:t\geq t_{o},\text{ }\forall t\in I\right\} $ that the
constant-entropy condition $S(f_{M}(t))=S(f_{M}(t_{o}))$ holds, requiring:%
\begin{equation}
\frac{\partial }{\partial t}S(f_{M}(t))\equiv \int\limits_{\Omega }d\mathbf{r%
}\frac{3}{2p_{1}}A=0.
\end{equation}%
Then, thanks to the identity%
\begin{equation}
\int\limits_{\Omega }d\mathbf{r}\frac{3\rho }{2p_{1}}A=\frac{\partial
p_{0}(t)}{\partial t}\frac{3}{2}\int\limits_{\Omega }d\mathbf{r}\frac{\rho }{%
p_{1}}+\frac{3}{2}\int\limits_{\Omega }d\mathbf{r}\frac{\rho }{p_{1}}\left[
\frac{\partial }{\partial t}\left( p-\phi +nT\right) +\mathbf{V}\cdot \nabla
\left( p-\phi +nT\right) \right] ,
\end{equation}%
it follows that $p_{0}(t)$ must satisfy the ODE%
\begin{equation}
\frac{dp_{0}(t)}{dt}=\frac{1}{\frac{3}{2}\int\limits_{\Omega }d\mathbf{r}%
\frac{\rho }{p_{1}}}S_{p}(t),
\end{equation}%
with $S_{p}(t)$ being given by Eq.(\ref{S_p}). Hence $p_{1}(\mathbf{r},t)$
is necessarily strictly positive in $\Omega \times I$. The same conclusion
manifestly follows imposing instead%
\begin{equation}
\frac{\partial }{\partial t}S(f_{M}(t))\equiv \int\limits_{\Omega }d\mathbf{r%
}\frac{3\rho }{2p_{1}}A=\frac{\partial }{\partial t}S_{T}(t)\geq 0,
\end{equation}%
which implies%
\begin{equation}
\frac{dp_{0}(t)}{dt}=\frac{1}{\frac{3}{2}\int\limits_{\Omega }d\mathbf{r}%
\frac{\rho }{p_{1}}}\left[ S_{p}(t)+\frac{\partial }{\partial t}S_{T}(t)%
\right] .
\end{equation}%
Analogous conclusion holds, thanks to the inequality (\ref{INEQUALITY FOR
Q(t)}), also in the case $f\neq f_{M}$ [see Eq.(\ref{constraint on p_0})].
Finally, let us impose the initial condition for the initial kinetic
pressure $p_{0}(t_{o})$. Denoting $M\equiv \int\limits_{\Omega }d\mathbf{r}%
\rho ,$ in view of Axiom \#2 this requires
\begin{equation*}
S(f_{M}(t_{o}))=\frac{3}{2}\int\limits_{\Omega }d\mathbf{r}\rho \ln p_{1}+M%
\left[ \frac{3}{2}+\ln \left( 2\pi \right) ^{3/2}\right] -\frac{5}{3}%
\int\limits_{\Omega }d\mathbf{r}\rho \ln \rho =0,
\end{equation*}%
which uniquely determines\ $p_{0}\left( t_{0}\right) $. \textbf{Q.E.D.}

Here it is worth noting that:

\begin{itemize}
\item Eqs.(\ref{FF-1})-(\ref{FF-4}) yield a realization of the mean-field $%
\mathbf{F}(\mathbf{x},t;f)$, of the type indicated above (see subsection
1.4) and holding for compressible NS thermofluids, which satisfies the CNSFE
problem [see Appendix A].

\item The expression of the vector field $\mathbf{F}(\mathbf{x},t;f)$ in
Eqs.(\ref{FF-1})-(\ref{FF-4}) is determined, up to the gauge field $\Delta
\mathbf{F}$, according to the form of the KDF as follows. In the case $%
f=f_{M}$ it is obtained by solving explicitly for $\mathbf{F}$ the equation%
\begin{equation}
\left( \frac{\partial }{\partial t}+\mathbf{v}\cdot \frac{\partial }{%
\partial \mathbf{r}}\right) \ln f_{M}-2\frac{\mathbf{u}}{v_{th}}\cdot
\mathbf{F}=-\left( \frac{\partial }{\partial \mathbf{v}}\mathbf{F}\right) ,
\end{equation}%
and imposing the validity of CNSFE. The procedure is analogous to that
outlined, for example, in Ref.\cite{Tessarotto2009}. Instead, in the general
case in which $f\neq f_{M}$, with $f$ being a strictly positive KDF
satisfying Axioms \#1-\#3, $\mathbf{F}(\mathbf{x},t;f)$ is determined by
requiring it is of the form (\ref{FF-1})-(\ref{FF-2}) with $\mathbf{F}_{a}(%
\mathbf{x},t;f)$ to be suitably prescribed. In particular, thanks to Axiom
\#3, $\mathbf{F}_{a}(\mathbf{x},t;f)$ is taken to be a polynomial of first
degree in the relative velocity $\mathbf{u}$. Therefore, it is necessarily
of the form $\mathbf{F}_{a}(\mathbf{r},\mathbf{u},t;f)=\mathbf{F}%
_{a}^{\left( 0\right) }(\mathbf{r},t;f)+\mathbf{u}F_{a}^{\left( 1\right) }(%
\mathbf{r},t;f)$, with $\mathbf{F}_{a}^{\left( 0\right) }(\mathbf{r},t;f)$
and $F_{a}^{\left( 1\right) }(\mathbf{r},t;f)$ being respectively two
suitable moments of the KDF $f$. Their precise form is obtained by imposing
Axiom \#1 and requiring that the velocity moments of IKE corresponding to
the weight functions $G(\mathbf{x},t)=\left( 1,\mathbf{v},u^{2}/3\right) $
coincide with CNSFE. For example, $\mathbf{F}_{a}^{\left( 0\right) }(\mathbf{%
r},t;f)$ follows by constructing the moment equation with respect to $G=%
\mathbf{v}$. From IKE, utilizing the definitions for $\mathbf{F}_{H}(\mathbf{%
r},t)$ and $\mathbf{F}_{H}(\mathbf{x},t;f)$, it follows that $\mathbf{F}%
_{a}^{\left( 0\right) }(\mathbf{r},t;f)$ must be prescribed so that the
equation%
\begin{equation}
\rho \frac{D}{Dt}\mathbf{V}+\left( \mathbf{\nabla \cdot }\underline{%
\underline{{\Pi }}}-\mathbf{\nabla }p_{1}\right) -\rho \mathbf{F}%
_{H}-\int\limits_{U}d\mathbf{v}f(\mathbf{x},t)\mathbf{F}_{a}^{\left(
0\right) }=0
\end{equation}%
coincides with the NS equation. This yields the unique solution $\mathbf{F}%
_{a}^{\left( 0\right) }(\mathbf{r},t;f)\equiv \frac{1}{\rho }\left[ \mathbf{%
\nabla \cdot }\underline{\underline{{\Pi }}}-\mathbf{\nabla }p_{1}\right] $.
Similarly, the expression for $F_{a}^{\left( 1\right) }(\mathbf{r},t;f)$
follows from the moment equation with respect to $G=u^{2}/3$. This yields
the solution $F_{a}^{\left( 1\right) }(\mathbf{r},t;f)\equiv \frac{1}{2p_{1}}%
\left[ \mathbf{\nabla \cdot Q}-\nabla \ln \left( \frac{p_{1}}{\rho }\right)
\cdot \mathbf{Q}\right] $. The resulting expression for $\mathbf{F}_{a}(%
\mathbf{r},\mathbf{u},t;f)$ coincides with Eq.(\ref{FF-4}).

\item If $f\equiv f_{M}(\mathbf{x},t)$ the extended fluid fields $\mathbf{Q}(%
\mathbf{r},t)$ and $\underline{\underline{{\Pi }}}(\mathbf{r},t)$ [see Eqs.(%
\ref{EXTENDED FLUID FIELDS})] vanish identically. As a consequence, in this
case $\mathbf{F}_{a}(\mathbf{r},t)\equiv 0$.

\item Eqs.(\ref{FF-1})-(\ref{FF-3}) apply also in the case of an
incompressible thermofluids [INSFE problem], and in particular for
isothermal fluids [INSE problem; see Appendix A]. In the case $f\equiv f_{M}(%
\mathbf{x},t)$ the corresponding mean-field $\mathbf{F}(\mathbf{x},t;f_{M})$
is consistent with Refs.\cite{Tessarotto2009,Ellero2005}. However, in the
case of non-Gaussian KDFs, the linearity condition here imposed as a kinetic
closure condition on $\mathbf{F}(\mathbf{x},t;f)$ (see \emph{Axiom \#3b})
actually leads to a representation of the mean-field in terms of the
extended fluid fields $\mathbf{Q}(\mathbf{r},t)$ and $\underline{\underline{{%
\Pi }}}(\mathbf{r},t)$ which is different from that adopted previously in
Refs.\cite{Ellero2005,Tessarotto2009}.
\end{itemize}

In addition, in agreement with the GENERIC dynamical model
\cite{Grmela1997,Oettnger1997}, it is possible to show that:

\begin{itemize}
\item Both in the case of CNSFE and INSFE the IKT-statistical description
permits to represent the reduced set of fluid fields $\left\{ Z_{1}\right\} $
[see Eq.(\ref{REDUCED-FLUID FIELDS})] in terms of velocity and phase-space
moments (bundle structure on $\Gamma $).

\item The functional setting for $f(\mathbf{x},t)$ can be suitably
prescribed so that $\left\{ f,\Gamma \right\} $ is \emph{compatible} with
the physical observables \cite{Grmela1997}. In other words, the predictions
for the fluid fields obtained in this way in terms of the statistical model
are in agreement with the experimental observations. This includes, in
particular, the prescription of appropriate kinetic boundary and initial
conditions \cite{Ellero2005}.

\item The NS-DS is a deterministic, non-conservative, irreversible and
non-canonical dynamical system. In fact, through the fluid fields, the
mean-field defined by Eq.(\ref{FF-1}) becomes generally explicitly
time-dependent, while generally $\frac{\partial }{\partial \mathbf{v}}\cdot
\mathbf{F}\neq 0$. Hence, the NS-DS [see Eq.(\ref{DYN-S})] defined in terms
of $\mathbf{F}$ is manifestly irreversible and non-conservative.
Furthermore, it is possible to show that Eq.(\ref{FF-1}) is non-variational,
and hence intrinsically non-symplectic, so that it cannot be cast in local
canonical form.
\end{itemize}

\section{Dynamics of NS ideal tracer particles in deterministic fluids}

A basic consequence of the previous theorem is that the functional form of $%
\Delta \mathbf{F}$ remains non-unique for ITPs. Its possible unique
prescription requires, therefore, the adoption of suitable additional
kinetic closure conditions\emph{. }For this purpose in this section we
intend to show that \emph{for all TTPs} $\Delta \mathbf{F}$ can be uniquely
determined in such a way to fulfill the requirement of the Gedanken
experiment (\emph{GDE-requirements \#1-\#6}).

\subsection{Preliminary Lemma}

Let us first show that due to the kinetic constraints (\ref{TANGENCY
CONDITION}) and (\ref{NO-PRECESSION}) for TTPs the pseudo-vector $\mathbf{%
\Omega }(\mathbf{r},t)$ entering the evolution equation for $\mathbf{n}(%
\mathbf{r},t)$ [see Eq.(\ref{EVOLUTION EQUATION for n})] is actually
uniquely determined. In fact the following result holds.\bigskip

\textbf{LEMMA to THM.2 - General form of }$\mathbf{\Omega }(\mathbf{r},t)$

\emph{If }$\Omega _{1}\equiv \Omega \times I$\emph{\ denotes the existence
domain of the fluid fields }$\left\{ Z\right\} $,\emph{\ let us assume that
in the subset }$\widehat{\Omega }_{1}\subseteq $\emph{\ }$\Omega _{1}$\emph{%
\ in which }$\left\vert \nabla \widehat{p}_{1}\right\vert \neq 0$\emph{:}

1) \emph{the real unit vectors} $\mathbf{n}(\mathbf{r},t),\mathbf{b}(\mathbf{%
r},t)$\emph{\ and the pseudo-vector }$\mathbf{\Omega }(\mathbf{r},t)$\emph{\
are all differentiable and suitably smooth;}

2) \emph{the} \emph{unit vector} $\mathbf{n}(\mathbf{r},t)$\emph{\ satisfies
the kinetic constraint (\ref{TANGENCY CONDITION}) and the initial-value
problem (\ref{EVOLUTION EQUATION for n});}

3) \emph{moreover, }$\mathbf{\Omega }(\mathbf{r},t)$\emph{\ satisfies the
constraint (\ref{NO-PRECESSION}) and is defined in the limit} $p_{1}(\mathbf{%
r},t)\rightarrow 0^{+}$ \emph{and also for arbitrary finite values of the
kinetic pressure }$p_{1}(\mathbf{r},t)$.

\emph{It follows that, in the domain }$\widehat{\Omega }_{1},$ $\mathbf{%
\Omega }(\mathbf{r},t)$\emph{\ necessarily takes the form:}%
\begin{equation}
\mathbf{\Omega }(\mathbf{r},t)=\mathbf{b}(\mathbf{\mathbf{r}},t)\mathbf{%
\times }\frac{d\mathbf{b}(\mathbf{r},t)}{dt}+c(\mathbf{r},t)\mathbf{b}(%
\mathbf{r},t),  \label{FINAL FORM}
\end{equation}%
\emph{where the GDE-requirement \#6 implies}%
\begin{equation}
c(\mathbf{r},t)=\mathbf{b}\times \mathbf{n}\cdot \frac{d\mathbf{n}}{dt}%
\equiv \mathbf{\Omega }(\mathbf{r},t)\cdot \mathbf{b}(\mathbf{r},t)=-\mathbf{%
\xi }(\mathbf{r},t)\cdot \mathbf{b}(\mathbf{r},t),  \label{pseudoscalar}
\end{equation}%
\emph{with }$\mathbf{\xi }(\mathbf{r},t)$ \emph{denoting the local fluid
vorticity.}

\emph{Proof - }In fact,$\ $by definition $\left\vert \mathbf{n}(\mathbf{r}%
,t)\right\vert =1$ and hence $\frac{d\mathbf{n}(\mathbf{r},t)}{dt}\cdot
\mathbf{n}(\mathbf{r},t)=0$, so that there must exist a pseudo-vector $%
\mathbf{\Omega }(\mathbf{r},t)$ such that Eq.(\ref{EVOLUTION EQUATION for n}%
) holds identically for arbitrary initial condition $\mathbf{n}(\mathbf{r}%
(t_{o}),t_{o})=\mathbf{n}(\mathbf{r}_{o},t_{o})$. Let us now impose the
validity of the kinetic constraint (\ref{TANGENCY CONDITION}), implying
\begin{equation}
\frac{d}{dt}\mathbf{n}(\mathbf{r},t)\cdot \mathbf{b}(\mathbf{r},t)=-\mathbf{n%
}(\mathbf{r},t)\cdot \frac{d}{dt}\mathbf{b}(\mathbf{r},t).
\label{CONSTRAINT II}
\end{equation}%
Hence, necessarily $\mathbf{\Omega }(\mathbf{r},t)$ satisfies the equation%
\begin{equation}
\mathbf{n}(\mathbf{r},t)\cdot \frac{d}{dt}\mathbf{b}(\mathbf{r},t)=-\left[
\mathbf{\Omega }(\mathbf{r},t)\times \mathbf{n}(\mathbf{r},t)\right] \cdot
\mathbf{b}(\mathbf{r},t),
\end{equation}%
i.e., due to the arbitrariness of the unit vector $\mathbf{n}(\mathbf{r},t)$%
, $\mathbf{b}(\mathbf{r},t)\times \mathbf{\Omega }(\mathbf{r},t)=-\frac{d}{dt%
}\mathbf{b}(\mathbf{r},t)$. This yields for $\mathbf{\Omega }(\mathbf{r},t)$
a general solution of the form%
\begin{equation}
\mathbf{\Omega }(\mathbf{r},t)=\mathbf{b}(\mathbf{r},t)\times \frac{d\mathbf{%
b}(\mathbf{r},t)}{dt}+c(\mathbf{r},t)\mathbf{b}(\mathbf{r},t).  \label{OMEGA}
\end{equation}%
Substituting this solution in Eq.(\ref{EVOLUTION EQUATION for n}) and taking
the scalar product of the resulting equation by $\mathbf{b}\times \mathbf{n}$%
, in validity of GDE-requirement \#6\emph{\ }Eq.(\ref{pseudoscalar}) follows
identically, which uniquely determines the form of $\mathbf{\Omega }(\mathbf{%
r},t)$.

\textbf{Q.E.D.}

\subsection{Construction of TTP solutions}

Let us now prove the existence of the TTPs, particular solutions of the
initial-value problem (\ref{eqq-2}).\ More precisely, we intend to prove
that for all $t\in I$ and an appropriate choice of the mean-field $\mathbf{F}%
+\Delta \mathbf{F}$, the NS-DS (\ref{DYN-S}) [or equivalent \emph{RD-NS-DS})
(\ref{RD-NS-DS})], necessarily maps an arbitrary TTP initial state $\mathbf{x%
}(t_{o})=\mathbf{x}_{o}$ into a TTP state $\mathbf{x}(t)=T_{t_{o},t}\mathbf{x%
}_{o}$. For this purpose we impose that, consistent with GDE-requirements
\#1-\#6, $\mathbf{F}$ is defined by Eqs.(\ref{FF-2})-(\ref{FF-3}). We intend
to show that, as a consequence, for all TTPs both $\mathbf{F}$ and the gauge
field $\Delta \mathbf{F}$ [see Eq.(\ref{gauge-condition})] are necessarily
\emph{uniquely determined}. On the other hand, in view of GDE, one expects
that the vector field $\mathbf{F}$ which characterizes TTP dynamics should
not depend on the form of the KDF. If true, the result would clearly be
conceptually important because it would imply the uniqueness of TTP dynamics
in all cases. For reference, let us first consider the case of the Gaussian
KDF $f=f_{M}$, leaving the extension to a non-Gaussian KDF to the discussion
below. Then, the following theorem holds.

\textbf{THEOREM 2 - Existence and uniqueness of TTP dynamics}

\emph{In validity of THM.1, let us require that }$\Delta \mathbf{F}$ \emph{%
and} $\mathbf{F}_{u}$\emph{\ are analytic functions in }$\Gamma $ \emph{%
which are defined also in the limit} $p_{1}(\mathbf{r},t)\rightarrow 0^{+}%
\emph{.}$\emph{\ Then} \emph{it follows that:}

T2$_{1})$\emph{\ The initial-value problem defined by (\ref{eqq-2}) admits
particular solutions fulfilling the GDE-requirements \#1-\#6, here denoted
as TTPs. In particular, for all }$t\in I$ \emph{they are characterized by a
relative kinetic velocity defined by Eq.(\ref{REPRESENTATION of u}) and is
such that: }

A) \emph{the local magnitude of the relative kinetic velocity} $\left\vert
\mathbf{u}(t)\right\vert $ \emph{is determined by the equation:}
\begin{equation}
\left\vert \mathbf{u}(t)\right\vert =u_{th}(\mathbf{r},t)\equiv \beta v_{th}(%
\mathbf{r},t),  \label{TTP-2}
\end{equation}%
\emph{with }$\beta \geq 0$ \emph{being independent of }$(\mathbf{r},t)$\emph{%
;}

B)\emph{\ }$\mathbf{n}(\mathbf{r},t)$ \emph{satisfies both the constraint
equation (\ref{TANGENCY CONDITION-2}) and the initial-value problem (\ref%
{EVOLUTION EQUATION for n});}

C) $\mathbf{\Omega }(\mathbf{r},t)$ \emph{is uniquely determined by Eqs.(\ref%
{FINAL FORM}) and (\ref{pseudoscalar}).}

T2$_{2})$\emph{\ For arbitrary TTPs,} \emph{the mean-field} $\mathbf{F}(%
\mathbf{r},\mathbf{u}_{th},t;f_{M})+\Delta \overline{\mathbf{F}}$ \emph{has
necessarily the unique representation}%
\begin{equation}
\mathbf{F}(\mathbf{r},\mathbf{u}_{th},t;f_{M})+\Delta \overline{\mathbf{F}}=%
\mathbf{F}_{H}+\mathbf{u}_{th}\cdot \nabla \mathbf{V}+\frac{\mathbf{u}_{th}}{%
2}\frac{D}{Dt}\ln \left( \widehat{p}_{1}\right) +\beta u_{th}\mathbf{\Omega }%
(\mathbf{r},t)\times \mathbf{n}.  \label{TTP-4}
\end{equation}%
\emph{This implies that, in the case of the Gausssian KDF }$f=f_{M}$\emph{,
and for arbitrary }$\beta \geq 0,$\emph{\ the vector field }$\Delta
\overline{\mathbf{F}}\equiv \Delta \mathbf{F}(\mathbf{r},\mathbf{u}_{th},t)$%
\emph{\ has the unique representation}%
\begin{equation}
\left\{
\begin{array}{c}
\Delta \overline{\mathbf{F}}=\Delta \overline{\mathbf{F}}_{0}+\Delta
\overline{\mathbf{F}}_{1}, \\
\Delta \overline{\mathbf{F}}_{0}(\mathbf{r},\mathbf{u}_{th},t)=-\left[ \frac{%
v_{th}^{2}}{2}\nabla \ln \rho +\frac{v_{th}^{2}}{2}\nabla \ln \left(
\widehat{p}_{1}\right) \left( \beta ^{2}-\frac{1}{2}\right) \right] , \\
\Delta \overline{\mathbf{F}}_{1}(\mathbf{r},\mathbf{u}_{th},t)=\beta u_{th}%
\mathbf{\Omega }(\mathbf{r},t)\times \mathbf{n}.%
\end{array}%
\right.  \label{TTP-3}
\end{equation}

T2$_{3})$\emph{\ Particular solutions of the form (\ref{REPRESENTATION of u}%
) which fulfill requirements A-C must satisfy the initial conditions}%
\begin{equation}
\left\{
\begin{array}{c}
\left. \mathbf{r}(t_{o})=\mathbf{r}_{o},\right. \\
\left. \mathbf{u}(t_{o})=\beta v_{th}(\mathbf{r}_{o},t_{o})\mathbf{n}(%
\mathbf{r}_{o},t_{o}),\right.%
\end{array}%
\right.  \label{TTP-5-6}
\end{equation}%
\emph{with }$\beta \geq 0$\emph{\ being an arbitrary real constant,} $p_{1}(%
\mathbf{r}_{o},t_{o})$ \emph{the initial kinetic pressure} \emph{and} $%
\mathbf{n}(\mathbf{r}_{o},t_{o})$ \emph{a unit vector satisfying the
orthogonality condition}%
\begin{equation}
\mathbf{n}(\mathbf{r}_{o},t_{o})\cdot \nabla \widehat{p}_{1}(\mathbf{r}%
_{o},t_{o})=0.  \label{TTP-7}
\end{equation}

\emph{Proof - }T2$_{1}-$T2$_{2})$ For generality let us assume that $%
\left\vert \nabla p_{1}\right\vert \neq 0$ everywhere in $\Omega \times I$.
Then, it is sufficient to prove the theorem in the subset $\widehat{\Omega }%
_{1}\subseteq $\ $\Omega _{1}\equiv \Omega \times I$\ in which $\left\vert
\nabla \widehat{p}_{1}\right\vert \neq 0$. Let us show that in $\widehat{%
\Omega }_{1},$ for an arbitrary non-negative constant $\beta \in
\mathbb{R}
^{+}$, a particular solution of the initial-value problem (\ref{eqq-2}) of
the type (\ref{REPRESENTATION of u}), which satisfies requirements A-C,
exists and is unique. In fact, let us assume that $\left\vert \mathbf{u}%
(t)\right\vert $ is of the form (\ref{TTP-2}), with $\beta \equiv \beta (%
\mathbf{r},t)\geq 0$ denoting now an arbitrary smooth real function of $(%
\mathbf{r},t)$ defined in $\overline{\Omega }\times I$. It is immediate to
show that necessarily $\beta $ must be everywhere constant with respect to $(%
\mathbf{r},t)$ in $\overline{\Omega }\times I$. Indeed, Eq.(\ref{eqq-2})
requires%
\begin{equation}
\left\{
\begin{array}{c}
\frac{d}{dt}\left( \beta v_{th}(\mathbf{r},t)\right) =\frac{v_{th}^{2}}{2}%
\mathbf{n}\cdot \nabla \ln \rho +\frac{1}{2}\beta v_{th}(\mathbf{r},t)\frac{D%
}{Dt}\ln \left( \widehat{p}_{1}\right) + \\
+\frac{v_{th}^{2}}{2}\mathbf{n}\cdot \nabla \ln \left( \widehat{p}%
_{1}\right) \left( \beta ^{2}-\frac{1}{2}\right) +\mathbf{n}\cdot \Delta
\overline{\mathbf{\mathbf{F}}}, \\
\frac{d}{dt}\mathbf{n}(\mathbf{r},t)=\frac{1}{\beta u_{th}}\left[ \frac{%
v_{th}^{2}}{2}\nabla \ln \rho +\frac{v_{th}^{2}}{2}\nabla \ln \left(
\widehat{p}_{1}\right) \left( \beta ^{2}-\frac{1}{2}\right) \right] \cdot
\left( \underline{\underline{\mathbf{1}}}-\mathbf{nn}\right) + \\
+\frac{1}{\beta u_{th}}\Delta \overline{\mathbf{\mathbf{F}}}\cdot \left(
\underline{\underline{\mathbf{1}}}-\mathbf{nn}\right) \equiv \mathbf{\Omega }%
(\mathbf{r},t)\times \mathbf{n}.%
\end{array}%
\right.  \label{TTP-10}
\end{equation}%
On the other hand, imposing the constraint (\ref{TANGENCY CONDITION})
requires necessarily, thanks to the Lemma, that $\mathbf{n}(\mathbf{r},t)$
must satisfy the initial-value problem (\ref{EVOLUTION EQUATION for n}).\ We
require for consistency%
\begin{eqnarray}
&&\left. \mathbf{n}\cdot \nabla \ln \left( \widehat{p}_{1}\right) =0,\right.
\\
&&\left. \frac{v_{th}^{2}}{2}\mathbf{n}\cdot \nabla \ln \rho +\mathbf{n}%
\cdot \Delta \overline{\mathbf{\mathbf{F}}}=0.\right.
\end{eqnarray}%
Hence, since by assumption $\Delta \overline{\mathbf{F}}$ is defined also in
the limit $p_{1}(\mathbf{r},t)\rightarrow 0^{+}$ (or equivalently $\beta
\rightarrow 0^{+})$, it follows necessarily that%
\begin{eqnarray}
&&\left. \Delta \overline{\mathbf{F}}=\Delta \overline{\mathbf{F}}_{0}(%
\mathbf{r},\mathbf{u}_{th},t)+\Delta \overline{\mathbf{F}}_{1}(\mathbf{r},%
\mathbf{u}_{th},t),\right.  \label{ALFA-1} \\
&&\left. \Delta \overline{\mathbf{F}}_{1}(\mathbf{r},\mathbf{u}_{th},t)\cdot
\left( \underline{\underline{\mathbf{1}}}-\mathbf{nn}\right) =\beta u_{th}%
\mathbf{\Omega }(\mathbf{r},t)\times \mathbf{n}(\mathbf{r},t),\right.
\end{eqnarray}%
with $\mathbf{\Omega }(\mathbf{r},t)$ being given by the Lemma and%
\begin{equation}
\frac{d\beta (\mathbf{r},t)}{dt}=0.  \label{betotal}
\end{equation}%
Therefore, due to the arbitrariness of $\mathbf{v}\equiv $ $\mathbf{V}(%
\mathbf{r},t)+u_{th}(\mathbf{r},t)\mathbf{n}(\mathbf{r},t)$, $\beta $ is
necessarily independent of $(\mathbf{r},t)$;\emph{\ }furthermore, $\Delta
\overline{\mathbf{F}}(\mathbf{r},\mathbf{u}_{th},t;f_{M})$ and $\mathbf{F}(%
\mathbf{r},\mathbf{u}_{th},t;f_{M})$ are necessarily of the form (\ref{TTP-3}%
) and (\ref{TTP-4}). T2$_{3})$ Finally, the initial conditions (\ref{TTP-5-6}%
) are an immediate consequence of Eq.(\ref{REPRESENTATION of u}) and the
requirements A-C.

\textbf{Q.E.D.}

\bigskip

Let us now consider the extension of the theorem to the case of a
non-Gaussian KDF. We notice that the gauge field $\Delta \overline{\mathbf{F}%
}$ evaluated for the state of a generic TTP, can always be identified with
the vector field
\begin{equation}
\Delta \overline{\mathbf{F}}=\Delta \overline{\mathbf{F}}_{0}+\Delta
\overline{\mathbf{F}}_{1}-\mathbf{F}_{a}(\mathbf{r},\mathbf{u}_{th},t;f),
\end{equation}%
where $\Delta \overline{\mathbf{F}}_{0}$ and $\Delta \overline{\mathbf{F}}%
_{1}$ are still given by Eq.(\ref{TTP-3}), while $\mathbf{F}_{a}(\mathbf{r},%
\mathbf{u}_{th},t;f)$ is prescribed according to Eq.(\ref{FF-4}) and
computed for $\mathbf{u}=\mathbf{u}_{th}$. As a consequence, it is immediate
to show that, for TTPs, this prescription of $\Delta \overline{\mathbf{F}}$
warrants the uniqueness of the vector field $\mathbf{F}(\mathbf{r},\mathbf{u}%
_{th},t;f_{M})$, in agreement with GDE. Therefore, for TTPs, its form is
independent of the form of the KDF, namely $\mathbf{F}(\mathbf{r},\mathbf{u}%
_{th},t;f_{M})=\mathbf{F}(\mathbf{r},\mathbf{u}_{th},t;f)$.

\subsection{Implications and physical interpretation}

Let us briefly analyze the implications of THM.2. \ First, we remark that by
construction for all TTPs the mean-field acceleration $\mathbf{F}+\Delta
\mathbf{F}$ is \emph{unique} and \emph{independent of the form of the KDF} $%
f(\mathbf{x},t)$. As a consequence TTP particular solutions [of the
initial-value problem (\ref{eqq-2})] realize a classical dynamical system
with existence domain $\Gamma _{1}\times I$, $\Gamma _{1}$ denoting a
suitable subset of the phase-space $\Gamma $ (see related discussion in
Section 6). This is defined by a\ homeomorphism of the form $%
T_{t_{o},t}^{(TTP)}:\mathbf{y}_{o}\rightarrow \mathbf{y}%
(t)=T_{t_{o},t}^{(TTP)}\mathbf{y}_{o}$, where for all $t\in I$,%
\begin{equation}
\mathbf{y}(t)\equiv \left\{ \mathbf{r},\mathbf{u}(t)\equiv \mathbf{u}%
_{th}(t)\right\}  \label{FORM}
\end{equation}%
and $\mathbf{u}_{th}(t)$ is given by Eq.(\ref{REPRESENTATION of u}).
Manifestly $T_{t_{o},t}^{(TTP)}$ is a subset of the RD-NS-DS defined by Eq.(%
\ref{RD-NS-DS}) [or equivalent of the NS-DS defined by Eq.(\ref{DYN-S})].

Furthermore, let us assume that the initial conditions for Eq.(\ref{eqq-2})
are of the form $\mathbf{y}(t_{o})\equiv \left\{ \mathbf{r}_{o}\mathbf{,u}%
_{o}\equiv \mathbf{u}_{th}(t_{o})\right\} $, where $\mathbf{u}_{th}(t_{o})$
is prescribed by Eq.(\ref{REPRESENTATION of u}) while the initial unit
vector $\mathbf{n}(\mathbf{r},t_{o})$ satisfies the constraint (\ref{INITIAL
DIRECTION}) at $\mathbf{r}\equiv \mathbf{r}_{o}$. Then, thanks to THM.2, it
follows that for all $t\in I$, $\mathbf{y}(t)$ it is necessarily of the form
(\ref{FORM}), i.e., it defines, for all $t\in I$, a TTP. In addition, thanks
to Eq.(\ref{betotal}) it follows that $\beta $ is constant and therefore is
uniquely determined for each TTP by the initial state $\mathbf{y}(t_{o})$,
namely $\beta =\frac{\left\vert \mathbf{u}_{o}\right\vert }{v_{th}\left(
\mathbf{r}_{o},t_{o}\right) }$.

We notice that, by construction, TTPs are uniquely associated to the local
state of the fluid. As a consequence, this permits to determine also for the
remaining NS-ITPs an explicit representation of the mean-field which is
necessarily of the form $\mathbf{F}(\mathbf{r},\mathbf{u},t;f)$, namely it
depends explicitly on the KDF. Similarly the gauge-field is of the type $%
\Delta \mathbf{F}(\mathbf{r},\mathbf{u},t;f)$. Both hold \emph{for arbitrary}
$\mathbf{u}\in U$ and satisfy at the same time the requirements posed by
THM.2, namely that for $\mathbf{u}\equiv \mathbf{u}_{th}$, the sum of the
two vectors $\mathbf{F}(\mathbf{r},\mathbf{u},t;f)+\Delta \mathbf{F}(\mathbf{%
r},\mathbf{u},t;f)$ must reduce to Eq.(\ref{TTP-4}). For definiteness, in
validity of THM.1, let us consider the case of a generally non-Gaussian KDF $%
f(\mathbf{r},\mathbf{u},t)$, with $f(\mathbf{r},\mathbf{u},t)$ denoting a
smooth, strictly positive function which satisfies at the same time Axioms
\#1-\#6, i.e., is a particular solution of Eq.(\ref{IKE}). Let us therefore
determine $\Delta \mathbf{F}(\mathbf{r},\mathbf{u},t;f)$ in such a way to
fulfill the constraint equations (\ref{TTP-4}) and the gauge condition (\ref%
{gauge-condition}). For this purpose it is sufficient to let%
\begin{equation}
\Delta \mathbf{F}(\mathbf{r},\mathbf{u},t;f)=\Delta \overline{\mathbf{F}}(%
\mathbf{r},\mathbf{u}_{th},t)\frac{f(\mathbf{r},\mathbf{u}_{th},t)}{f(%
\mathbf{r},\mathbf{u},t)},  \label{DELTA-F-i}
\end{equation}%
with $\Delta \overline{\mathbf{F}}(\mathbf{r},\mathbf{u}_{th},t)$ being
defined by Eq.(\ref{TTP-3}).

Finally, an interesting issue concerns the physical interpretation
of the evolution equation for the unit vector
$\mathbf{n}(\mathbf{r},t)$ [i.e., the direction of the particle
relative velocity] and related pseudo-vector $\mathbf{\Omega }$.\
In fact, that Eqs.(\ref{EVOLUTION EQUATION for n}) are similar to
the Euler equations for a rigid body
rotating with angular velocity $\mathbf{\Omega }_{r}\equiv -\mathbf{\Omega }$%
. This suggests that due to the GDE-requirements \#5 and \#6, two different
physical effects contribute to $\mathbf{\Omega }$. These are due both to the
rotation of the unit vector $\mathbf{n}(\mathbf{r},t)$ as determined by Eq.(%
\ref{EVOLUTION EQUATION for n}) as a consequence of non-uniform specific
kinetic pressure, and the contribution of fluid vorticity specified by Eq.(%
\ref{NO-PRECESSION}). Indeed, from Eq.(\ref{OMEGA}), denoting by $\frac{D%
\mathbf{b}(\mathbf{r},t)}{Dt}\equiv \frac{\partial }{\partial t}+\mathbf{V}(%
\mathbf{r},t)\cdot \nabla $ the fluid convective derivative and since by
construction $\nabla \mathbf{u=}-\nabla \mathbf{V}$, it follows
\begin{equation}
\mathbf{\Omega }_{r}\equiv -\mathbf{\Omega }=\frac{d\mathbf{b}(\mathbf{r},t)%
}{dt}\times \mathbf{b}(\mathbf{r},t)-c(\mathbf{r},t)\mathbf{b}(\mathbf{r},t)=%
\frac{D\mathbf{b}(\mathbf{r},t)}{Dt}\times \mathbf{b}(\mathbf{r},t)+\left(
\mathbf{u}\cdot \nabla \right) \mathbf{b}(\mathbf{r},t)\times \mathbf{b}(%
\mathbf{r},t)-c(\mathbf{r},t)\mathbf{b}(\mathbf{r},t),  \label{AAA-1}
\end{equation}%
where%
\begin{eqnarray}
&&\left. \left( \mathbf{u}\cdot \nabla \right) \mathbf{b}(\mathbf{r}%
,t)\times \mathbf{b}(\mathbf{r},t)-c(\mathbf{r},t)\mathbf{b}(\mathbf{r}%
,t)=\right.  \label{AAA-2} \\
&&\left. =\mathbf{\xi }-\frac{1}{\left\vert \nabla p_{1}\right\vert }\left[
\mathbf{b}\times \nabla (\nabla p_{1}\cdot \mathbf{V})+\mathbf{b}\times
\left( \nabla p_{1}\cdot \nabla \right) \mathbf{V}\right] \right. .  \notag
\end{eqnarray}%
In the last equation the first term on the r.h.s. denotes the vorticity $%
\mathbf{\xi }\equiv \nabla \times \mathbf{V}(\mathbf{r},t)$. This means that
near a vortex the motion of TTPs is qualitatively similar to that of a
rotating rigid body. However, by inspection of the remaining terms in Eqs.(%
\ref{AAA-1}) and (\ref{AAA-2}), it is evident that more complex
particle-acceleration effects may be present, which are driven by
time-dependent pressure and velocity-gradients contributions.

\section{The TTP-statistical description}

In this section we introduce a\emph{\ statistical description} associated to
the ensemble of TTPs,\emph{\ }denoted as \emph{TTP-statistical model,} which
is represented by the couple $\left\{ f_{1},\Gamma _{1}\right\} $. We intend
to show that, like the IKT-statistical model $\left\{ f,\Gamma \right\} $,
also $\left\{ f_{1},\Gamma _{1}\right\} $ determines uniquely the time
evolution of the complete set of fluid fields $\left\{ Z\right\} $. However,
the result is conceptually important because $\Gamma _{1}$ is a \emph{%
reduced-dimension subset of} $\Gamma $. For this purpose, we notice that if $%
\mathbf{u}_{th}$ denotes the relative velocity of an arbitrary TTP endowed
with a relative-Newtonian state $\mathbf{y}\equiv \left( \mathbf{r},\mathbf{u%
}_{th}\right) $, then $\mathbf{u}_{th}$ spans the subset of velocity space $%
U $:
\begin{equation}
U_{1}=\left\{ \left. \mathbf{u}\right\vert \text{ }\mathbf{u}\in U,\mathbf{u}%
=\mathbf{u}_{th}=\beta v_{th}(\mathbf{r},t)\mathbf{n}(\mathbf{r},t),\text{ }%
\mathbf{n}(\mathbf{r},t)\cdot \mathbf{b}(\mathbf{r},t)=0,\beta \in
\mathbb{R}
^{+}\right\} .  \label{U1}
\end{equation}%
Here by construction $\beta ^{2}=\frac{u_{th}^{2}}{v_{th}^{2}(\mathbf{r},t)}$
is a constant independent of $(\mathbf{r},t)$. As a consequence, it follows
that $\Gamma _{1}$ is the subset of $\Gamma ,$ $\Gamma _{1}=\Omega \times
U_{1}$, with $U_{1}\subset U\equiv
\mathbb{R}
^{3}$ and $dim(U_{1})=2$.

To define a KDF $f_{1}(\mathbf{r,u}_{th},t)$ on $\Gamma _{1}$ let us first
consider the PDF defined on $U$ in terms of the KDF $f(\mathbf{r,u},t)$. For
a prescribed IKT-statistical model $\left\{ f,\Gamma \right\} $, the
corresponding \emph{conditional velocity PDF} on $U_{1}$ is defined as%
\begin{equation}
\widehat{f}_{1}(\mathbf{r,u}_{th},t)\equiv \frac{f(\mathbf{r,u}_{th},t)}{%
\int\limits_{U_{1}}d\eta f(\mathbf{r,u},t)}.  \label{POSITION-0}
\end{equation}%
Here, introducing for $\mathbf{u}_{th}$ a representation in terms
of the spherical coordinates ($u\equiv u_{th},\varphi ,\vartheta
$) and requiring the $\widehat{u}_{z}\equiv
\mathbf{b}(\mathbf{r},t)$ it follows by construction that both the
conditional PDF and KDF defined by
Eqs.(\ref{POSITION-0}) and (\ref{POSITION-1}) are independent of the angle $%
\vartheta $. In particular it follows that by definition $%
\int\limits_{U_{1}}d\eta f(\mathbf{r},\mathbf{u},t)=\int\limits_{U}d\mathbf{v%
}\delta \left( \vartheta -\pi /2\right) f(\mathbf{r},\mathbf{u},t)$ and
hence $d\eta =u^{2}dud\varphi ,$ while the corresponding phase-space measure
is $d\mathbf{x}_{1}\equiv d^{3}rd\eta $. Thus, the \emph{conditional KDF} on
$\Gamma _{1}$ is defined as%
\begin{equation}
f_{1}(\mathbf{r},\mathbf{u}_{th},t)=\rho (\mathbf{r},t)\widehat{f}_{1}(%
\mathbf{r},\mathbf{u}_{th},t).  \label{POSITION-1}
\end{equation}%
In particular, if $f(\mathbf{r},\mathbf{u}_{th},t)$ coincides with the
Gaussian KDF (\ref{Maxwellian}), it follows that%
\begin{equation}
f_{1}(\mathbf{r},\mathbf{u}_{th},t)=\frac{2\rho (\mathbf{r},t)}{\pi
^{3/2}v_{th}^{3}}\exp \left\{ -\beta ^{2}\right\} \equiv f_{1M}(\mathbf{r},%
\mathbf{u}_{th},t).  \label{GAUSSIAN f_1}
\end{equation}%
The main result can be summarized by the following theorem.

\textbf{THEOREM 3 - TTP-statistical model for CNSFE}

\emph{Let us require that the IKT-statistical model satisfies THMs. 1 and 2}$%
.$\emph{\ Then it follows that the conditional KDF} $f_{1}(t)\equiv f_{1}(%
\mathbf{r},\mathbf{u}_{th},t)$ \emph{defined by Eqs.(\ref{POSITION-0}) and (%
\ref{POSITION-1}) has the following properties:}

T3$_{1})$\emph{\ It is a particular solution of IKE which is
independent of the angle }$\vartheta $\\emph{.}

T3$_{2})$\emph{\ It satisfies the functional constraints:}
\begin{equation}
\left\{
\begin{array}{c}
\int\limits_{U_{1}}d\eta f_{1}(\mathbf{r},\mathbf{u}_{th},t)=\rho (\mathbf{%
\mathbf{r}},t), \\
\frac{1}{\rho (\mathbf{\mathbf{r}},t)}\int\limits_{U_{1}}d\eta \mathbf{v}%
f_{1}(\mathbf{r},\mathbf{u}_{th},t)=\mathbf{V}(\mathbf{\mathbf{r}},t), \\
\int\limits_{U_{1}}d\eta \frac{1}{3}u^{2}f_{1}(\mathbf{r},\mathbf{u}%
_{th},t)=p_{1}(\mathbf{r},t), \\
S\left( f_{1}(t)\right) \equiv -\int\limits_{\Gamma _{1}}d\mathbf{x}%
_{1}f_{1}(\mathbf{r},\mathbf{u}_{th},t)\ln f_{1}(\mathbf{r},\mathbf{u}%
_{th},t)=S_{T}\left( t\right)%
\end{array}%
\right.  \label{MOMENTS-reduced}
\end{equation}%
\emph{(correspondence principle).}

T3$_{3})$\emph{\ Its velocity moment equations,} \emph{determined from IKE
in terms of the weight-functions} $\left\{ G(\mathbf{x},t)\right\} =\left\{
1,\mathbf{v},\frac{1}{3}u^{2}\right\} $,\emph{\ imply again Eqs.(\ref%
{MOMENTS}) and therefore, together with (\ref{POSITION-0}) and the
constraint equation (\ref{constraint on p_0}),\ they coincide again with
CNSFE, so that in particular} $f_{1}(t)$ \emph{satisfies the entropy law.}

\emph{Proof - }T3$_{1})$ In fact, due to the hypothesis and the definitions (%
\ref{POSITION-0}) and (\ref{POSITION-1}), it follows that in $U_{1}$%
\begin{equation}
f_{1}(\mathbf{r},\mathbf{u}_{th},t)=2f(\mathbf{r},\mathbf{u}_{th},t),
\label{NEW}
\end{equation}%
and is therefore by construction independent of the angle $\vartheta $.
Hence, in the subset $\Gamma _{1}\times I$, $f_{1}(\mathbf{r},\mathbf{u}%
_{th},t)$ is a particular solution of IKE, in the sense that it satisfies by
construction the statistical equation%
\begin{eqnarray}
&&\frac{\partial }{\partial t}f_{1}(\mathbf{r},\mathbf{\mathbf{u}}_{th},t)+%
\mathbf{v}\cdot \nabla f_{1}(\mathbf{r},\mathbf{\mathbf{u}}_{th},t)+\mathbf{F%
}(\mathbf{\mathbf{r}},\mathbf{\mathbf{\mathbf{\mathbf{u}}}}_{th},t)\cdot
\frac{\partial }{\partial \mathbf{\mathbf{\mathbf{\mathbf{u}}}}_{th}}f_{1}(%
\mathbf{r},\mathbf{\mathbf{\mathbf{\mathbf{\mathbf{u}}}}}_{th},t)+
\label{IKE-2} \\
&&\left. +f_{1}(\mathbf{r},\mathbf{u}_{th},t)\left[ \frac{\partial }{%
\partial \mathbf{v}}\cdot \mathbf{F}(\mathbf{\mathbf{r}},\mathbf{\mathbf{%
\mathbf{\mathbf{u}}}},t,f)f(\mathbf{r},\mathbf{u},t)\right] _{\mathbf{u=u}%
_{th}}=0.\right.  \notag
\end{eqnarray}

T3$_{2})$\ The proof follows by noting that, thanks to Eq.(\ref{NEW}) and
the fact that $f(\mathbf{r},\mathbf{u}_{th},t)$ satisfies by construction
the correspondence principle (see THM.1), the conditional KDF $f_{1}(\mathbf{%
r},\mathbf{\mathbf{u}}_{th},t)$ fulfills identically the correspondence
principle (\ref{MOMENTS-reduced}) too.

T3$_{3})$ Due to T3$_{1})$,T3$_{2})$ and Eq.(\ref{IKE-2}) the moment
equations coincide necessarily with CNSFE and hence, in particular,
consistent with the second principle of thermodynamics [see Eq.(\ref{5B})], $%
f_{1}(t)$ satisfies the weak H-theorem%
\begin{equation}
\frac{\partial }{\partial t}S\left( f_{1}(t)\right) \geq 0.
\label{H-THEOREM FOR f_1}
\end{equation}%
\textbf{Q.E.D.}

We remark that here:

\begin{itemize}
\item $f_{1}(\mathbf{r},\mathbf{u}_{th},t)$ denotes generally a \emph{%
non-Gaussian conditional KDF}. A particular solution is provided by the
\emph{Gaussian} \emph{conditional KDF }defined by\emph{\ }Eq.(\ref{GAUSSIAN
f_1}).

\item $f_{1}(\mathbf{r},\mathbf{u}_{th},t)$ satisfies the inverse kinetic
equation (\ref{IKE-2}).

\item $\left\{ f_{1},\Gamma _{1}\right\} $ is a reduced-dimension
statistical model for CNSFE problem. The fluid fields $\left\{ Z_{1}\right\}
$ are uniquely advanced in time by means of Eq.(\ref{IKE-2}), or equivalent,
by means of the integral Lagrangian IKE\emph{\ }(\ref{INTEGRAL LIOUVILLE EQ})%
$.$

\item $f_{1}(\mathbf{r},\mathbf{u}_{th},t)$ determines uniquely the
time-evolution of the fluid fields $\left\{ Z_{1}(\mathbf{r},t)\right\} $.

\item In view of the discussion presented above after THM.1, $\left\{
f_{1},\Gamma _{1}\right\} $ applies also to incompressible fluids described
either by the INSFE or INSE problems.
\end{itemize}

In the following we analyze basic implications of this result.

\section{TTP-dynamics in stochastic fluids}

A remarkable application of the TTP dynamics concerns the
modelling of stochastic tracer-particle dynamics in stochastic
fluids, such as for example due to temperature and pressure
fluctuations (\emph{thermal fluctuations}). \ Thermal fluctuations
are important in a wide variety of mesoscopic flows (see for
example Refs.\cite{Barrat,Zwanzig,Hohenberg}). Theoretically, they
are usually treated within the framework of fluctuating
hydrodynamics, an approach pioneered by Landau and Lifshitz
\cite{Landau1959,Reichl1998}. In this framework extra (stochastic)
terms are added to the fluid equations to model possible
stochastic effects, so that generally the functional form of the
corresponding fluid equations is actually \emph{modified }with
respect to the customary fluid equations. For example, in this
case the latter may typically include higher-order spatial
derivatives of the fluid fields. The numerical solution of
fluctuating hydrodynamic equations may present, as a consequence,
serious difficulties (which are nevertheless also present in the
case of the incompressible NS equations for isothermal fluids). A
possible alternative is represented by particle simulation methods
based on kinetic theory. In such an approach, unlike fluctuating
hydrodynamics:

1) Requirement \#1: the functional form of the corresponding fluid equations
and of the related initial-boundary value problem is left \emph{unchanged }%
\cite{Tessarotto20081,Tessarotto20082,Tessarotto20083}, i.e., the
differential operators appearing in the fluid stochastic equations are the
same ones entering the customary fluid equations in the absence of
stochasticity.

2) Requirement \#2:\ the stochastic fluid fields are assumed to be strong
solutions of the stochastic CNSFE problem (see Appendix A).

An approach of this type can be achieved by means of the TTP-statistical
model. \ A convenient representation of stochastic fluid fields of this
type, and fulfilling Requirements \#1 and \#2, is provided by Eq.(\ref%
{stochastic functions}). In this case the fluid fields are assumed to depend
on a suitable set of stochastic variables $\mathbf{\alpha }=\left\{ \alpha
_{i},i=1,k\right\} \in V_{\mathbf{\alpha }}\subseteq \mathbf{R}^{k}$, with $%
k\geq 1$, by assumption all independent of $(\mathbf{r,}t)$ and endowed with
a stochastic probability density $g(\mathbf{r},t,\alpha )$ on $V_{\mathbf{%
\alpha }}$ \cite{Tessarotto20081,Tessarotto20082,Tessarotto20083}. It must
be remarked that \textquotedblleft a priori\textquotedblright\ the
parameters $\mathbf{\alpha \equiv }\left( \alpha _{1},...,\alpha _{n}\right)
$ and the related probability density $g(\mathbf{r},t,\alpha )$ can be set
arbitrarily. Thus, they can in principle be chosen to provide prescribed
mathematical models of stochasticity. In the following we shall assume in
particular that the $\mathbf{\alpha }^{\prime }s$ are also independent of $(%
\mathbf{r},t)$. As a consequence, introducing the stochastic averaging
operator (\ref{stochastic averaging operator}), the fluid fields can be
represented in terms of the \emph{stochastic decomposition}%
\begin{equation}
\emph{\ }Z(\mathbf{r},t,\alpha )=\left\langle Z(\mathbf{r},t)\right\rangle _{%
\mathbf{\alpha }}+\delta Z(\mathbf{r},t,\alpha ),
\label{STOCHASTIC DECOMPOSITION-2}
\end{equation}%
$\left\langle Z(\mathbf{r},t)\right\rangle _{\mathbf{\alpha }}$ and $\delta {%
Z(\mathbf{r}},{t,\alpha \mathbb{)}}$ denoting respectively the corresponding
stochastic-averages and stochastic fluctuations of the fluid fields.

In particular, in contrast to fluctuating hydrodynamics, in the present
approach the functional form of the fluid equations is left unchanged. It
follows that the precise form of the stochastic-averaged fluid fields $%
\left\langle Z(\mathbf{r},t)\right\rangle _{\mathbf{\alpha }}$ and of the
stochastic fluctuations $\delta Z(\mathbf{r},t,\alpha )$ \emph{depends solely%
} on the model of stochasticity adopted, i.e., the choice of the set $%
\left\{ g\mathbf{(r},t,\alpha ),V_{\mathbf{\alpha }}\right\} $. This means
that its realization may generally depend on the possible \emph{sources of
stochasticity} adopted,\emph{\ }namely: 1) \emph{Stochastic initial
conditions:} in this case the initial fluid fields $Z(\mathbf{r}%
,t_{o})\equiv Z_{o}(\mathbf{r})$ are assumed stochastic, i.e. of the form, $%
Z_{o}(\mathbf{r},\alpha )=\left\langle Z_{o}(\mathbf{r},\alpha
)\right\rangle _{\mathbf{\alpha }}+\delta Z_{o}(\mathbf{r},\alpha ),$ with $%
\left\langle Z_{o}(\mathbf{r},\alpha )\right\rangle _{\mathbf{\alpha }}$ and
$\delta Z_{o}(\mathbf{r},\alpha )$ being suitable vector fields. 2) \emph{%
Stochastic boundary conditions:} this occurs if the boundary fluid fields $%
\left. Z_{w}(\mathbf{r},t)\right\vert _{\delta \Omega }$ are prescribed in
terms of a suitable stochastic vector field of the form $\left. Z_{w}(%
\mathbf{r},t,\alpha )_{\mathbf{\alpha }}\right\vert _{\delta \Omega }=\left.
\left\langle Z_{w}(\mathbf{r},t,\alpha )\right\rangle _{\mathbf{\alpha }%
}\right\vert _{\delta \Omega }+\left. \delta Z_{w}(\mathbf{r},t,\alpha
)\right\vert _{\delta \Omega }$. 3) \emph{Stochastic forcing:} in this case
the volume force density acting on the fluid is assumed stochastic, i.e., of
the form $\mathbf{f}(\mathbf{r},t,\alpha )=\left\langle \mathbf{f}(\mathbf{r}%
,t,\alpha )\right\rangle _{\mathbf{\alpha }}+\delta \mathbf{f}(\mathbf{r}%
,t,\alpha ),$ being $\left\langle \mathbf{f}(\mathbf{r},t,\alpha
)\right\rangle _{\mathbf{\alpha }}$ and $\delta \mathbf{f}(\mathbf{r}%
,t,\alpha )$ suitable vector fields.

\subsection{Langevin dynamics in fluctuating fluids}

As stated above, a fundamental consequence of THM.2 is the \emph{uniqueness }%
of the deterministic equations of motion for arbitrary TTPs belonging to a
compressible or incompressible, thermal or isothermal NS fluid. It is
immediate to show that the dynamics of TTPs is unique also when the same
fluids are considered stochastic. It follows that the relative state $%
\mathbf{y}$ of a generic TTP advances in time by means of a stochastic
dynamical system (DS), namely the flow generated by the initial value
problem associated to the stochastic equations of motion for TTPs (\emph{%
Langevin equations}):%
\begin{equation}
\left\{
\begin{array}{c}
\frac{d\mathbf{r}}{dt}=\beta v_{th}(\mathbf{r},t,\mathbf{\alpha })\mathbf{n}(%
\mathbf{r},t,\mathbf{\alpha })+\mathbf{V}(\mathbf{r},t,\mathbf{\alpha }), \\
\frac{d\mathbf{u}}{dt}=\mathbf{F}_{u}(\mathbf{r},\beta v_{th}\mathbf{n}(%
\mathbf{r},t),t,\mathbf{\alpha }), \\
\mathbf{y}(t_{o},\mathbf{\alpha })=\mathbf{y}_{o}(\mathbf{\alpha }),%
\end{array}%
\right.  \label{stochastic eqq-1}
\end{equation}%
where the unit vector $\mathbf{n}(\mathbf{r},t,\mathbf{\alpha })$ satisfies
the stochastic initial-value problem%
\begin{equation}
\left\{
\begin{array}{l}
\frac{d\mathbf{n}(\mathbf{r},t,\mathbf{\alpha })}{dt}=\mathbf{\Omega }(%
\mathbf{r},t,\mathbf{\alpha })\times \mathbf{n}(\mathbf{r},t,\mathbf{\alpha }%
), \\
\mathbf{n}(\mathbf{r}_{o},t_{o},\mathbf{\alpha })=\mathbf{n}_{o}(\mathbf{r},%
\mathbf{\alpha }).%
\end{array}%
\right.  \label{stochastic-initial-value problem}
\end{equation}%
In particular, it follows that the stochastic mean-field $\mathbf{F}$ is
provided by Eq.(\ref{FF-1}), with $\mathbf{F}_{u}\equiv \mathbf{F}_{u}(%
\mathbf{r},\beta v_{th}\mathbf{n}(\mathbf{r,}t),t,\mathbf{\alpha })$ being
identified with%
\begin{equation}
\mathbf{F}_{u}(\mathbf{r},\beta v_{th}\mathbf{n}(\mathbf{r},t),t,\mathbf{%
\alpha })=\beta v_{th}(\mathbf{r},t,\mathbf{\alpha })\mathbf{\Omega }(%
\mathbf{r},t,\mathbf{\alpha })\times \mathbf{n}(\mathbf{r},t,\mathbf{\alpha }%
)+\beta \mathbf{n}(\mathbf{r},t,\mathbf{\alpha })\frac{Dv_{th}(\mathbf{r},t,%
\mathbf{\alpha })}{Dt},  \label{STOCHASTIC MEAN-FIELD}
\end{equation}%
where $\mathbf{n}(\mathbf{r},t,\mathbf{\alpha })$ is given by Eq.(\ref%
{EVOLUTION EQUATION for n}) (see also Lemma to THM.2). The flow associated
to the Eqs.(\ref{stochastic eqq-1})\ and (\ref{stochastic-initial-value
problem})%
\begin{equation}
T_{t_{o},t}:\mathbf{y}_{o}\rightarrow \mathbf{y}(t,\mathbf{\alpha }%
)=T_{t_{o},t}\mathbf{y}_{o},  \label{STOCH-DYN-S}
\end{equation}%
is referred to as \emph{stochastic TTP dynamical system} (TTP-DS). Here $%
T_{t_{o},t}$ is a measure-preserving evolution operator associated to the
Newtonian vector field $\mathbf{X}(\mathbf{x},t,\mathbf{\alpha })\equiv
\left\{ \mathbf{v},\mathbf{F}(\mathbf{x},t,\mathbf{\alpha })\right\} $ and $%
\mathbf{F}(\mathbf{x},t,\mathbf{\alpha })$ the stochastic mean-field\emph{\ }%
satisfying the initial-value problem (\ref{STOCHASTIC MEAN-FIELD}). Thus, by
definition, the TTP-DS is uniquely prescribed by the instantaneous state of
a generic TTP $\mathbf{x}(t,\mathbf{\alpha })\equiv \mathbf{x}=(\mathbf{r,v}%
) $. Therefore, we conclude that:

\begin{itemize}
\item The initial-value problem (\ref{stochastic eqq-1}) can be viewed as a
stochastic model of particle motion in compressible/incompressible NS
fluids, describing the dynamics of TTPs in stochastic thermofluids.

\item Based on the TTP-statistical model developed in the previous section
(see THM.3) the time-evolution of the stochastic KDF $f_{1}(\mathbf{r},%
\mathbf{u}_{th},t,\mathbf{\alpha })$ is uniquely prescribed. In particular,
independent of the choice of the stochastic model $\left\{ g\mathbf{(r}%
,t,\alpha ),V_{\mathbf{\alpha }}\right\} $, the dynamics prescribed by (\ref%
{STOCH-DYN-S}) preserves the exact form of the stochastic fluid equations%
\emph{\ }and is unique.

\item The stochastic KDF $f_{1}(\mathbf{r},\mathbf{u}_{th},t,\mathbf{\alpha }%
)$ determined by Eq.(\ref{INTEGRAL LIOUVILLE EQ}) is necessarily a
particular solution of Eq.(\ref{IKE-2}).

\item The stochastic KDF $f_{1}(\mathbf{r},\mathbf{u}_{th},t,\mathbf{\alpha }%
)$ prescribes uniquely the time-evolution of the stochastic fluid fields $%
\left\{ Z(\mathbf{r},t,\mathbf{\alpha })\right\} \emph{.}$
\end{itemize}

\subsection{Fokker-Planck description in strong turbulence}

Let now analyze the time evolution of the stochastic KDF $f_{1}(\mathbf{r},%
\mathbf{u}_{th},t,\mathbf{\alpha })$ represented in terms of the stochastic
decomposition%
\begin{equation}
f_{1}(\mathbf{r},\mathbf{u}_{th},t,\mathbf{\alpha })=\left\langle f_{1}(%
\mathbf{r},\mathbf{u}_{th},t,\mathbf{\alpha })\right\rangle _{\mathbf{\alpha
}}+\delta f_{1}(\mathbf{r},\mathbf{u}_{th},t,\mathbf{\alpha }),
\label{dec-2}
\end{equation}%
with $\left\langle f_{1}(\mathbf{r},\mathbf{u}_{th},t,\mathbf{\alpha }%
)\right\rangle _{\mathbf{\alpha }}$ denoting the stochastic-average defined
by Eq.(\ref{stochastic averaging operator}). Then, requiring that the
stochastic PDF $g\mathbf{(r},t,\mathbf{\alpha })$ is homogeneous and
stationary [see Appendix B], i.e., that $g\equiv g(\mathbf{\alpha })$, it is
immediate to obtain from IKE [see Eq.(\ref{IKE})] the stochastic statistical
equations advancing in time $\left\langle f_{1}(\mathbf{r},\mathbf{u}_{th},t,%
\mathbf{\alpha })\right\rangle _{\mathbf{\alpha }}$ and $\delta f_{1}(%
\mathbf{r},\mathbf{u}_{th},t,\mathbf{\alpha })$. These are explicitly%
\begin{equation}
\left\langle L\right\rangle _{\mathbf{\alpha }}\left\langle
f_{1}\right\rangle _{\mathbf{\alpha }}=-\left\langle \delta L\delta
f_{1}\right\rangle _{\mathbf{\alpha }}\equiv \left\langle C\right\rangle _{%
\mathbf{\alpha }},  \label{stochasrtic-averaged IKE}
\end{equation}%
\begin{equation}
\left\langle L\right\rangle _{\mathbf{\alpha }}\delta f_{1}=-\delta L\left\{
\left\langle f_{1}\right\rangle _{\mathbf{\alpha }}+\delta f_{1}\right\}
+\left\langle \delta L\delta f_{1}\right\rangle _{\mathbf{\alpha }},
\label{stochastic
IKE}
\end{equation}%
where the streaming operator $L$ has been similarly represented as $%
L=\left\langle L\right\rangle _{\mathbf{\alpha }}+\delta L$. Eqs.(\ref%
{stochasrtic-averaged IKE}) and (\ref{stochastic IKE}) are
formally similar to the Vlasov equation arising in the kinetic
theory of quasi-linear and strong turbulence for Vlasov-Poisson
plasmas
\cite{Tessarotto20081,Dupree1966,Kraichnan1967,Weinstock1969}. As
is well-known, the construction of the precise form of the
operator $\left\langle C\right\rangle $ appearing in the
stochastic-averaged kinetic equation [i.e.,
Eq.(\ref{stochasrtic-averaged IKE})] represents a task of
formidable difficulty. The reason is that it requires constructing
a formal perturbative solution of the equation (\ref{stochastic
IKE}) for the stochastic perturbation $\delta f$. To obtain a
convergent perturbative theory, however, this usually requires the
adoption of a suitable renormalization scheme in order to obtain a
consistent statistical (kinetic) equation for $\left\langle
f_{1}\right\rangle _{\mathbf{\alpha }}$ [see earlier approaches
developed in
Refs.\cite{Martin-Siggia1973,Thompson1973,Krommes1979} which
pertain to the statistical treatment of particle dynamics and the
specific application to Vlasov-Poisson plasmas]. Nonetheless, in
the case of weak-turbulence, the stochastic-averaged kinetic
equation [i.e., Eq.(\ref{stochasrtic-averaged IKE})] is known to
be amenable to an approximate Fokker-Planck kinetic equation
advancing in time $\left\langle f_{1}\right\rangle
_{\mathbf{\alpha }}$ alone. Analogous suggestions are provided by
phenomenologically-based Markovian Fokker-Planck models of
small-scale fluid turbulence [see for example
Refs.\cite{Naert1997,Friedrich1999,Renner2002}].

This raises the issue of (the construction of) a possible representation of
this type for the stochastic-averaged operator $\left\langle C\right\rangle
_{\mathbf{\alpha }}$ which has the following properties:

\begin{itemize}
\item Property \#1: it holds at least \emph{locally} in the velocity space $%
U_{1}$, in a suitable space to be specified.

\item Property \#2: it holds in the case of \textquotedblleft \emph{strong\
turbulence}\textquotedblright , namely when there are fluctuating quantities
such that their stochastic fluctuation is comparable in order of magnitude
with their corresponding stochastic averages. In particular, denoting by $%
\zeta $ a dimensionless infinitesimal parameter, in the following the strong
turbulence regime is defined in such a way that \footnote{%
Notice that in the case of homogeneous, isotropic and stationary turbulence
(HIST) the precise definition of the stochastic fluctuations of the relevant
fluid fields is independent of the specific definition adopted for the
stochastic variables $\mathbf{\alpha .}$}%
\begin{equation}
\left\vert \delta \mathbf{V}\right\vert \sim \left\vert \left\langle \mathbf{%
V}\right\rangle _{\mathbf{\alpha }}\right\vert \left[ 1+O\left( \zeta
\right) \right] ,
\end{equation}%
\begin{equation}
\left\vert \delta p_{1}\right\vert \sim \left\vert \left\langle
p_{1}\right\rangle _{\mathbf{\alpha }}\right\vert \left[ 1+O\left( \zeta
\right) \right] .
\end{equation}%
The two requirements are mutually consistent and, for arbitrary choices of
the velocity stochastic fluctuations $\delta \mathbf{V}$, are required by
the Navier-Stokes equation.

\item Property \#3: it is applicable also in the case in which $\left\langle
f_{1}\right\rangle _{\mathbf{\alpha }}$ is \emph{generally non-Gaussian}.

\item Property \#4: it does not rely on renormalization theory.

\item Property \#5: the KDF is assumed of the form $f_{1}=f_{1}(\mathbf{r}%
,\zeta ^{1/2}\mathbf{u}_{th},t,\mathbf{\alpha })$, namely it exhibits slow
dependence with respect to the velocity $\mathbf{u}_{th}$.
\end{itemize}

Notice that the previous properties are assumed to hold for arbitrary smooth
non-Gaussian KDFs. In the specific case of a Gaussian KDF Property \#5
requires necessarily that $\frac{\mathbf{v}-\mathbf{V}}{v_{th}}\sim O\left(
\zeta ^{1/2}\right) $. For TTPs this implies $\frac{\mathbf{u}_{th}}{v_{th}}%
=\beta \mathbf{n}\sim O\left( \zeta ^{1/2}\right) $, namely $\beta \sim
O\left( \zeta ^{1/2}\right) $ and also $\frac{\delta \mathbf{u}_{th}\cdot
\left\langle \mathbf{u}_{th}\right\rangle _{\mathbf{\alpha }}}{v_{th}^{2}}%
\sim O\left( \zeta \right) $. Therefore, the required slow velocity
dependence effectively limits the validity of strong turbulence theory to
the subset of velocity space in which such an ordering holds.

Regarding, in particular, the form of the KDF $f_{1}(\mathbf{r},\zeta ^{1/2}%
\mathbf{u}_{th},t,\mathbf{\alpha })$, it must be noted that, even if the
latter coincides locally with a Gaussian KDF $f_{M1}(\mathbf{r},\zeta ^{1/2}%
\mathbf{u}_{th},t,\mathbf{\alpha })$, its stochastic average $\left\langle
f_{1}\right\rangle _{\mathbf{\alpha }}$ still remains generally
non-Gaussian. Here we intend to prove that an explicit representation of $%
\left\langle C\right\rangle _{\mathbf{\alpha }}$ fulfilling the previous
properties can be determined, based on the IKT approach earlier pointed out
in Ref.\cite{Tessarotto20082}. In this case, the previous problem is exactly
solvable provided:

\begin{enumerate}
\item $f_{1}$ depends on the stochastic variables $\mathbf{\alpha }$ only
through the fluid fields, i.e., $\rho \left( \mathbf{r},t,\mathbf{\alpha }%
\right) $, $p_{1}(\mathbf{r},t,\mathbf{\alpha })$ and\ $\mathbf{V}(\mathbf{r}%
,t,\mathbf{\alpha })$, and hence also $u_{th}\equiv \beta v_{th}(\mathbf{r}%
,t,\mathbf{\alpha })$ and the unit vector $\mathbf{n}=\mathbf{n}\left(
\mathbf{r},t,\mathbf{\alpha }\right) $, namely is of the form%
\begin{equation}
f_{1}\equiv f_{1}(\mathbf{r},\zeta ^{1/2}\mathbf{u}_{th},t,\rho ,p_{1}).
\label{form of f_1}
\end{equation}

\item $f_{1}$ is generally a non-Gaussian KDF which is analytic in $\mathbf{u%
}_{th}$, $\rho $ and $p_{1}$.

\item The kinetic pressure $p_{1}$ and the pseudo-pressure $p_{0}\equiv
\left\langle p_{0}\right\rangle _{\mathbf{\alpha }}$ satisfy the ordering%
\begin{equation}
\frac{p_{1}-p_{0}}{p_{0}}\sim O(\zeta ).  \label{finite amplitude}
\end{equation}

\item The mass-density perturbations are weak, in the sense%
\begin{equation}
\frac{\left\vert \delta \rho \right\vert }{\left\langle \rho \right\rangle
_{\alpha }}\sim O(\zeta ).  \label{velocity
ordering}
\end{equation}

\item The stochastic fluctuation of the KDF $\delta f_{1}(\mathbf{r},\zeta
^{1/2}\mathbf{u}_{th},t,\mathbf{\alpha })$ is such that%
\begin{equation}
\frac{\delta f_{1}(\mathbf{r},\zeta ^{1/2}\mathbf{u}_{th},t,\mathbf{\alpha })%
}{\left\langle f_{1}(\mathbf{r},\zeta ^{1/2}\mathbf{u}_{th},t,\mathbf{\alpha
})\right\rangle _{\mathbf{\alpha }}}\sim O(\zeta ).
\label{strong
turbulence}
\end{equation}
\end{enumerate}

Invoking for $\delta \mathbf{u}_{th}$ the representation $\delta \mathbf{u}%
_{th}\equiv \delta \rho \frac{\partial \mathbf{u}_{th}}{\partial
\left\langle \rho \right\rangle _{\alpha }}+\delta p_{1}\frac{\partial
\mathbf{u}_{th}}{\partial p_{o}}-\delta \mathbf{V}$, as a consequence of the
previous requirements, $f_{1}$ can be Taylor-expanded with respect to $%
\delta \rho $, $\delta p_{1}$ and $\delta \mathbf{u}_{th}$ yielding%
\begin{eqnarray}
&&\left. f_{1}(\mathbf{r},\zeta ^{1/2}\mathbf{u}_{th},t,\rho ,p_{1})=f_{1}(%
\mathbf{r},\left\langle \zeta ^{1/2}\mathbf{u}_{th}\right\rangle _{\mathbf{%
\alpha }},t,\left\langle \rho \right\rangle _{\alpha },\left\langle
p_{1}\right\rangle _{\alpha })+\right.  \notag \\
&&+\left[ \overset{\infty }{\sum\limits_{n=1}}\frac{1}{n!}\left( \delta \rho
\frac{\partial }{\partial \left\langle \rho \right\rangle _{\alpha }}+\delta
p_{1}\frac{\partial }{\partial p_{o}}+\delta \mathbf{u}_{th}\cdot \frac{%
\partial }{\partial \mathbf{u}}\right) ^{n}f_{1}(\mathbf{r},\zeta ^{1/2}%
\mathbf{u},t,\left\langle \rho \right\rangle _{\alpha },\left\langle
p_{1}\right\rangle _{\alpha })\right] _{\mathbf{u=}\left\langle \mathbf{u}%
_{th}\right\rangle _{\mathbf{\alpha }}}.
\end{eqnarray}%
Due to the previous ordering assumptions and the requirement of analyticity,
the series converges uniformly in $\Gamma _{1}$ and permits the explicit
determination of both $\left\langle f_{1}(\mathbf{r},\zeta ^{1/2}\mathbf{u}%
_{th},t,p_{1})\right\rangle _{\alpha }$ and $\delta f_{1}(\mathbf{r},\zeta
^{1/2}\mathbf{u}_{th},t,p_{1})$. As a result, after straightforward algebra
the operator $\left\langle C\right\rangle _{\mathbf{\alpha }}$ becomes
explicitly \cite{Tessarotto20082}
\begin{equation}
\left\langle C\right\rangle _{\mathbf{\alpha }}=C_{FP}\equiv \left[
\sum_{i,j,k=1}^{\infty }\frac{\partial }{\partial \mathbf{u}}\cdot \left(
\mathbf{C}_{i,j,k}\otimes \frac{\partial ^{n}}{\partial ^{i}\left\langle
\rho \right\rangle _{\alpha }\partial ^{i}p_{0}\partial ^{k}\mathbf{u}}f_{1}(%
\mathbf{r},\zeta ^{1/2}\mathbf{u},t,\left\langle p_{1}\right\rangle _{\alpha
})\right) \right] _{\mathbf{u=}\left\langle \mathbf{u}_{th}\right\rangle _{%
\mathbf{\alpha }}},  \label{Fokker-Planck operator}
\end{equation}%
where $\mathbf{C}_{i,j,k}$ are the tensor Fokker--Planck (or Kramers-Moyal)
coefficients%
\begin{equation}
\mathbf{C}_{i,j,k}=\frac{1}{n!}\left\langle \delta \mathbf{F}\left( \delta
\rho \right) ^{i}\left( \delta p_{1}\right) ^{j}\left( \delta \mathbf{u}%
_{th}\right) ^{k}\right\rangle _{\mathbf{\alpha }},
\label{F-P
COEFFICIENTS}
\end{equation}%
$\otimes $ denotes the tensor product and the summation is carried out on $%
i,j,l$ from $0$ to $\infty $, with $n=i+j+k\geq 2$. Therefore $\left\langle
C\right\rangle _{\mathbf{\alpha }}$ takes the form of a \emph{generalized
Fokker-Planck} (F-P) \emph{operator }($C_{FP}$). Remarkably, Eq.(\ref%
{Fokker-Planck operator}) satisfies by construction Properties \#2-\#5.

As a final point, we comment on the implications of the GDE-requirement \#6
which concern TTP dynamics and hold in validity of the strong turbulence
formulation developed here. In fact, the kinetic constraint (\ref%
{NO-PRECESSION}) implies that time evolution of the TTP relative velocity
depends on the stochastic fluctuations of both kinetic pressure $\delta
p_{1} $ and fluid velocity $\delta \mathbf{V}$. Qualitatively this means
that, even in the case in which pressure fluctuations are negligible, TTP
dynamics can still exhibit turbulent motion through the fluctuations $\delta
\mathbf{V}$ entering in the fluid vorticity.

\subsection{Statistical irreversibility}

We first notice that due to the integral IKE (\ref{INTEGRAL LIOUVILLE EQ}),
if $f_{1}(t_{o})$ is strictly positive then, for all $t\geq t_{o}$ in the
existence time interval $I$, also $f_{1}(t)$ is necessarily so. This
manifestly implies, in turn, that in such a case the stochastic-averaged KDF
$\left\langle f_{1}(\mathbf{x},t)\right\rangle _{\mathbf{\alpha }}\equiv
\left\langle f_{1}(\mathbf{r},\mathbf{u}_{th},t,\mathbf{\alpha }%
)\right\rangle _{\mathbf{\alpha }}$, solution of the stochastic-averaged
statistical equation (\ref{stochasrtic-averaged IKE}), is necessarily
strictly positive. It is immediate to show that, thanks to THMs.1 and 3, $%
\left\langle f_{1}(t)\right\rangle _{\mathbf{\alpha }}$ must satisfy in $%
\Gamma _{1}\times I$ also a weak H-theorem of the form:%
\begin{equation}
\frac{\partial S(\left\langle f_{1}(t)\right\rangle _{\mathbf{\alpha }})}{%
\partial t}\geq 0.  \label{H-THEOREM FOR F_1-averaged2}
\end{equation}%
For definiteness, let us assume that at the initial time $t_{o}$ both $f_{1}(%
\mathbf{x,}t)\equiv f_{1}(\mathbf{r},\mathbf{u}_{th},t,\mathbf{\alpha })$
and $\left\langle f_{1}(\mathbf{x},t)\right\rangle _{\mathbf{\alpha }}$ are
strictly positive and admit the BS entropies $S(f_{1}(\mathbf{x},t_{o}))$
and $S(\left\langle f_{1}(\mathbf{x},t_{o})\right\rangle _{\mathbf{\alpha }%
}) $. Then, we notice that if $g(\mathbf{x},t\mathbf{)}$ is an arbitrary
strictly positive function such that $\int\limits_{\Gamma _{1}}d\mathbf{x}%
_{1}g(\mathbf{x},t)=\mu (\Omega )$, the majorization%
\begin{equation}
\int\limits_{\Gamma _{1}}d\mathbf{x}_{1}f_{1}(\mathbf{x},t)\ln f_{1}(\mathbf{%
x,}t)\geq \int_{\Gamma _{1}}d\mathbf{x}_{1}f_{1}(\mathbf{x},t)\ln g(\mathbf{x%
},t)
\end{equation}%
necessarily holds (Brillouin Lemma \cite{Brillouin1959}). On the other hand,
assuming that the stochastic-averaging operator $\left\langle \cdot
\right\rangle _{\mathbf{\alpha }}$ commutes with the phase-space integral
operator $\int_{\Gamma _{1}}d\mathbf{x}_{1}$,%
\begin{equation}
\int_{\Gamma _{1}}d\mathbf{x}_{1}\left\langle f_{1}(\mathbf{x}%
,t)\right\rangle _{\mathbf{\alpha }}=\left\langle \int_{\Gamma _{1}}d\mathbf{%
x}_{1}f_{1}(\mathbf{x},t)\right\rangle _{\mathbf{\alpha }},
\end{equation}%
it follows%
\begin{equation}
\int_{\Gamma _{1}}d\mathbf{x}_{1}f_{1}(\mathbf{x},t)\ln f_{1}(\mathbf{x}%
,t)\geq \int_{\Gamma _{1}}d\mathbf{x}_{1}f_{1}(\mathbf{x},t)\ln \left\langle
f_{1}(\mathbf{x},t)\right\rangle _{\alpha }.
\end{equation}%
This yields in turn
\begin{equation}
\left\langle \int_{\Gamma _{1}}d\mathbf{x}_{1}f_{1}(\mathbf{x},t)\ln f_{1}(%
\mathbf{x},t)\right\rangle _{\mathbf{\alpha }}\geq \int_{\Gamma _{1}}d%
\mathbf{x}_{1}\left\langle f_{1}(\mathbf{x},t)\right\rangle _{\mathbf{\alpha
}}\ln \left\langle f_{1}(\mathbf{x},t)\right\rangle _{\alpha }.
\end{equation}%
The last inequality implies manifestly that%
\begin{equation}
S(\left\langle f_{1}(\mathbf{x},t)\right\rangle _{\mathbf{\alpha }})\geq
\left\langle S(f_{1}(\mathbf{x},t)\right\rangle _{\mathbf{\alpha }}.
\label{inequality}
\end{equation}%
Therefore, in view of the entropy inequality (\ref{H-THEOREM FOR f_1}), $%
\left\langle f_{1}(t)\right\rangle _{\mathbf{\alpha }}$ satisfies
necessarily the weak H-theorem (\ref{H-THEOREM FOR F_1-averaged2}). This
assures that for all $t\geq t_{o}$ in the time interval $I,$ $\left\langle
f_{1}(t)\right\rangle $ admits the BS entropy integral $S\left( \left\langle
f_{1}(t)\right\rangle \right) $, i.e., that $S\left( \left\langle
f_{1}(t)\right\rangle \right) $ is defined for all\ $t\geq t_{o}$ in $I$. As
a consequence $\left\langle f_{1}(t)\right\rangle $ exhibits an irreversible
behavior.

\subsection{Comparisons with the HRE statistical model and solution of the
Closure Problem}

In this section we display the relationship between the
statistical models obtained here - i.e., both the IKT and TTP
statistical models $\left\{ f,\Gamma \right\} $ and $\left\{
f_{1},\Gamma _{1}\right\} $, prescribed respectively by means of
THMs 1 and 3 - and the customary statistical treatment of
turbulence due to Hopf, Rosen and Edwards (\emph{HRE approach},
\cite{Hopf1952,Rosen1960,Edwards1964}; see also
\cite{Novikov,Kollmann,Pope1983,Givi,Dopazo}; for a review see
\cite{Monin1975,Pope2000}).\ Since the latter is usually developed
in the case of incompressible fluids, we shall restrict the
following analysis to such a case.

The HRE approach, which in its original form applies only to incompressible
and isothermal NS fluids, is based on the introduction of a suitable
statistical model \cite{Hopf1952,Rosen1960,Edwards1964} $\left\{
f_{H},\Gamma \right\} $, here referred to as the \textit{HRE statistical
model, }with $\Gamma \equiv \Omega \times U$. In this case, denoting by $%
\mathbf{u}\equiv \mathbf{u}(t)\equiv \mathbf{v}-\mathbf{V}(\mathbf{r},t)$, $%
f_{H}$ is identified with the velocity PDF%
\begin{equation}
f_{H}(\mathbf{r},\mathbf{u},t)\equiv \rho _{o}\delta \left( \mathbf{u}\right)
\label{FH-1}
\end{equation}%
(\emph{HRE velocity KDF}), $\rho _{o}$ denoting the constant mass density
characterizing an incompressible fluid\emph{. }Upon identifying the
mean-field $\mathbf{F}(\mathbf{\mathbf{x}},t)$ with the fluid acceleration $%
\mathbf{F}_{H}(\mathbf{r},t)$ [see definition given by Eq.(\ref%
{fluid-acceleration}) in Appendix A], it follows that by construction $f_{H}$
is a particular solution of Eq.(\ref{IKE}). As a consequence, its velocity
moment equations, evaluated with respect to the weight functions $\left\{
G\right\} =\left\{ 1,\mathbf{v}\right\} $, coincide respectively with the
continuity and the NS equations (\ref{1A}), (\ref{1B}) (see Appendix A). It
is important to remark, instead, that by construction the fluid pressure $p(%
\mathbf{\mathbf{r}},t)$ cannot be determined as velocity moment of $f_{H}(%
\mathbf{r},\mathbf{u},t)$ of the form $\int d\mathbf{v}G(\mathbf{\mathbf{r}},%
\mathbf{\mathbf{v}},t)f_{H}(\mathbf{r},\mathbf{u},t)$ in terms of a weight
function $G(\mathbf{\mathbf{r}},\mathbf{\mathbf{v}},t)$ which is independent
of the same fluid field. Hence, in the HRE approach the fluid pressure has
to be suitably prescribed. Thus, for incompressible NS fluids it is
identified with a solution of the boundary-value problem associated to the
corresponding Poisson equation.

Let us now pose the problem of the connection existing between the
HRE velocity KDF\emph{\ }$f_{H}(\mathbf{r},\mathbf{u},t)$ and the
IKT approach. For this purpose we first remak that the
representation of the HRE velocity KDF is actually
\emph{non-unique}. The proof of the statement is immediate. In
fact, let us notice - preliminarily - that INSE can be
equivalently represented in terms of the stochastic fluid fields:
\begin{equation}
\left\{ Z_{1}(\mathbf{r},t,\Delta \mathbf{V})\right\} =\left\{ \mathbf{V}%
_{1}(\mathbf{r},t)\equiv \mathbf{V}(\mathbf{r},t)+\Delta \mathbf{V},p_{1}(%
\mathbf{r},t),S_{T}\right\} ,  \label{STOCHASTIC
REPRESENTATION}
\end{equation}%
where%
\begin{equation}
\left\{ Z_{1}(\mathbf{r},t)\right\} \equiv \left\{ \mathbf{V}(\mathbf{r}%
,t),p_{1}(\mathbf{r},t),S_{T}\right\}  \label{REGULAR NS FLUIDS}
\end{equation}%
and $\Delta \mathbf{V}\equiv (\Delta V_{1},\Delta V_{2},\Delta V_{3})\in
U\equiv
\mathbb{R}
^{3}$ denote an arbitrary particular solution of the related
initial-boundary value problems [i.e., Eqs.(\ref{1})-(\ref{2}) with the
constant-entropy equation (\ref{isoentropic fluid})] and an arbitrary
stochastic vector independent of $(\mathbf{r},t)$ respectively. In
particular, the NS equation for $\mathbf{V}_{1}(\mathbf{r},t)$ is manifestly
\begin{eqnarray}
&&\left. \frac{\partial }{\partial t}\mathbf{V}_{1}(\mathbf{r},t)+\mathbf{V}%
_{1}(\mathbf{r},t)\cdot \nabla \mathbf{V}_{1}(\mathbf{r},t)\equiv \right.
\notag \\
&&\left. \equiv \frac{\partial }{\partial t}\mathbf{V}(\mathbf{r},t)+\left(
\mathbf{V}(\mathbf{r},t)+\Delta \mathbf{V}\right) \cdot \nabla \mathbf{V}(%
\mathbf{r},t)=\mathbf{F}_{H}+\Delta \mathbf{F}_{H},\right.  \label{NS-eq-2}
\end{eqnarray}%
with\emph{\ }$\Delta \mathbf{F}_{H}$ being defined as the stochastic vector
field $\Delta \mathbf{F}_{H}=\Delta \mathbf{V}\cdot \nabla \mathbf{V}(%
\mathbf{\mathbf{r}},t).$ As a result, for an incompressible fluid the HRE
velocity KDF, corresponding to Eq.(\ref{NS-eq-2})\emph{\ }becomes $f_{H}(%
\mathbf{r},\mathbf{u}-\Delta \mathbf{\mathbf{V}},t)\equiv \rho
_{o}\delta \left( \mathbf{u}-\Delta \mathbf{\mathbf{V}}\right) $,
which proves the statement. In view of these considerations let us
now introduce the
stochastic model defined by the set $\left\{ \Delta \mathbf{\mathbf{V}}\in U%
\mathbf{\mathbf{,}}g(\mathbf{r},t,\Delta \mathbf{\mathbf{V}})\right\} $,
with $g(\mathbf{r},t,\Delta \mathbf{V})$ being a suitable stochastic PDF
[see Appendix B]. Due to its arbitrariness, it can always be identified with
\begin{equation}
g(\mathbf{r},\Delta \mathbf{\mathbf{V}},t)\equiv \widehat{f}(\mathbf{r}%
,\Delta \mathbf{V},t),  \label{G1}
\end{equation}%
with $\widehat{f}$ being defined by Eq.(\ref{PDF}) in terms of the
IKT-statistical model $\left\{ f,\Gamma \right\} $. This means that the
corresponding KDF $f$\ is a particular solution of Eq.(\ref{IKE}) obtained
by prescribing the mean-field $\mathbf{F}$ in accordance with THM.1. As a
result the following identity holds:%
\begin{equation}
\left\langle f_{H}(\mathbf{r},\mathbf{u}-\Delta \mathbf{\mathbf{V}}%
,t)\right\rangle _{\Delta \mathbf{V}}=f(\mathbf{r},\mathbf{u},t),
\label{HRE-RELATIONSHIP}
\end{equation}%
where $\left\langle \cdot \right\rangle _{\Delta \mathbf{V}}$ is the
stochastic average defined in Appendix B [see Eq.(\ref{stochastic averaging
operator})] with the stochastic vector $\mathbf{\alpha }$ being identified
with $\Delta \mathbf{\mathbf{V}}$. In particular this implies, thanks to the
correspondence principle (\ref{MOMENTS}), that the variance of the
stochastic velocity $\Delta \mathbf{\mathbf{V}}$ is prescribed as%
\begin{equation}
\left\langle \frac{1}{3}\Delta \mathbf{\mathbf{V}}^{2}\right\rangle _{\Delta
\mathbf{V}}=\widehat{p}_{1}(\mathbf{r},t).  \label{VARIANCE}
\end{equation}%
As a further implication, when Eq.(\ref{HRE-RELATIONSHIP}) is evaluated in
the subspace of TTPs $U_{1}$, in view of Eq.(\ref{NEW}) and THM.3, it
requires%
\begin{equation}
\left\langle f_{H}(\mathbf{r},\mathbf{u}_{th}-\Delta \mathbf{\mathbf{V}}%
,t)\right\rangle _{\Delta \mathbf{V}}=f(\mathbf{r},\mathbf{u}_{th},t)=\frac{1%
}{2}f_{1}(\mathbf{r},\mathbf{u}_{th},t).  \label{HRE-RELATIONSHIP-2}
\end{equation}%
For all $(\mathbf{r},\mathbf{u}_{th},t)\in \Gamma _{1}\times I$, this yields
also the \emph{relationship between the HRE velocity KDF and the conditional
KDF} $f_{1}(\mathbf{r},\mathbf{u}_{th},t)$ \emph{characterizing the
statistical model of TTPs} $\left\{ f_{1},\Gamma _{1}\right\} $.

A preliminary summary is in order. In the case of a NS fluid obeying the
INSE problem, the following conclusions are reached:

\begin{itemize}
\item In view of the constraint (\ref{VARIANCE}), $g(\mathbf{r},\Delta
\mathbf{\mathbf{V}},t)$ can be interpreted as the stochastic PDF taking into
account the \emph{stochastic \textquotedblleft thermal\textquotedblright\
motion of ITPs} produced in a compressible thermal fluid by the kinetic
pressure $p_{1}(\mathbf{r},t)$.

\item Eqs.(\ref{HRE-RELATIONSHIP}) and (\ref{HRE-RELATIONSHIP-2}) permit a
comparison between the HRE, IKT and TTP statistical models, $\left\{
f_{H},\Gamma \right\} $, $\left\{ f,\Gamma \right\} $ and $\left\{
f_{1},\Gamma _{1}\right\} $ respectively. \emph{We remark that such a
comparison is always possible (and hence it holds also in the case of INSE).
}In particular, Eq.(\ref{HRE-RELATIONSHIP}) determines the relationship
between the KDFs $f_{H}$ and $f$ prescribed by the statistical model $%
\left\{ f,\Gamma \right\} $. In addition, Eq.(\ref{HRE-RELATIONSHIP-2})
yields the analogous relationship with the KDF $f_{1}$ characterizing the
TTPs statistics.

\item Remarkably, both conclusions follow by invoking a single suitable
stochastic model. In both cases, in fact, the stochastic-averaging operator
[see Eq.(\ref{stochastic averaging operator})] which enters the l.h.s. of
Eqs.(\ref{HRE-RELATIONSHIP}) and (\ref{HRE-RELATIONSHIP-2}) is defined with
respect to the same stochastic probability density $g(\mathbf{r},\Delta
\mathbf{\mathbf{V}},t)$ prescribed according to Eq.(\ref{G1}).

\item Eqs.(\ref{G1}) and (\ref{HRE-RELATIONSHIP}) - or equivalent (\ref%
{HRE-RELATIONSHIP-2}) - do not imply any restriction on the flow dynamics,
i.e., on the (strong) solutions of the INSE problem.

\item Denoting $\left\langle \left\langle {}\right\rangle \right\rangle
_{\Omega }$ the configuration-space average $\left\langle \left\langle
{}\right\rangle \right\rangle _{\Omega }\equiv \frac{1}{\mu (\Omega )}%
\int\limits_{\Omega }d\mathbf{r}$, Eq.(\ref{HRE-RELATIONSHIP-2}) uniquely
prescribes also the relationship between the corresponding spatial averages,
i.e.,%
\begin{equation}
\left\langle \left\langle \left\langle f_{H}(\mathbf{r},\mathbf{u}%
_{th}-\Delta \mathbf{\mathbf{V}},t)\right\rangle _{\Delta \mathbf{V}%
}\right\rangle \right\rangle _{\Omega }=\left\langle \left\langle f(\mathbf{r%
},\mathbf{u}_{th},t)\right\rangle \right\rangle _{\Omega }=\frac{1}{2}%
\left\langle \left\langle f_{1}(\mathbf{r},\mathbf{u}_{th},t)\right\rangle
\right\rangle _{\Omega }.
\end{equation}%
Hence, assuming that the initial frequency $\left\langle \left\langle
\left\langle f_{H}(\mathbf{r},\mathbf{u}_{th}-\Delta \mathbf{\mathbf{V}}%
,t_{o})\right\rangle _{\Delta \mathbf{V}}\right\rangle \right\rangle
_{\Omega }$\ is prescribed, for the initial velocity PDF $f_{1}(\mathbf{r},%
\mathbf{u}_{th},t_{o})$ the average $\left\langle \left\langle f_{1}(\mathbf{%
r},\mathbf{u}_{th},t_{o})\right\rangle \right\rangle _{\Omega }$ remains
uniquely determined too.
\end{itemize}

Let us now address in detail the issues of the comparison between the HRE,
IKT and TTP statistical approaches according to Eqs.(\ref{HRE-RELATIONSHIP})
and (\ref{HRE-RELATIONSHIP-2}) and the related closure condition problem
arising in the HRE approach.

For this purpose it is worth recalling that the aim of the HRE approach is
actually to predict the time evolution, in the presence of turbulence, of
suitable ensemble-averages of the KDF $f_{H}$ and of the NS fluid fields $%
\left\{ \mathbf{V},p\right\} $, i.e., respectively $\left\langle f_{H}(%
\mathbf{r},\mathbf{u},t)\right\rangle $ and $\left\langle \mathbf{V}(\mathbf{%
r},t)\right\rangle $, $\left\langle p(\mathbf{r},t)\right\rangle $. Here the
brackets $\left\langle \cdot \right\rangle $ denote a suitable \emph{%
ensemble-averaging operator }(see for example \cite{Pope2000}). In the case
of so-called homogeneous, isotropic and stationary turbulence (HIST), this
is required to commute with the differential and integral operators $\left\{
Q\right\} \equiv \left\{ \frac{\partial }{\partial t},\frac{\partial }{%
\partial \mathbf{r}},\frac{\partial ^{2}}{\partial \mathbf{r}\cdot \partial
\mathbf{r}},\frac{\partial }{\partial \mathbf{v}},\ \int\limits_{\Omega }d%
\mathbf{r,}\int\limits_{U}d\mathbf{v}\right\} $\textbf{.} As pointed out
above, in the context of the statistical description of turbulence, the
operator $\left\langle \cdot \right\rangle $ may be equivalently intended as
a mean value \textit{in the probabilistic sense} \footnote{%
This viewpoint is also adopted to describe turbulence in plasmas (see for
example \cite{Dupree1966,Kraichnan1967,Weinstock1969}).}, namely $%
\left\langle \cdot \right\rangle \equiv \left\langle \cdot \right\rangle _{%
\mathbf{\alpha }}$. Here $\left\langle \cdot \right\rangle _{\mathbf{\alpha }%
}$ denotes again the stochastic average of the form (\ref{stochastic
averaging operator}) [see Appendix B], prescribed in terms of a suitable
stochastic probability density defined on the space $V_{\mathbf{\alpha }}$
of the stochastic parameters $\mathbf{\alpha }$. Thus, in this context the
fluid fields are considered as stochastic functions dependent on $\mathbf{%
\alpha }$. In the case of INSE this requires letting $\left\{ Z_{1}(\mathbf{r%
},t,\mathbf{\alpha })\right\} \equiv \left\{ \mathbf{V}(\mathbf{r},t,\mathbf{%
\alpha }),p_{1}(\mathbf{r},t,\mathbf{\alpha }),S_{T}(\mathbf{\alpha }%
)\right\} $. In particular, in validity of HIST this implies that
necessarily the stochastic PDF $g(\mathbf{r},t,\mathbf{\alpha })$ must be
taken of the form $g\equiv g(\mathbf{\alpha })$. The corresponding
statistical evolution equation for $\left\langle f_{H}\right\rangle $ is
well-known and has been investigated by several authors (see for example
\cite{Dopazo}). In the case of the unforced NS equation, its explicit
solution involves the construction of an infinite set of \textit{continuous }%
many-point PDFs, coupled via the fluid pressure, which obey a
hierarchy of statistical equations, the so-called ML
(Monin-Lundgren \cite{Monin1967,Lundgren1967}) hierarchy.

The search of possible \textquotedblleft closure
conditions\textquotedblright\ for the ML hierarchy (\emph{Closure Problem }%
for the statistical description of HT) remains - to date - one of the
outstanding unsolved theoretical problems in fluid dynamics. Its solution
involves in principle the search of possible alternative statistical models
with the following features:

Requirement \#1: it should be characterized by a finite number of
(multi-point) PDFs.

Requirement \#2: it should be determined in such a way that the complete set
of fluid fields can be uniquely represented in terms of the same PDFs.

Requirement \#3: the time evolution of the said PDFs is solely determined by
a finite number velocity moments of the same PDFs (\emph{closure conditions}%
).

It is immediate to show that a possible candidate satisfying all of these
features is provided by the TTP statistical model $\left\{ f_{1},\Gamma
_{1}\right\} $. In fact, on the basis of the theory developed above (see in
particular THMs. 1 and 3), we conclude that, in the case of INSE:

\begin{itemize}
\item The statistical set $\left\{ f_{1},\Gamma _{1}\right\} $ is realized
in terms of the stochastic 1-point conditional PDF $\widehat{f_{1}}(\mathbf{r%
},\mathbf{u}_{th},t,\mathbf{\alpha })$.

\item The fluid fields $\left\{ Z_{1}(\mathbf{r},t,\mathbf{\alpha })\right\}
$ are all uniquely prescribed in terms of the same stochastic PDF $\widehat{f%
}_{1}(\mathbf{r},\mathbf{u}_{th},t,\mathbf{\alpha })$ (correspondence
principle).

\item The time-evolution of the stochastic PDF $\widehat{f}_{1}(\mathbf{r},%
\mathbf{u}_{th},t,\mathbf{\alpha })$ is uniquely prescribed by means of
Liouville statistical equation [see the inverse kinetic equation (\ref{IKE-2}%
)].

\item By assumption, such a\ kinetic equation depends functionally [see
Axiom \#3 - Kinetic closure conditions] only on a finite number of velocity
moments of the same PDF.
\end{itemize}

Let us analyze how, in practice, the previous conclusions can be implemented
in order to avoid the closure problem.

In the customary HRE approach one is faced with the formidable issue of
prescribing the multi-point PDFs which enter the Monin-Lundgren hierarchy.
It is immediate to recognize that this problem arises specifically because
of the treatment adopted for the fluid pressure in such approaches. In fact,
because the pressure is not represented by a PDF velocity moment, but rather
its contribution enters by means of a Green-function convolution integral,
it follows that non-local (i.e., multi-point) contributions are necessarily
introduced. The precise prescription of such contributions, however, remains
undetermined, giving rise to the closure problem. Although several attempts
have been suggested (see for example Ref.\cite{Hoso}), no definite solution
exists to date.

In contrast, based either on the IKT or TTP statistical descriptions, the
problem can be given a consistent solution. This is reached as follows:

1) By replacing the ensemble-averaged HRE-KDF $\left\langle f_{H}(\mathbf{r},%
\mathbf{v},t)\right\rangle \equiv \left\langle f_{H}(\mathbf{r},\mathbf{v},t,%
\mathbf{\alpha })\right\rangle _{\mathbf{\alpha }}$ either with the
corresponding IKT or TTP KDFs, namely $\left\langle f(\mathbf{r},\mathbf{u}%
,t)\right\rangle \equiv \left\langle f(\mathbf{r},\mathbf{u},t,\mathbf{%
\alpha })\right\rangle _{\mathbf{\alpha }}$\ or $\left\langle f_{1}(\mathbf{r%
},\mathbf{u}_{th},t)\right\rangle \equiv \left\langle f_{1}(\mathbf{r},%
\mathbf{u}_{th},t,\mathbf{\alpha })\right\rangle _{\mathbf{\alpha }}$.

2) In both cases the kinetic pressure is determined as a velocity moment of
the corresponding KDF. This permits to overcome the closure problem arising
in the HRE approach. In fact, non-local contributions due to multi-point
PDFs do not appear anymore.

3) By requiring that the KDF $f(\mathbf{r},\mathbf{u},t,\mathbf{\alpha })$,
and respectively $f_{1}(\mathbf{r},\mathbf{u}_{th},t,\mathbf{\alpha })$,
satisfy the Liouville equations (\ref{IKE}) and (\ref{IKE-2}). These
equations are determined imposing the requirement that they depend
functionally only on a finite set of velocity moments of the same KDFs.
Their numerical solution involves, at most, the determination of the local
gradients of the corresponding fluid fields. For comparison, the HRE-KDF
obeys, instead, a Fokker-Planck statistical equation which contains
non-local contributions due to the fluid pressure which depend by
higher-order multi-point PDFs (see for example Ref.\cite{Pope2000}).

4) Unlike the HRE approach, both IKT and TTP statistical approaches allow
the stochastic fluid pressure to be uniquely determined as a velocity moment
of the relevant KDFs. Its stochastic average and fluctuating part follow
simply by applying the stochastic average operator on the resulting
expression.

5) The choice of the stochastic variables $\mathbf{\alpha }$ remains in
principle arbitrary. Thus, they can be identified in accordance with the
specific stochastic model adopted (e.g., stochastic initial conditions,
stochastic boundary conditions or stochastic volume force).

6) Let us consider the comparison between the customary HRE statistical
evolution equation (see again for example Ref.\cite{Pope2000}) and the
corresponding stochastic-averaged Liouville equations following from Eqs.(%
\ref{IKE}) and (\ref{IKE-2}) upon applying the averaging operator $%
\left\langle \cdot \right\rangle _{\mathbf{\alpha }}$. As pointed out in
Section 6 and in Ref.\cite{TexRGD}, the latter equation can be approximated,
locally in velocity space, in terms of a Fokker-Planck equation. The
corresponding Kramers-Moyal coefficients however are different from those
entering in the HRE equation. The remarkable features of our equation is,
first, that unlike the HRE one, it recovers exactly the correct velocity
moment equations obtained for the weight functions $G=\left( 1,\mathbf{v}%
\right) $ (see Ref.\cite{TexRGD} on this issue) which also follow from the
corresponding stochastic-averaged Liouville equation (Eqs.(\ref{IKE}) and (%
\ref{IKE-2})). Second, the Kramers-Moyal coefficients depend explicitly also
on the stochastic pressure fluctuations $\delta p$, while the strict
positivity of the KDF is warranted by the weak H-theorem (\ref{H-THEOREM FOR
F_1-averaged2}) following from the exact stochastic-averaged Liouville
equations (\ref{IKE}) or (\ref{IKE-2}).

7) Another remarkable difference with respect to the HRE statistical
equation is that the Liouville equations as well as the corresponding
Fokker-Planck approximations presented here hold both for Gaussian and
suitably-smooth non-Gaussian KDFs. Therefore, both IKT and TTP statistical
models appear suitable to describe the non-Gaussian behavior arising in the
statistical description of turbulence.

\section{CONCLUSIONS}

A fundamental issue for Navier-Stokes fluids, is their characterization in
terms of the dynamics of ideal tracer particles (ITPs), and in particular of
the sub-set of thermal tracer particles (TTPs). Based on the formulation of
an inverse kinetic theory for compressible/incompressible NS thermofluids,
in this paper properties of TTP dynamics and a mathematical model for their
description have been investigated. It is found that TTP dynamics can be
\emph{uniquely} determined both for incompressible (isothermal or
non-isothermal) and compressible NS fluids described respectively by the
INSE, INSFE and CNSFE problems. In addition it has been proved that the
discovery of TTP dynamics allows for the construction of a reduced-dimension
statistical model, to be identified with the TTP-statistical model. The
latter is defined by the set $\left\{ f_{1},\Gamma _{1}\right\} $, in terms
of which the \emph{self-consistent} time evolution of the fluid fields is
determined.

Basic consequences of the theory mentioned in this paper concern the
treatment of stochastic fluid fields arising in compressible/incompressible
NS fluids by means of the TTP-statistical model. Here we have pointed out in
particular:

\begin{enumerate}
\item The formulation of TTP dynamics for the stochastic CNSFE problem,
represented in two possible forma. The first one is provided by Langevin
equations, which describe the dynamics of TTPs in fluctuating
compressible/compressible NS fluids. These provide a mathematical model of
tracer-particle motion in stochastic fluids. The second one is given by the
corresponding Fokker-Planck description.

\item The comparison with the statistical treatment of turbulence due to
Hopf, Rosen and Edwards \cite{Hopf1952,Rosen1960,Edwards1964}, which holds
in the case of incompressible isothermal NS fluids, has been carried out. As
a result, the relationship between the HRE, IKT and TTP velocity PDFs has
been pointed out.

\item Finally, based on the statistical model $\left\{ f_{1},\Gamma
_{1}\right\} $, a solution of the closure problem for the statistical
description of HT has been proposed.
\end{enumerate}

Applications of the present theory are in principle several. They concern,
in general, the dynamics of small particles (such as solid particles or
droplets, commonly found in natural\ phenomena and industrial applications)
in compressible/incompressible thermofluids. The accurate description of
particle dynamics, as they are pushed along erratic trajectories by
fluctuations of the fluid fields, is essential, for example, in combustion
processes, in the industrial production of nanoparticles as well as in
atmospheric pollutant transport, cloud formation and air-quality monitoring
of the atmosphere.

\section*{Acknowledgments}

This work was developed in the framework of current PRIN research projects
(2008 and 2009, Italian Ministry for Universities and Research, Italy), the
research projects of the Consortium for Magnetofluid Dynamics (University of
Trieste, Italy) and the GDRE (Groupe des Recherches Europ\'{e}enne) GAMAS,
C.N.R.S., France.





\section{Appendix A: Deterministic/stochastic NS thermofluids}

Let us consider for definiteness a viscous and generally non-isentropic
thermofluid either incompressible or compressible, described in both cases by%
{\ the fluid fields }(\ref{fluid fields}){. In the following, we shall
assume that the fluid fields are defined and suitably smooth in the
existence domain }$\Omega \times I$, with $\Omega $ and $I${\ denoting
respectively an open subset of the 3-dimensional Euclidean space }$%
\mathbb{R}
^{3}$ and a subset of the real axis $%
\mathbb{R}
$.

\subsection{A.1 - Case of a compressible fluid: CNSFE}

Let us consider the case of a compressible thermofluid. Denoting by $\frac{D%
}{Dt}=\frac{\partial }{\partial t}+\mathbf{V}\cdot \nabla $ the fluid
convective derivative, its fluid equations are identified with {the so-called%
} {\emph{compressible Navier-Stokes-Fourier equations} (\emph{CNSFE}) }
\begin{eqnarray}
&&\left. \frac{D\rho }{Dt}+\rho \nabla \cdot \mathbf{V}=0,\right.  \label{1A}
\\
&&\left. \frac{D}{Dt}\mathbf{V}=\mathbf{F}_{H},\right.  \label{1B} \\
&&\left. \frac{DT}{Dt}=K,\right.  \label{2B} \\
&&\left. \frac{\partial }{\partial t}S_{T}\geq 0,\right.  \label{5B} \\
&&\left. p(\mathbf{r},t)=p(\mathbf{r},t,\rho ,V,T),\right.  \label{6B}
\end{eqnarray}%
where Eqs.(\ref{1A})-(\ref{2B}) denote the \emph{mass continuity, forced
Navier-Stokes} and \emph{Fourier} \emph{equations }respectively;
furthermore,\ {the inequality (\ref{5B}) identifies\ the \emph{entropy law},
customarily known as the \emph{2nd principle of thermodynamics }}and Eq.(\ref%
{6B}) is the{\emph{\ equation of state }}for the fluid pressure{. }Here the
notation is standard. Thus, in particular in Eq.(\ref{1B}) $\mathbf{F}_{H}$
denotes the NS acceleration
\begin{equation}
\left. \mathbf{F}_{H}\equiv -\frac{1}{\rho }\left[ \nabla p-\mathbf{f}\right]
+\nabla \cdot \underline{\underline{\mathbf{\sigma }}}^{\prime }(\mathbf{r}%
,t)\right. ,  \label{fluid acceleration}
\end{equation}%
where the \emph{viscous stress tensor} $\underline{\underline{\mathbf{\sigma
}}}^{\prime }(\mathbf{r},t)$ is assumed of the form%
\begin{equation}
\underline{\underline{\mathbf{\sigma }}}^{\prime }(\mathbf{r},t)=\mu \left(
\nabla \mathbf{V+}\left( \nabla \mathbf{V}\right) ^{T}-\frac{2}{3}\underline{%
\underline{\mathbf{1}}}\nabla \cdot \mathbf{V}\right) +\lambda \underline{%
\underline{\mathbf{1}}}\nabla \cdot \mathbf{V},
\label{VISCOUS
STRESS TENSOR}
\end{equation}%
\noindent with $\mu $, $\lambda >0$ to be denoted as \emph{first} and \emph{%
second viscosity coefficients}. As a consequence
\begin{equation}
\nabla \cdot \underline{\underline{\mathbf{\sigma }}}^{\prime }(\mathbf{r}%
,t)=\nabla \cdot \left[ \mu \left( \nabla \mathbf{V+}\left( \nabla \mathbf{V}%
\right) ^{T}\right) \right] +\nabla \left[ \left( \lambda -\frac{2}{3}\mu
\right) \nabla \cdot \mathbf{V}\right] \mathbf{.}  \label{VISCOUS TERM-NS}
\end{equation}%
In addition, $\mathbf{f}$ denotes {the volume force density, which is
assumed to be a suitable smooth vector field of the general form }$\mathbf{f}%
=-\nabla \phi +\mathbf{f}_{R}$, with $\phi (\mathbf{r},t)$ and $\mathbf{f}%
_{R}(\mathbf{r},t;\left\{ Z(\mathbf{r},t)\right\} )$ denoting respectively a
scalar function (potential) and an additional non-conservative vector
generally dependent of the fluid fields and in particular the temperature $T(%
\mathbf{r},t)$. As an example, $\mathbf{f}$ can be identified with $\mathbf{f%
}=-\rho \mathbf{g}\left[ 1-k_{\rho }T\right] +\mathbf{f}_{R1}$, where $%
\mathbf{g}$ and\textbf{\ }$k_{\rho }$ are real constants denoting
respectively the local acceleration of gravity and the density
thermal-dilatation coefficient, $-\rho \mathbf{g}k_{\rho }T$ is the
temperature-dependent gravitational force density and $\mathbf{f}_{R1}(%
\mathbf{r},t;\left\{ Z(\mathbf{r},t)\right\} )$ is a possible additional
volume force density. Finally, the source term $K$ in Eq.(\ref{2B}) denotes
the heat production rate, defined as%
\begin{eqnarray}
K(\mathbf{r},t) &=&\frac{1}{n\left( c_{p}-\frac{\alpha }{n}p\right) }\left\{
nJ_{T}-\left( \beta _{T}n-\alpha T\right) \frac{Dp}{Dt}-p\nabla \cdot
\mathbf{V}\right. +  \label{TEMPERATURE SOURCE TERM} \\
&&\left. +\frac{\mu }{2}\left( \frac{\partial V_{i}}{\partial r_{k}}+\frac{%
\partial V_{k}}{\partial r_{i}}-\frac{2}{3}\delta _{ij}\frac{\partial V_{i}}{%
\partial r_{i}}\right) ^{2}+\lambda \left( \nabla \cdot \mathbf{V}\right)
^{2}+\nabla \cdot \left[ k\nabla T\right] \right\} ,  \notag
\end{eqnarray}%
where $c_{p}$ is the \emph{heat capacity at constant pressure} and $\alpha
,\beta _{T}$ are suitable (dimensional) \emph{phenomenological parameters}.
In the case of an incompressible fluid, denoting $\rho _{o}$ the constant
mass density, $\nu \equiv \mu /\rho _{o}$ the kinematic viscosity and
requiring $\lambda =\alpha =\beta _{T}=0$, the previous equations reduce to
the{\ \emph{incompressible Navier-Stokes-Fourier equations} (\emph{INSFE}):}%
\begin{eqnarray}
&&\left. \rho =\rho _{o}>0,\right.  \label{1} \\
&&\left. \nabla \cdot \mathbf{V}=0,\right.  \label{1b} \\
&&\left. \frac{D}{Dt}\mathbf{V}=\mathbf{F}_{H},\right.  \label{2} \\
&&\left. \frac{DT}{Dt}=K,\right.  \label{3} \\
&&\left. \frac{\partial }{\partial t}S_{T}\geq 0,\right.  \label{4b}
\end{eqnarray}%
where now%
\begin{equation}
\left. \mathbf{F}_{H}\equiv -\frac{1}{\rho _{o}}\left[ \nabla p-\mathbf{f}%
\right] +\nu \nabla ^{2}\mathbf{V}\right. ,  \label{fluid-acceleration}
\end{equation}%
\begin{equation}
K=\frac{1}{nc_{p}}\left[ nJ_{T}+\frac{\mu }{2}\left( \frac{\partial V_{i}}{%
\partial r_{k}}+\frac{\partial V_{k}}{\partial r_{i}}-\frac{2}{3}\delta _{ij}%
\frac{\partial V_{i}}{\partial r_{i}}\right) ^{2}+\lambda \left( \nabla
\cdot \mathbf{V}\right) ^{2}+\nabla \cdot \left[ k\nabla T\right] \right] .
\label{TEMPERATURE SOURCE INSFE}
\end{equation}%
Furthermore the equation of state (\ref{6B}) is replaced by the Poisson
equation for the fluid pressure. This is obtained by taking the divergence
of {the NS equation (\ref{2}), yielding}%
\begin{equation}
\nabla ^{2}p=-\rho \nabla \cdot \left( \mathbf{V}\cdot \nabla \mathbf{V}%
\right) +\nabla \cdot \mathbf{f},  \label{4B}
\end{equation}%
with $p$ to be assumed non negative and bounded in $\overline{\Omega }\times
{I}$.{\ }Finally, {we remark that Eqs.(\ref{1})-(\ref{4b}) include, as a
particular case, the treatment of \emph{isothermal fluids}. This is obtained}
assuming an initial spatially uniform temperature%
\begin{equation}
T(\mathbf{r},t_{o})=T_{o},  \label{ISOTHERMAL CONDITION}
\end{equation}%
requiring that for all $\left( \mathbf{r},t\right) \in \overline{\Omega }%
\times I$ the heat production rate\emph{\ }$K(\mathbf{r},t)$ is identically
zero in $\overline{\Omega }\times I$ and imposing, at the same time, the
\emph{isentropic law}%
\begin{equation}
\left. \frac{\partial }{\partial t}S_{T}=0\right. .
\label{isoentropic fluid}
\end{equation}%
Eqs.{(\ref{1})-(\ref{4b})} with the constraints (\ref{ISOTHERMAL CONDITION})
and (\ref{isoentropic fluid}) are denoted as (isothermal and) \emph{%
incompressible NS equations} (INSE).

\subsubsection{Equivalent forms of the Fourier equation}

To prove Eq.(\ref{2B}) with (\ref{TEMPERATURE SOURCE TERM}) let us start
from the law of energy conservation equation. For a compressible viscous
fluid this is \cite{Landau1959}:

\begin{equation}
\frac{\partial }{\partial t}\left( n\varepsilon +\frac{1}{2}\rho \mathbf{V}%
^{2}\right) +\nabla \cdot \left[ \mathbf{V}\left( n\varepsilon +\frac{1}{2}%
\rho \mathbf{V}^{2}+p\right) -\mathbf{V}\cdot \mathbf{\sigma }^{\prime
}-k\nabla T\right] =0,
\end{equation}%
with $n(\mathbf{r},t)$ and $\rho (\mathbf{r},t)$ denoting respectively the
number and mass densities, while $\varepsilon (\mathbf{r},t),k>0$ and $T(%
\mathbf{r},t)\geq 0$ are the \emph{internal energy density}, the \emph{%
thermal conductivity} and the \emph{temperature}.\ In terms of the
convective derivative this delivers%
\begin{equation}
n\frac{D}{Dt}\varepsilon =-\rho \frac{D}{Dt}\left( \frac{1}{2}\mathbf{V}%
^{2}\right) +\nabla \cdot \left[ -\mathbf{V}p+\mathbf{V}\cdot \mathbf{\sigma
}^{\prime }+k\nabla T\right] .  \label{AA-1}
\end{equation}%
On the other hand from the NS equation [see Eq.(\ref{1B})] it follows

\begin{equation}
\rho \frac{D}{Dt}\left( \frac{1}{2}\mathbf{V}^{2}\right) =-\mathbf{V}\cdot
\nabla p+\mathbf{V}\cdot \mathbf{f}+\mathbf{V}\cdot \nabla \cdot \underline{%
\underline{\mathbf{\sigma }}}^{\prime }(\mathbf{r},t),  \label{AA-3}
\end{equation}%
so that Eq.(\ref{AA-1}) recovers immediately\ the internal energy-transfer
equation%
\begin{eqnarray}
&&\left. \frac{D\varepsilon }{Dt}=S_{\varepsilon },\right.
\label{TOTAL HEAT PRODUCTION RATE} \\
&&\left. S_{\varepsilon }\equiv J_{T}+\frac{1}{n}\left\{ -p\nabla \cdot
\mathbf{V}+\nabla \cdot \left[ k\nabla T\right] +\left( \nabla \mathbf{V}%
\right) :\underline{\underline{\mathbf{\sigma }}}^{\prime }(\mathbf{r}%
,t)\right\} \right. ,  \notag
\end{eqnarray}%
where%
\begin{equation}
\left. J_{T}\equiv -\frac{1}{n}\mathbf{V}\cdot \mathbf{f}\right. ,
\label{AA-0}
\end{equation}%
and%
\begin{equation}
\left( \nabla \mathbf{V}\right) :\underline{\underline{\mathbf{\sigma }}}%
^{\prime }(\mathbf{r},t)=\frac{\mu }{2}\left( \frac{\partial V_{i}}{\partial
r_{k}}+\frac{\partial V_{k}}{\partial r_{i}}-\frac{2}{3}\delta _{ij}\frac{%
\partial V_{i}}{\partial r_{i}}\right) ^{2}+\lambda \left( \nabla \cdot
\mathbf{V}\right) ^{2}  \label{viscous energy production rate}
\end{equation}%
carry respectively the contributions to the due to the $\varepsilon $%
-production rate generated by the volume force $\mathbf{f}$ (i.e., external
sources) and viscous energy dissipation.

This equation, thanks to Eqs.(\ref{AA-2})-(\ref{AA-3}), can be equivalently
cast into an equation for the temperature $T$ of the form (\ref{2B}). In
fact, for generally non-isothermal and compressible fluids $\varepsilon (%
\mathbf{r},t)$ can be taken such that%
\begin{equation}
\frac{D}{Dt}\varepsilon =\left( c_{p}-\frac{\alpha }{n}p\right) \frac{DT}{Dt}%
+\left( \beta _{T}-\frac{\alpha T}{n}\right) \frac{Dp}{Dt}.  \label{AA-2}
\end{equation}%
Eq.(\ref{AA-2}) requires%
\begin{eqnarray}
&&\left. n\left( c_{p}-\frac{\alpha }{n}p\right) \frac{DT}{Dt}=-n\left(
\beta _{T}-\frac{\alpha T}{n}\right) \frac{Dp}{Dt}+\right.  \notag \\
&&-\rho \frac{D}{Dt}\left( \frac{1}{2}\mathbf{V}^{2}\right) +\nabla \cdot %
\left[ -\mathbf{V}p+\mathbf{V}\cdot \mathbf{\sigma }^{\prime }+k\nabla T%
\right] .
\end{eqnarray}%
Introducing the definition (\ref{AA-0}) and invoking Eq.(\ref{viscous energy
production rate}) this yields%
\begin{eqnarray}
&&\left. n\left( c_{p}-\frac{\alpha }{n}p\right) \frac{DT}{Dt}=nJ_{T}-\left(
\beta _{T}n-\alpha T\right) \frac{Dp}{Dt}-p\nabla \cdot \mathbf{V}+\right.
\notag \\
&&\left. +\frac{\mu }{2}\left( \frac{\partial V_{i}}{\partial r_{k}}+\frac{%
\partial V_{k}}{\partial r_{i}}-\frac{2}{3}\delta _{ij}\frac{\partial V_{i}}{%
\partial r_{i}}\right) ^{2}+\lambda \left( \nabla \cdot \mathbf{V}\right)
^{2}+\nabla \cdot \left[ k\nabla T\right] ,\right.
\end{eqnarray}%
and hence Eqs.(\ref{2B}) and (\ref{TEMPERATURE SOURCE TERM}). If we
introduce, instead, the representation of $\varepsilon (\mathbf{r},t)$ in
terms of the \emph{local entropy} $s(\mathbf{r},t)$:%
\begin{equation}
\frac{D}{Dt}\varepsilon =T\frac{Ds}{Dt}+\frac{p}{n}\frac{D\ln n}{Dt},
\end{equation}%
Eq.(\ref{AA-1}) delivers the \emph{local entropy equation}:%
\begin{eqnarray}
&&\left. nT\frac{Ds}{Dt}=S_{s},\right.  \label{ENTROPY DENSITY EQUATION} \\
&&S_{s}(\mathbf{r},t)\equiv nJ_{T}+\nabla \cdot \left[ k\nabla T\right] +%
\frac{\mu }{2}\left( \frac{\partial V_{i}}{\partial r_{k}}+\frac{\partial
V_{k}}{\partial r_{i}}-\frac{2}{3}\delta _{ij}\frac{\partial V_{i}}{\partial
r_{i}}\right) ^{2}+\lambda \left( \nabla \cdot \mathbf{V}\right) ^{2},
\label{ENTROPY DENSITY PRODUCTION RATE}
\end{eqnarray}%
with $S_{s}$ denoting the \emph{local entropy production rate.}

\subsubsection{Entropy law - Externally heated thermofluid}

Denoting the \emph{global thermodynamic entropy} $S_{T}(t)$ as%
\begin{equation}
S_{T}(t)\equiv \int\limits_{\Omega }d\mathbf{r}ns,
\label{THERMODYNAMIC ENTROPY}
\end{equation}%
it follows identically%
\begin{eqnarray}
&&\left. \frac{\partial }{\partial t}S_{T}(t)=\int\limits_{\Omega }d\mathbf{r%
}\frac{\partial }{\partial t}\left( ns\right) =\int\limits_{\Omega }d\mathbf{%
r}\left[ n\frac{\partial }{\partial t}s+s\frac{\partial }{\partial t}n\right]
=\right.  \notag \\
&&\left. =\int\limits_{\Omega }d\mathbf{r}\left[ -n\mathbf{V\cdot \nabla }%
s+S-s\nabla \cdot \left( n\mathbf{V}\right) \right] \equiv
\int\limits_{\Omega }d\mathbf{r}S_{s}.\right.
\end{eqnarray}%
Hence, subject to the requirement:%
\begin{equation}
\int\limits_{\Omega }d\mathbf{r}\frac{1}{T}J_{T}\geq 0,
\label{fluid
with external heating-0}
\end{equation}%
- usually referred to as (condition of) \emph{externally heated thermofluid }%
- the entropy law (\ref{5B}) follows, with
\begin{equation}
\frac{\partial }{\partial t}S_{T}(t)\equiv \int\limits_{\Omega }d\mathbf{r}%
\left[ \frac{1}{T}J_{T}+\frac{k\left( \nabla T\right) ^{2}}{T^{2}}+\frac{\mu
}{2T}\left( \frac{\partial V_{i}}{\partial r_{k}}+\frac{\partial V_{k}}{%
\partial r_{i}}-\frac{2}{3}\delta _{ij}\frac{\partial V_{i}}{\partial r_{i}}%
\right) ^{2}+\frac{\lambda }{T}\left( \nabla \cdot \mathbf{V}\right) ^{2}%
\right]  \label{THERMODYNAMIC ENTROPY PRODICTION RATE}
\end{equation}%
denoting the \emph{global thermodynamic entropy production rate}.

\subsection{A.2 - Deterministic and stochastic initial-boundary value
problems}

{The fluid equations defined by CNSFE are required to satisfy
initial-boundary value problems defined by appropriate }initial and
Dirichlet boundary conditions%
\begin{equation}
\left\{ \left.
\begin{array}{l}
Z(\mathbf{r},t_{o})=Z_{o}(\mathbf{r}), \\
\left. Z(\mathbf{r},t)\right\vert _{\partial \Omega }=\left. Z_{w}(\mathbf{r}%
,t)\right\vert _{\partial \Omega }.%
\end{array}%
\right. \right.  \label{initial-boundary conditions}
\end{equation}%
{\ In particular, denoting }$I\subseteq
\mathbb{R}
$ a suitable subset of the real axis, {we shall require that }$S_{T}(t)$ is
defined and smooth for all $t\in I$ and that {a smooth (strong) solution
exists for the previous initial-boundary value problem in} $\Gamma \times I$
(existence domain).

Finally, we shall assume that the fluid fields $\left\{ Z\right\} $,
together with the volume force density $\mathbf{f}$ and the initial and
boundary fields $Z_{o}$ and $Z_{w}$ \textit{are all stochastic functions} of
the form (see Appendix B)%
\begin{equation}
\left\{
\begin{array}{l}
\left. Z=Z(\mathbf{r},t,\mathbf{\alpha }),\right. \\
\left. \mathbf{f}=\mathbf{f}(\mathbf{r},t,\mathbf{\alpha }),\right. \\
\left. Z_{o}=Z_{o}(\mathbf{r},\mathbf{\alpha }),\right. \\
\left. Z_{w}=\left. Z_{w}(\mathbf{r},t,\mathbf{\alpha })\right\vert
_{\partial \Omega },\right.%
\end{array}%
\right.  \label{FUNCTIONS-2}
\end{equation}%
with $\mathbf{\alpha }\in V_{\mathbf{\alpha }}$ stochastic variables
independent of $\left( \mathbf{r},t\right) $. Depending whether the previous
functions (\ref{FUNCTIONS-2}) are considered deterministic or stochastic,
the previous initial-boundary-value problems{\ defined either by:}

\begin{enumerate}
\item \emph{Eqs.(\ref{1A})-(\ref{5B}) with the initial-boundary conditions (%
\ref{initial-boundary conditions}), }

or:

\item \emph{Eqs.(\ref{1})-(\ref{4b}) with (\ref{4B}) and the
initial-boundary conditions (\ref{initial-boundary conditions}),}

\item \emph{Eqs.(\ref{1})-(\ref{2}) with (\ref{4B}) and together with the
constraints (\ref{ISOTHERMAL CONDITION}), (\ref{isoentropic fluid}) and the
initial-boundary conditions (\ref{initial-boundary conditions}), }

will be denoted, respectively,{\ as \emph{deterministic }}or{\ }\emph{%
stochastic CNSFE, INSFE and INSE problems}{.}
\end{enumerate}

\subsection{A.3 - Equivalent stochastic fluid equations}

For a prescribed stochastic model $\left\{ g,V_{\mathbf{\alpha }}\right\} $,
with $g=g(\mathbf{r},t,\mathbf{\alpha })$ being a stochastic probability
density on $V_{\mathbf{\alpha }}$, in terms of stochastic decomposition (\ref%
{STOCHASTIC DECOMPOSITION-2}) it is immediate to obtain the equations for
the average and stochastic fluid fields $\left\langle Z\right\rangle _{%
\mathbf{\alpha }}$, $\delta Z$ and the corresponding initial-boundary value
problem. For example, in the case of INSE, requiring that $\left\langle
\cdot \right\rangle _{\mathbf{\alpha }}$ commutes with the nabla operator $%
\nabla $, Laplacian $\nabla ^{2}$ and partial time derivative $\frac{%
\partial }{\partial t}$ operators, the fluid equations for $\left\langle
Z\right\rangle _{\alpha }$ and $\delta Z$, to be referred to as \emph{%
stochastic incompressible NS equations},\ become respectively%
\begin{eqnarray}
&&\left. \left\langle \frac{D}{Dt}\mathbf{V}\right\rangle _{\alpha
}=\left\langle \mathbf{F}_{H}\right\rangle _{\alpha }-\frac{1}{\rho _{o}}%
\left[ \nabla \left\langle p\right\rangle _{\alpha }-\left\langle \mathbf{f}%
\right\rangle \right] +\nu \nabla ^{2}\left\langle \mathbf{V}\right\rangle
_{\alpha },\right.  \label{average NS} \\
&&\left. \nabla \cdot \left\langle \mathbf{V}\right\rangle _{\alpha
}=0\right. ,  \label{physical realiz.2-average}
\end{eqnarray}%
\begin{eqnarray}
&&\left. \frac{D}{Dt}\mathbf{V}-\left\langle \frac{D}{Dt}\mathbf{V}%
\right\rangle _{\alpha }=\delta \mathbf{F}_{H}-\frac{1}{\rho _{o}}\left[
\nabla \delta p-\delta \mathbf{f}\right] +\nu \nabla ^{2}\delta \mathbf{V}%
,\right.  \label{stoch. INSE} \\
&&\left. \nabla \cdot \delta \mathbf{V}=0.\right.
\label{physical
realiz. 2 - stoch}
\end{eqnarray}%
In particular Eqs.(\ref{average NS})-(\ref{physical realiz.2-average})
identify the so-called \emph{stochastic-averaged INSE}.

\section{Appendix B - \textbf{Stochastic/deterministic variables and
stochastic models}}

Let $(S,\Sigma ,P)$ be a probability space; a measurable function $\mathbf{%
\alpha }:S\longrightarrow V_{\mathbf{\alpha }}$, where $V_{\mathbf{\alpha }%
}\subseteq
\mathbb{R}
^{k}$ and $k\geq 1$, is called \textit{stochastic} (or \textit{random})
\textit{variable}.

A stochastic variable $\mathbf{\alpha }$\ is called \textit{continuous} if%
\textit{\ }it is endowed with a \textit{continuous} \textit{stochastic model}
$\left\{ \mathbf{\alpha }\in V_{\mathbf{\alpha }},g(\mathbf{r},t,\mathbf{%
\alpha })\right\} $,\textit{\ }namely a real continuous function\textit{\ }$%
g $, called \textit{PDF\ on the set} $V_{\mathbf{\alpha }}$, such that:

1) $g$ is measurable, non-negative and of the form%
\begin{equation}
\left. g=g(\mathbf{r},t,\cdot );\right.  \label{stochastic PDF}
\end{equation}

2) if $A\subseteq V_{\mathbf{\alpha }}$ is an arbitrary Borelian subset of $%
V_{\mathbf{\alpha }}$ (written $A\in \mathcal{B}(V_{\mathbf{\alpha }})$),
the integral
\begin{equation}
\left. P_{\mathbf{\alpha }}(A)=\int\limits_{A}d\mathbf{\alpha }g(\mathbf{r}%
,t,\mathbf{\alpha })\right.  \label{dist-of-alpha}
\end{equation}%
exists and is the probability that $\mathbf{\alpha }\in A$; in particular,
since $\mathbf{\alpha }\in V_{\mathbf{\alpha }}$, $g$ admits the
normalization%
\begin{equation}
\left. \int\limits_{V_{\mathbf{\alpha }}}d\mathbf{\alpha }g(\mathbf{r},t,%
\mathbf{\alpha })=P_{\mathbf{\alpha }}(V_{\mathbf{\alpha }})=1\right. .
\label{normalization}
\end{equation}

The set function $P_{\mathbf{\alpha }}:\mathcal{B}(V_{\mathbf{\alpha }%
})\rightarrow \lbrack 0,1]$ defined by (\ref{dist-of-alpha}) is a
probability measure on $V_{\mathbf{\alpha }}$. Consequently, if a function $%
f:V_{\mathbf{\alpha }}\longrightarrow V_{f}\subseteq
\mathbb{R}
^{m}$ is measurable, $f$ is a stochastic variable too.

Then we define the \textit{stochastic-averaging operator
}$\left\langle \cdot \right\rangle _{\mathbf{\alpha }}$(see also
Refs.\cite{Tessarotto20081,Tessarotto2009}) as%
\begin{equation}
\left\langle f\right\rangle _{\mathbf{\alpha }}=\left\langle f(\mathbf{y}%
,\cdot )\right\rangle _{\mathbf{\alpha }}\equiv \int\limits_{V_{\mathbf{%
\alpha }}}d\mathbf{\alpha }g(\mathbf{r},t,\mathbf{\alpha })f(\mathbf{y},%
\mathbf{\alpha }),  \label{stochastic averaging operator}
\end{equation}%
for any $P_{\mathbf{\alpha }}$-integrable function $f(\mathbf{y},\cdot ):V_{%
\mathbf{\alpha }}\rightarrow
\mathbb{R}
$, where the vector $\mathbf{y}$ is some parameter.

The ensemble $\left\{ g(\mathbf{r},t,\mathbf{\alpha }),V_{\mathbf{\alpha }%
}\right\} $ is denoted as stochastic model. Examples of stochastic model are
represented by statistical models. In such a case the stochastic variables $%
\mathbf{\alpha }$ are identified with \emph{hidden variables, }i.e.,
variables from which the fluid fields $\left\{ Z\right\} $ depend only
implicitly. For example, in the IKT statistical models $\left\{ f,\Gamma
\right\} $ and $\left\{ f_{1},\Gamma _{1}\right\} $ (see THMs.1 and 3, in
Sections 4 and 6) the stochastic variables and PDF are prescribed letting $%
\mathbf{\alpha }\equiv \mathbf{u}$ and $g(\mathbf{r},t,\mathbf{\alpha }%
)\equiv \widehat{f}(\mathbf{r},\mathbf{u},t)$ or $g(\mathbf{r},t,\mathbf{%
\alpha })\equiv \widehat{f}_{1}(\mathbf{r},\mathbf{u},t)$, with $\widehat{f}(%
\mathbf{r},\mathbf{u},t)$ and $\widehat{f}_{1}(\mathbf{r},\mathbf{u},t)$
being defined, respectively, by Eqs. (\ref{PDF}) and (\ref{POSITION-0}).
Furthermore, in the two cases the set $V_{\mathbf{\alpha }}$ is identified
with the velocity spaces $U$ or $U_{1}$.

A classification of stochastic models can be given in terms of the defining
PDF $g$ as follows.

\textbf{Definition - Homogeneous, stationary, deterministic and stochastic
PDF}

The PDF $g$ is denoted:

a) \textit{homogeneous }if $g$ is independent of $\mathbf{r},$ namely $g=g(t,%
\mathbf{\alpha })$;

b) \textit{isotropic} if $g$ is a function of the form $g=g(\left\vert
\mathbf{r}\right\vert ,t,\mathbf{\alpha })$;

c) \textit{stationary} if $g$ is independent of $t$, i.e., $g=g(\mathbf{r},%
\mathbf{\alpha })$;

d) \textit{deterministic} if $g$ is a distribution on $V_{\mathbf{\alpha }}$
of the form$g(\mathbf{r},t,\mathbf{\alpha })=\delta ^{(k)}(\mathbf{\alpha }-%
\mathbf{\alpha }_{o})$, with $\delta ^{(k)}(\mathbf{\alpha }-\mathbf{\alpha }%
_{o})$ denoting the $k$-dimensional Dirac delta on the space $V_{\mathbf{%
\alpha }}$;

e) \textit{stochastic} if $g$ is an ordinary function on the space $V_{%
\mathbf{\alpha }}$.

\end{document}